\documentclass[
amsmath,amssymb,
prd,nofootinbib,floatfix,12pt,
]{revtex4}

\usepackage{amsmath,amssymb,amsfonts}
\usepackage[dvips]{color}
\usepackage{xcolor}
\usepackage{dsfont}
\usepackage{mathrsfs}
\usepackage{float}
\usepackage{hyperref}
\usepackage{graphicx}
\usepackage{graphics}
\usepackage{setspace}
\usepackage{lmodern}

\usepackage{scalerel}

\hypersetup{colorlinks,citecolor=blue,urlcolor=blue, linkcolor=black}
\newcommand\beq{\begin{eqnarray}}
\newcommand\eeq{\end{eqnarray}}

\def\gsim{\mathrel{\rlap{\lower4pt\hbox{$\sim$}}\raise1pt\hbox{$>$}}}

\def\CLexcl{{\rm CL}_{\rm excl}}
\def\CLexclA{{\rm CL}_{\rm excl}^A}
\def\CLdisc{{\rm CL}_{\rm disc}}
\def\CLdiscA{{\rm CL}_{\rm disc}^A}
\def\lsim{\mathrel{\rlap{\lower4pt\hbox{$\sim$}}
    \raise1pt\hbox{$<$}}}                
\def\gsim{\mathrel{\rlap{\lower4pt\hbox{$\sim$}}
    \raise1pt\hbox{$>$}}}

\allowdisplaybreaks
\begin{document}
\renewcommand{\theequation}{\arabic{section}.\arabic{equation}}
\renewcommand{\thefigure}{\arabic{section}.\arabic{figure}}
\renewcommand{\thetable}{\arabic{section}.\arabic{table}}

\title{\large Statistical significances and projections\\for proton decay experiments}

\author{Prudhvi N. Bhattiprolu$^{\dag}$,
        Stephen P. Martin$^{\ddag}$,
        James D. Wells$^{\dag}$
}

\affiliation{\it \baselineskip=18pt
  $^{\dag}$ Leinweber Center for Theoretical Physics, University of Michigan, Ann Arbor MI 48109, USA\\
  $^{\ddag}$ Department of Physics, Northern Illinois University, DeKalb IL 60115, USA
}

\begin{abstract}\normalsize \baselineskip=17pt
We study the statistical significances for exclusion and discovery of proton decay at current and future neutrino detectors. Various counterintuitive flaws associated with frequentist and modified frequentist statistical measures of significance for multi-channel counting experiments are discussed in a general context and illustrated with examples.
We argue in favor of conservative Bayesian-motivated statistical measures,
and as an application we employ these measures to obtain the
current lower limits on proton partial lifetime at various confidence levels, based on
Super-Kamiokande's data, generalizing the 90\% CL published limits.
Finally, we present projections for exclusion and discovery reaches for proton partial lifetimes in $p \rightarrow \overline \nu K^+$ and $p \rightarrow e^+ \pi^0$ decay channels at Hyper-Kamiokande, DUNE, JUNO,  and THEIA.
\end{abstract}

\maketitle

\tableofcontents

\baselineskip=17pt

\newpage

\section{Introduction\label{sec:intro}}
\setcounter{equation}{0}
\setcounter{figure}{0}
\setcounter{table}{0}
\setcounter{footnote}{1}

In order to account for the observed matter-antimatter asymmetry in our universe,
baryon number must be violated as required by the
Sakharov conditions \cite{Sakharov:1967dj}.
Although baryon number is a global symmetry of the (renormalizable) Standard Model (SM) Lagrangian,
it may be violated by non-perturbative electroweak sphaleron effects (as yet unconfirmed by experiment)
that are heavily suppressed at temperatures much lower than the electroweak
scale \cite{tHooft:1976rip,Kuzmin:1985mm}.
The sphaleron effects, however, together with the CP-violation in the electroweak sector are not
sufficient to explain the observed baryon asymmetry, and therefore provide a key motivation for
theories beyond the SM with additional $B$-violation.
Grand unified theories (GUTs), with or without supersymmetry, are well-motivated
and generically predict baryon number violation, and therefore can lead to proton
decay \cite{Pati:1973rp,Georgi:1974sy,Langacker:1980js,Dorsner:2019vgf,FileviezPerez:2016sal,Dorsner:2005fq,Bajc:2006ia,Saad:2019vjo,ArbelaezRodriguez:2013kxw,Altarelli:2013aqa,Babu:2015bna,Bertolini:2012im,Preda:2022izo,Chakrabortty:2019fov,Dimopoulos:1981zb,Dimopoulos:1981dw,Sakai:1981pk,Hisano:1992jj,Ellis:2019fwf,Babu:2012pb,Nath:1985ub,Nath:1998kg,Liu:2013ula,Pati:2003qia,Ellis:2002vk,Ellis:2021vpp,Ellis:2020qad,Arkani-Hamed:2004zhs,Hebecker:2002rc,Alciati:2005ur,Babu:2018tfi,Babu:2018qca,Mohapatra:2018biy,BhupalDev:2010he,Shafi:1999vm,Lucas:1996bc,Babu:1997js,Babu:1998wi,Pati:2000wu,BhupalDev:2012nm}.
After integrating out the heavy fields, the non-renormalizable operators built out of the SM fields that allow
proton decay are of dimension-six or higher, with the suppression scale of order the GUT breaking scale.

In this paper, we consider proton decay
in the $p \rightarrow \overline{\nu} K^+$ and $p \rightarrow e^+ \pi^0$ decay channels that are
typically predicted to be the leading modes in supersymmetric \cite{Dimopoulos:1981zb,Dimopoulos:1981dw,Sakai:1981pk,Hisano:1992jj,Ellis:2019fwf,Babu:2012pb,Nath:1985ub,Nath:1998kg,Liu:2013ula,Pati:2003qia,Ellis:2002vk,Ellis:2021vpp,Ellis:2020qad,Arkani-Hamed:2004zhs,Hebecker:2002rc,Alciati:2005ur,Babu:2018tfi,Babu:2018qca,Mohapatra:2018biy,BhupalDev:2010he,Shafi:1999vm,Lucas:1996bc,Babu:1997js,Babu:1998wi,Pati:2000wu,BhupalDev:2012nm} and non-supersymmetric \cite{Georgi:1974sy,Langacker:1980js,Dorsner:2019vgf,FileviezPerez:2016sal,Dorsner:2005fq,Bajc:2006ia,Saad:2019vjo,ArbelaezRodriguez:2013kxw,Altarelli:2013aqa,Babu:2015bna,Bertolini:2012im,Preda:2022izo,Chakrabortty:2019fov} GUTs, respectively. At present, the strongest constraints on these proton partial lifetimes
are from the Super-Kamioka neutrino detection experiment
(Super-Kamiokande), where the most stringent published 90\% CL lower limits are $5.9 \times 10^{33}$ years for the $p \rightarrow \overline{\nu} K^+$ mode \cite{Super-Kamiokande:2014otb} and
$2.4 \times 10^{34}$ years for the $p \rightarrow e^+ \pi^0$ mode \cite{Super-Kamiokande:2020wjk}.
We will make projections for the exclusion and discovery reaches for these proton modes decays at future neutrino detectors at 
Deep Underground Neutrino Experiment (DUNE) \cite{DUNE:2015lol}, 
Jiangmen Underground Neutrino Observatory (JUNO) \cite{JUNO:2015zny},
Hyper-Kamiokande (the successor to Super-Kamiokande, and an order of magnitude larger) \cite{Hyper-Kamiokande:2018ofw},
and THEIA (a novel detector concept with water based liquid scintillator, 10\% liquid scintillator and 90\% water, that can detect and distinguish
between Cerenkov and the scintillation light) \cite{Theia:2019non}. 

In order to project the exclusion and discovery reaches, it is necessary to make choices regarding the statistical tools to be employed. Indeed, the results for such projections are only meaningful in the context of those choices. Here, we are interested in counting experiments with multiple independent channels with different signal rates and backgrounds, with uncertainties. 

Our statistical analysis choices are guided by several requirements. 
\begin{itemize}
\item We aim for statistical measures that avoid reporting an exclusion or discovery when the experiment is actually not sensitive to the physics signal hypothesis under investigation. As we will discuss, pure frequentist statistics can suffer from this problem. 
\item We choose statistical measures such that the presence of a non-informative channel (one with a much higher background and/or a much lower signal rate than other channels) does not unduly affect the exclusion or discovery conclusion. 
\item We avoid statistical measures that contain the subtle flaw that they could counterintuitively imply a greater sensitivity for an experiment if it increases its background. 
\end{itemize}
Regarding this last point, in a previous paper \cite{Bhattiprolu:2020mwi}, we have discussed the fact that the median expected significance for discovery or exclusion has just such a counterintuitive flaw in the context of frequentist $p$-values for a single-channel counting experiment. We proposed a solution to that problem. As we will see below, this type of problem also occurs in the case of multi-channel counting experiments, and can be avoided using Bayesian-motivated statistical measures. 

For these reasons, section \ref{sec:statistics} of this paper is devoted to a rather extensive discussion of the statistical issues associated with multi-channel counting experiments with background and nuisance parameter uncertainties, in which we highlight some of the problems that can occur and explain our choices of statistical tools in a general context. In Section \ref{sec:protondecay} we apply these statistical measures to discuss the present exclusions from Super-Kamiokande, and we project exclusion and discovery prospects for proton decay at DUNE, JUNO, Hyper-Kamiokande, and THEIA, for the proton decay modes $p \rightarrow e^+ \pi^0$ and $p \rightarrow \overline \nu K^+$. Section \ref{sec:outlook} summarizes our findings for exclusion and discovery prospects for run-times of 10 and 20 years.

\section{Statistics for discovery and exclusion\label{sec:statistics}}
\setcounter{equation}{0}
\setcounter{figure}{0}
\setcounter{table}{0}
\setcounter{footnote}{1}

\subsection{Basic definitions}

In this paper we are concerned with new physics signals and backgrounds, which are both assumed to occur as random discrete events governed by Poisson statistics, possibly in multiple independent channels. In general, given data resulting from an experiment, the significance of a possible exclusion or discovery can be given in terms of a $p$-value, defined as the probability of obtaining a result of equal or greater incompatibility with a null hypothesis $H_0$. In high-energy physics, the $p$-value is often conventionally reported as a significance, defined by
\beq
Z &\equiv& \sqrt{2} \ {\rm erfc}^{-1}\left(2 p\right),
\label{eq:Zfromp}
\eeq
which in the special case of a Gaussian distribution would coincide with the number of standard deviations. 

The assumption for discovery is that the null hypothesis is a background-only hypothesis $H_0 = H_{b}$, while for exclusion the null hypothesis is a signal plus background model $H_0 = H_{s+b}$. Consider a test-statistic $Q$ defined in such a way that larger $Q$ is more signal-like and smaller $Q$ is more background-like. In a single-channel counting experiment,  for example, $Q$ is simply the number of observed events. Then, for an experimental outcome $Q_{\rm obs}$, one has the $p$-value for discovery:
\beq
p_{\rm disc} &=& P(Q \geq Q_{\rm obs} | H_{b})
,
\eeq
and the $p$-value for exclusion:
\beq
p_{\rm excl} &=& P(Q \leq Q_{\rm obs} | H_{s+b})
.
\eeq
In a frequentist approach, the $p$-value for a given data outcome is often used to provide a quantitative measure of the credence we give to $H_0$. However, the $p$-value cannot be directly interpreted as the probability that the null hypothesis is true, given the data. Nevertheless, small $p$-values are considered a measure of evidence against $H_0$ in frequentist statistics. In particle physics, two popular standards for exclusion are to require that $p_{\rm excl} < 0.10$ or $0.05$, commonly referred to as 90\% or 95\% exclusion. For rejection of the background-only hypothesis in favor of some new model, a higher standard is almost always required, with either $Z_{\rm disc} > 3$ $(p_{\rm disc} < 0.001350)$ for ``evidence", or $Z_{\rm disc}>5$ $(p_{\rm disc} < 2.867 \times 10^{-7})$ for ``discovery".

In high energy physics experiments in the 21st century, starting with the Higgs boson searches at the LEP $e^-e^+$ collider and for all kinds of searches for new phenomena at the Large Hadron Collider (LHC), it has become very common to use a modified frequentist statistical measure for exclusion, called the CL$_s$ method. This is a more conservative approach to assigning exclusion significances than $p_{\rm excl}$. 
The idea of CL$_s$ \cite{Zech:1988un,Junk:1999kv,Read:2000ru,Read:2002hq} is to divide the usual $p$-value for exclusion by the $p$-value that would be obtained with the signal assumed absent:
\beq
{\rm CL}_s (Q_{\rm obs}) &=& 
\frac{P(Q \leq Q_{\rm obs} | H_{s+b})}{P(Q \leq Q_{\rm obs} | H_{b})}.
\label{eq:CLs_general}
\eeq
A specific motivation for using CL$_s$ rather than $p_{\rm excl}$ is to avoid reporting an exclusion in cases for which the experiment is actually not sensitive to the purported signal hypothesis, but the observed data has a small $p$-value anyway. This can occur, for example, in a counting experiment if the observed number of events is significantly smaller than the background estimate, as we will discuss in detail shortly.

Note that, by design, CL$_s$ is not a $p$-value or even a probability, but rather a ratio of probabilities. Nevertheless, the exclusion is reported using CL$_s$ in place of the exclusion $p$-value, so that one reports 95\% (or 90\%) exclusion if ${\rm CL}_s < 0.05$ (or $0.1$). Because the denominator is always less than 1, the modified frequentist measure CL$_s$ is always more conservative in reporting exclusions than the frequentist $p$ value, in the sense that using it reduces the false exclusion rate compared to using $p_{\rm excl}$. In particle physics literature, ${\rm CL}_s$ was introduced in ref.~\cite{Zech:1988un} and detailed (along with its advantages, reviewed and illustrated below) in refs.~\cite{Junk:1999kv, Read:2000ru,Read:2002hq}.

It is also useful to have a counterpart to the $p_{\rm disc}$ statistic that similarly guards against claiming discovery in situations where the experiment is not sensitive to the signal model. In ref.~\cite{Berger:2011fuz}, an approach to discovery significance was proposed using the Bayes factor \cite{Cowan:BayesFactors,Kass:1995loi,Jeffreys:BayesFactors} of the null hypothesis $H_0 = H_{b}$ to the alternative hypothesis $H_1 = H_{s+b}$. For an experiment investigating a putative signal with strength $s$, the Bayes factor $B_{01}$ is (using the probabilities in place of the likelihoods, to which they are proportional):
\beq
B_{0 1}  &=& \frac{P(Q_{\rm obs} | H_b)}{\displaystyle 
\displaystyle\scaleobj{0.85}{\int_{\scaleobj{1.2}{0}}^{\scaleobj{1.2}{\infty}}} ds^\prime 
\, \pi(s^\prime) 
\, P(Q_{\rm obs} |H_{s^\prime + b} )} ,
\label{eq:Bayesfactor}
\eeq
where $\pi(s')$ is a Bayesian prior probability distribution for the signal strength.
As mentioned in \cite{Berger:2011fuz}, this expression is only meaningful in the case of a prior that is proper, i.e. $\int_0^\infty ds^\prime \, \pi(s^\prime) = 1$, since otherwise the arbitrary normalization of an improper prior would make the Bayes factor $B_{01}$ also arbitrary. This precludes the use of a flat prior, for example. For a single-channel counting experiment with background mean $b$, that reference argues in favor of the proper prior $\pi(s^\prime) = 
{b}/{(s^\prime + b)^2}$, referred to as the objective signal prior. However, we find it counterintuitive to use a prior for the signal that depends on the background. Instead, we choose simply $\pi(s') = \delta(s' - s)$, expressing certainty in the prediction of the signal model. If the signal model prediction is not perfectly well known, it is straightforward to generalize this with an appropriate $\pi(s')$.
We therefore define the simple likelihood ratio statistic for the confidence level in  the discovery,
\beq
\CLdisc(Q_{\rm obs}) &=& \frac{P(Q_{\rm obs}|H_b)}{P(Q_{\rm obs}|H_{s+b})}
.
\label{eq:CLdisc_general}
\eeq
While various scales have been proposed (see e.g.~Jeffreys' in \cite{Jeffreys:BayesFactors} and
Kass and Raftery's in \cite{Kass:1995loi}) to
interpret the Bayes factor as a measure of evidence in favor of or against
a null hypothesis, we propose to use $\CLdisc$ in place of $p$ in eq.~(\ref{eq:Zfromp}) to obtain a discovery significance $Z$, in exactly the same way that a frequentist $p_{\rm disc}$ would be used. As we will illustrate below, our choice gives results that
are always more conservative than the significances obtained from $p_{\rm disc}$.
This is very similar to the way the modified frequentist measure ${\rm CL}_s$ is now commonly used in place of $p$ in eq.~(\ref{eq:Zfromp}) to report an exclusion significance that is always
more conservative than that of the standard frequentist method, even though ${\rm CL}_s$,
like $\CLdisc$, is not a probability.

\subsection{Single-channel counting experiments\label{subsec:singlechannel}}
 
To illustrate the statistical methods discussed above let us consider the special case of a simple experiment that counts the number of events $n$, with signal and background modeled as independent Poisson processes with means $s$ and $b$ respectively. For a mean $\mu$, the Poisson probability to observe $n$ events is
\beq
P(n | \mu) &=& \frac{\mu^n e^{-\mu}}{n!}.
\eeq
Therefore, in the idealized case of perfectly known background, the 
$p$-value for discovery is the probability that data generated under hypothesis $H_0 = H_{b}$
is equally or more signal-like than the actual observed number of events $n$:
\beq
p_{\rm disc} (n, b) = \sum_{k=n}^\infty P(k|b) = {\gamma(n, b)}/{\Gamma(n)}.
\label{eq:pdisc_singlechannel}
\eeq
The $p$-value for exclusion is the probability that data generated under hypothesis $H_0 = H_{s+b}$ is equally or more background-like
than the actual observed number of events $n$:
\beq
p_{\rm excl} (n, b, s) = \sum_{k=0}^n P(k |s + b) = {\Gamma(n + 1, s + b)}/{\Gamma(n + 1)}.
\label{eq:pexcl_singlechannel}
\eeq
In these equations, $\gamma(z, x)$ and $\Gamma(z, x)$ are the lower and upper incomplete gamma functions, respectively,
defined by
\beq
\gamma (z,x) \>=\> \int_0^x dt \> t^{z-1} e^{-t},
\qquad
\Gamma (z,x) \>=\> \int_x^\infty dt \> t^{z-1} e^{-t}, 
\eeq
so that $\Gamma(z) = \gamma(z,x) + \Gamma(z,x)$ is the ordinary gamma function.

The CL$_s$ statistic for exclusion in this case is
\beq
{\rm CL}_s (n,b,s) &=& 
\frac{p_{\rm excl}(n,b,s)}{p_{\rm excl}(n,b,0)}
\>=\>
\frac{\Gamma(n+1, s+b)}{\Gamma(n+1, b)}
.
\label{eq:CLs_singlechannel}
\eeq
This is larger than $p_{\rm excl}(n,b,s)$ by a factor $\Gamma(n+1)/\Gamma(n+1,b)$.

\begin{figure}
\includegraphics[width=12.5cm]{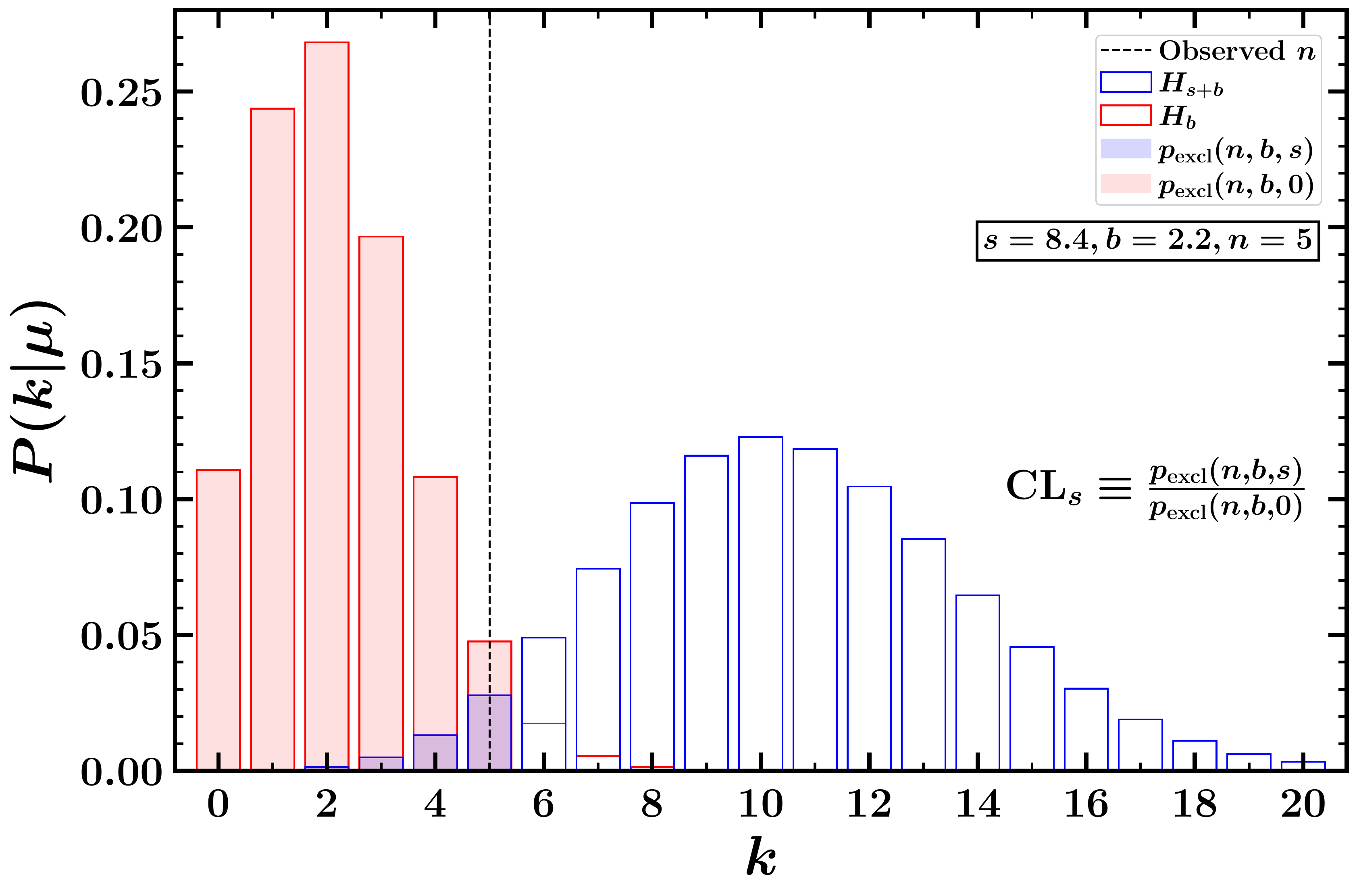}
\includegraphics[width=12.5cm]{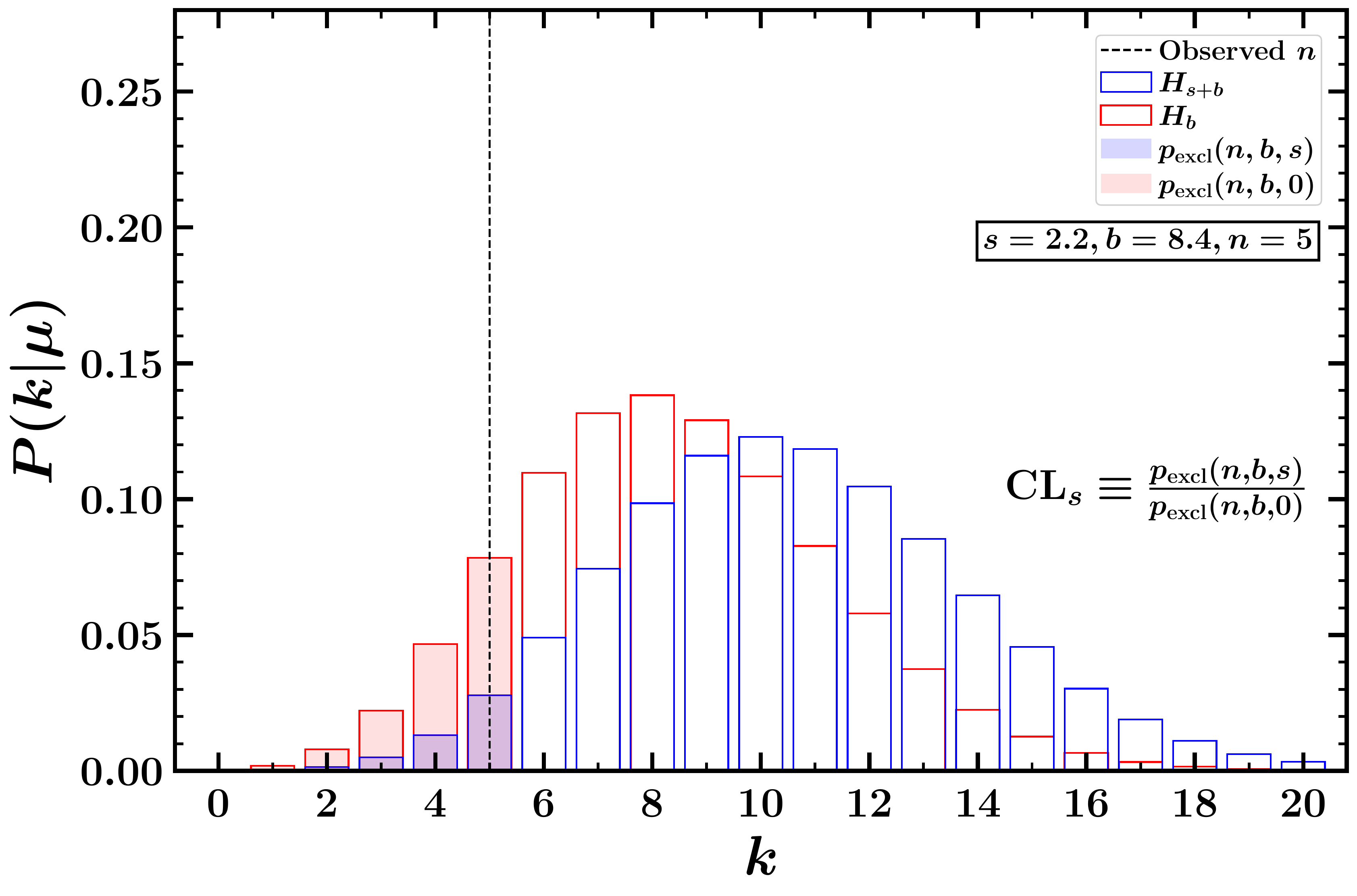}
\begin{minipage}[]{0.96\linewidth}
\caption{Illustration of the idea of the ${\rm CL}_s$ statistic for exclusion as an improvement over $p_{\rm excl}$. 
The Poisson distributions $P(k|\mu)$ are generated under the hypotheses that signal and background are both present $\mu = s + b$ (blue histograms) and that the signal is absent $\mu = b$ (red histograms). For the observed number of events $n$, $p_{\rm excl}(n,b,s)$ [from eq.~(\ref{eq:pexcl_singlechannel})] is shown by
the shaded area part under the blue histogram, and $p_{\rm excl}(n,b,0)$ is the shaded area part
under the red histogram, while CL$_s$ [from eq.~(\ref{eq:CLs_singlechannel})] is their ratio. 
In the first plot, the Poisson means of the signal and background are taken to be $s=8.4$ and $b=2.2$, respectively, while in the second plot they are $s=2.2$ and $b=8.4$. In both plots,
the observed number of events is $n=5$. In the first plot, there is little overlap between the
distributions from the $H_b$ and $H_{s+b}$ hypotheses, and $p_{\rm excl} = 0.0475$ and CL$_s = 0.0487$, so one would report better than 95\% exclusion using either criterion. In the second plot, the overlap is much larger. Although $p_{\rm excl} = 0.0475$ is the same (since $s+b$ and $n$ did not change), one finds CL$_s = 0.3022$, and one refrains from reporting an exclusion of the hypothesis $H_{s+b}$.
\label{fig:CLs_illustration}}
\end{minipage}
\end{figure}

Figure~\ref{fig:CLs_illustration} illustrates the idea of the ${\rm CL}_s$ method
\cite{Zech:1988un,Junk:1999kv, Read:2000ru,Read:2002hq}. In the figure,
$p_{\rm excl}(n,b,s)$ (the shaded area under the blue histograms) is divided by 
$p_{\rm excl}(n,b,0)$ (the shaded area under the red histograms) to give ${\rm CL}_s$. 
The first panel shows the case $b=2.2$, $s=8.4$, and $n=5$. In situations like this, where the 
$H_b$ and $H_{s+b}$ hypothesis distributions do not have much overlap, $p_{\rm excl}$ and CL$_{\rm s}$ evaluate to very similar results due to the denominator of the CL$_{\rm s}$ definition being close to 1.  
For this particular case, one finds $p_{\rm excl} = 0.0475$ and CL$_s = 0.0487$, and by either criterion one would report a better than 95\% exclusion.

The second panel of Figure~\ref{fig:CLs_illustration} illustrates the case $b = 8.4$, $s=2.2$, and $n=5$, so that the overlap between the distributions for $H_b$ and $H_{s+b}$ is much larger.
In cases like this with a larger overlap (i.e. the signal regions get polluted by the background) statistical conclusions based on $p_{\rm excl}$  alone can be too aggressive.
Since we engineered this example to have the same $b+s$ and $n$ as for the first panel, we get the same\footnote{The general fact that $p_{\rm excl}(n,b,s)$ depends only on the sum $s+b$, and not on $s$ or $b$ separately, is a clear reason to reject it as a measure of confidence in the presence of the signal model, because it says that any exclusion for signal $s$ and background $b$ would imply an equally strong exclusion for the case that the signal is $s=0$ if the background $b$ were increased by the numerical value of $s$.} $p_{\rm excl} = 0.0475$, which taken at face value would again give a better than 95\% exclusion. However, proponents of the CL$_s$ criteria point out that here it must be recognized that for $b = 8.4$, the outcome $n\leq5$ would have been a low-probability occurrence no matter what\footnote{Here we are taking it as a requirement that $s\geq 0$, although in some situations quantum interference with the background could allow for $s<0$. See, for example, the case of a digluon resonance at the LHC \cite{Bhattiprolu:2020zoq}.} the signal mean $s$ was. Thus, the frequentist $p_{\rm excl}$ is really telling us more about the observed data than making a useful statement about the signal hypothesis. One finds that CL$_s = 0.3022$, and using this one would, sensibly and conservatively, refrain from excluding the signal hypothesis.

In fact, no matter the outcome for $n$, the experiment with $b=8.4$ simply lacks the statistical ability to exclude the $s=2.2$ signal model at 90\% confidence, according to the ${\rm CL}_s$ statistic. This can be seen by computing it for the least signal-like outcome, $n=0$, which gives CL$_s = 0.1108$. One possible practical interpretation of the very small $p_{\rm excl}$ in such cases with $n$ significantly less than $b$ might be that the background estimate could be wrong for reasons unknown, while another is that the background simply fluctuated low from its true mean. In any case, the intuitive interpretation of the ${\rm CL}_s$ statistic is that the quoted significance for exclusion should be reduced from the usual frequentist value, due to the large overlap between the signal+background region and the background-only region.

Indeed, if the number of events is sufficiently small, one finds that the usual frequentist $p$-value would correspond to an exclusion even in cases that defy sensible practical interpretation. Considering the case $n=0$ more generally, one finds
$p_{\rm excl}(n\!=\!0,s,b) = e^{-(s+b)}$, which becomes arbitrarily small for any fixed $s$, if $b$ is sufficiently large. One could use this to make an absurd claim of exclusion for a model that predicted $s = 10^{-500}$ or even $s=0$ exactly, simply by observing a smaller than expected number of events, if the background is large enough! In contrast, usage of the statistic ${\rm CL}_s(n=0,b,s) = e^{-s}$ conforms to the intuitively reasonable idea that, as an absolute prerequisite for excluding a signal hypothesis, the expected signal strength must not be too small. Specifically, only models that predict $s>-\ln(0.05) \approx 2.996$ can be excluded at 95\% confidence according to the ${\rm CL}_s$ measure, for any $b$ and for any possible experimental outcome $n$. Similarly, 90\% exclusion by the ${\rm CL}_s$ method requires $s > -\ln(0.1) \approx 2.303$. 

\begin{figure}
\begin{minipage}[]{0.5\linewidth}
\includegraphics[width=7.7cm]{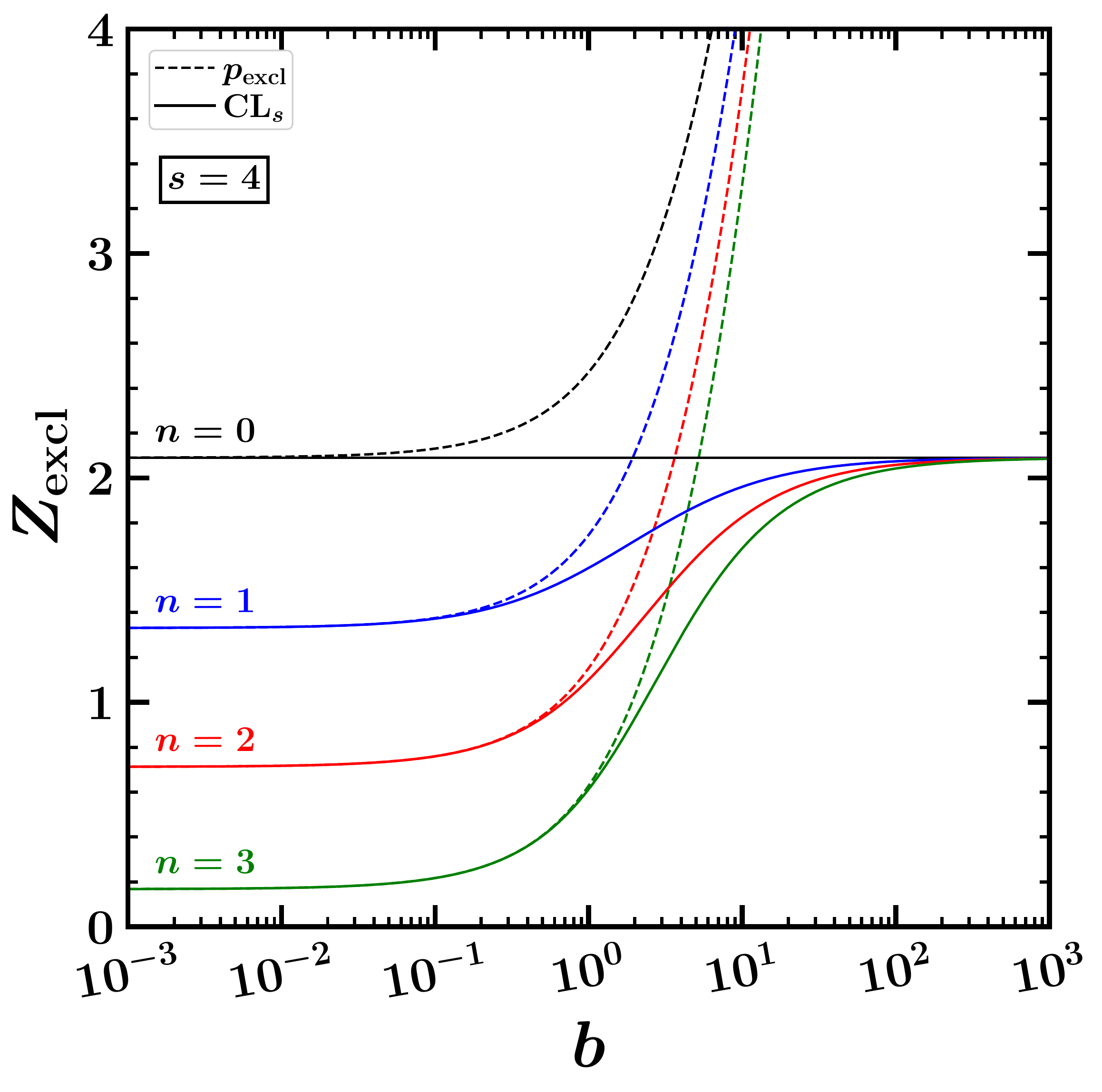}
\end{minipage}
\hspace{0.001\linewidth}
\begin{minipage}[]{0.48\linewidth}
\caption{Comparison of significances $Z$ obtained using eq.~(\ref{eq:Zfromp}) from $p_{\rm excl}$ [dashed lines, from eq.~(\ref{eq:pexcl_singlechannel})] and CL$_s$ [solid lines, from eq.~(\ref{eq:CLs_singlechannel})], for fixed $s=4$ as a function of varying $b$,
for $n=0,1,2,$ and 3. For very small $b$, the two statistics are nearly equal, $p_{\rm excl} \simeq {\rm CL}_s$. In the limit of large $b$ one has CL$_s = e^{-s}$, independent of $n$,
while $p_{\rm excl}$ becomes absurdly small in comparison. 
\label{fig:pexclvsCLs}}
\end{minipage}
\end{figure}

The dependence of the exclusion significance on $b$ is shown for fixed $s = 4$ and $n=0,1,2,3$
in Figure \ref{fig:pexclvsCLs}. For very small $b$, the two statistics are nearly equal, $p_{\rm excl} \simeq {\rm CL}_s$. For any fixed $n$, in the limit of large $b$ one has CL$_s = e^{-s}$,
while $p_{\rm excl}$ becomes absurdly small in comparison, which would imply an absurdly large $Z_{\rm excl}$. 

\begin{figure}
  \begin{minipage}[]{0.495\linewidth}
    \includegraphics[width=8cm]{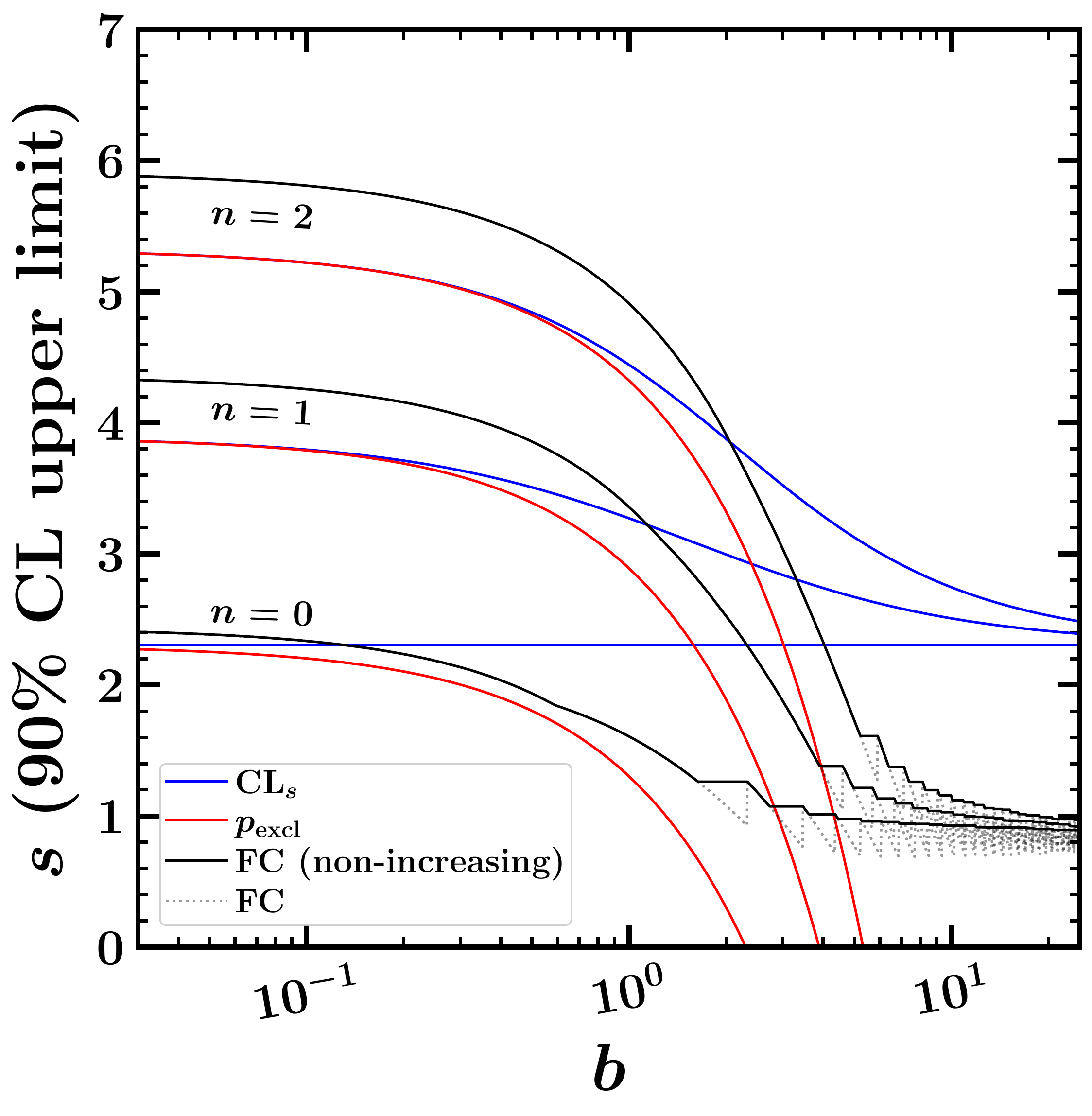}
  \end{minipage}
  \begin{minipage}[]{0.495\linewidth}
    \includegraphics[width=8cm]{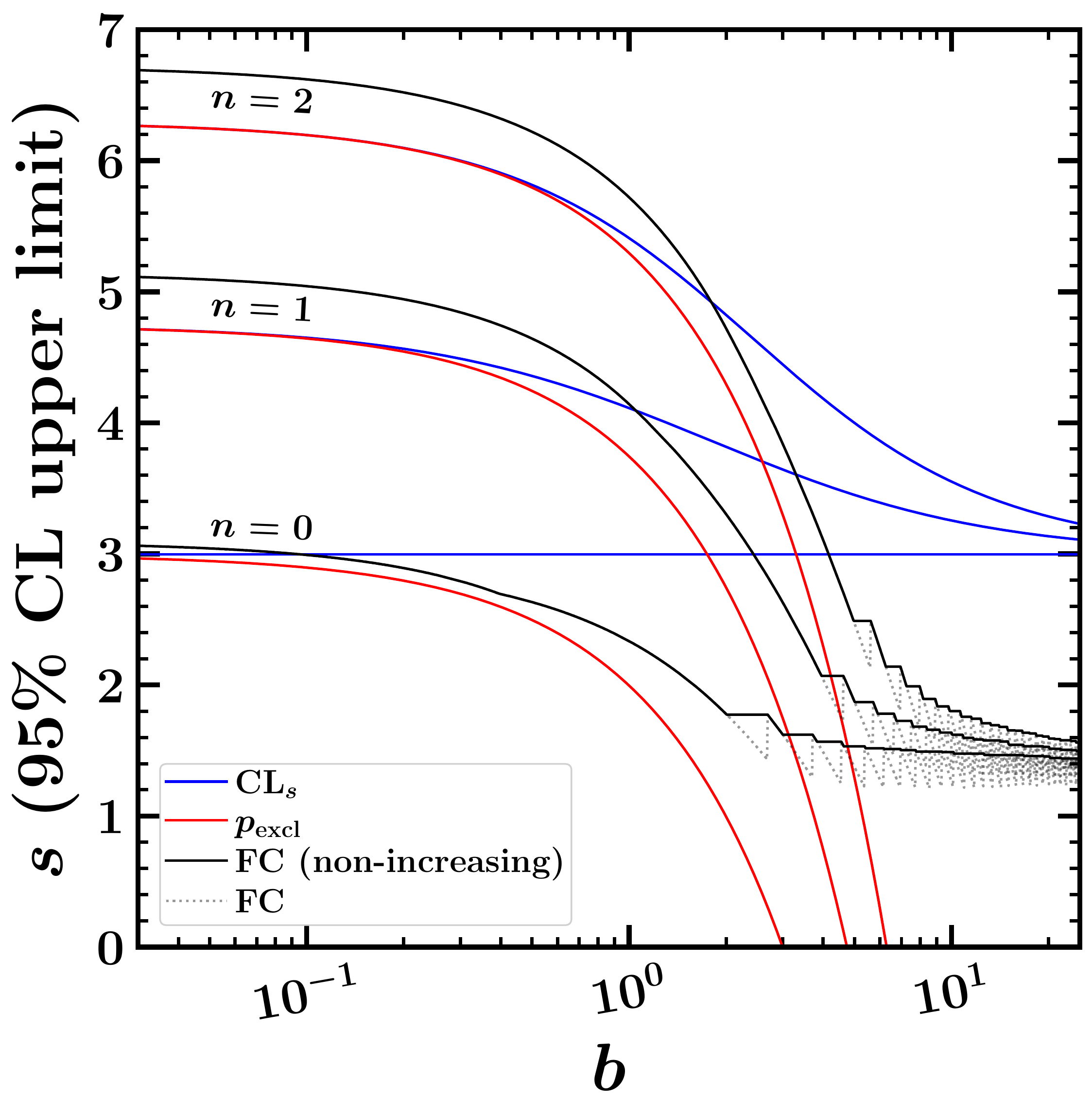}
  \end{minipage}
 \caption{The 90\% CL (left panel) and 95\% CL (right panel) upper limits on signal as
functions of the background mean $b$, for a fixed number of observed events $n = 0, 1, 2$,
using the ${\rm CL}_s$ technique [blue lines, from eq.~(\ref{eq:sUL_CLs})], standard frequentist $p$-value approach [red lines, from eq.~(\ref{eq:sUL_pexcl})],
and Feldman-Cousins method [solid black lines, from ref.~\cite{Feldman:1997qc}].
The dotted black lines show the results obtained by the Feldman-Cousins method before requiring them to be
non-increasing as a function of background mean.
\label{fig:sULexcl_vs_b_fix_n}}
\end{figure}

Non-observation of a significant excess above background expectations can be used to constrain new physics. In particular, for a single-channel counting experiment, the minimum signal needed to claim an exclusion at a given confidence level $1-\alpha$, equivalent to significance $Z = \sqrt{2}\, {\rm erfc}^{-1}(2 \alpha)$, for a perfectly known background mean $b$, is obtained \cite{ParticleDataGroup:2006fqo,ParticleDataGroup:2020ssz} by solving for $s$ in either
\beq
\alpha &=& \frac{\Gamma(n + 1, s + b)}{\Gamma(n + 1)} \qquad (\text{$p_{\rm excl}$ method})
\label{eq:sUL_pexcl}
\eeq
in the standard frequentist approach, or 
\beq
\alpha &=& \frac{\Gamma(n + 1, s + b)}{\Gamma(n + 1, b)} \qquad (\text{CL$_s$ method})
\label{eq:sUL_CLs}
\eeq
in the modified frequentist approach.
Figure~\ref{fig:sULexcl_vs_b_fix_n} shows the 90\% CL ($\alpha = 0.1$, left panel) and 
95\% CL ($\alpha = 0.05$, right panel) upper limits
on signal as functions of the background
mean, for a fixed number of observed events $n = 0, 1, 2$, using the 
$p_{\rm excl}$ (red lines) 
and CL$_s$ (blue lines) criteria.
Also shown in the figure are the 90\% CL and 95\% CL upper limits on $s$ that are obtained using
the Feldman-Cousins (FC) method
based on an ordering principle introduced in ref.~\cite{Feldman:1997qc}.
The upper limits obtained by the FC method
for a fixed $n$ do not always decrease with increasing $b$; instead they have a sawtooth pattern,
as can be seen from the dotted lines in the figure.
This behavior is because of the discreteness of Poisson distributions.
The solid black lines in Figure \ref{fig:sULexcl_vs_b_fix_n} show the results obtained by the FC method after requiring them to be non-increasing as a function of background mean.

It is clear from the figure that the upper limits on $s$ obtained using the standard
frequentist $p_{\rm excl}$ approach are the least conservative,
and can even go negative in the case where the number of observed events $n$
is small compared to the expected background mean.
For a fixed $n$, despite the upper limits given by the ${\rm CL}_s$ and FC methods
being very different from each other, we note that they are both almost flat at very small backgrounds and then 
decrease slowly (or stay constant) as a function of background, always remaining positive.
For small $b$ the FC upper limits are more conservative,
and for large $b$, the ${\rm CL}_s$ upper limits are more conservative.
The other striking difference between these two upper limits is that, for $n=0$,
the FC upper limits decrease with $b$, but the ${\rm CL}_s$ upper limits are independent of $b$.
In particular, at a chosen confidence level $1 - \alpha$, for $n=0$ the ${\rm CL}_s$ upper limit 
on $s$ is $-\ln(\alpha)$.
The same result also holds for any $n$ in the limit that the background is extremely large.
At 90\% (95\%) CL, the upper limit given by ${\rm CL}_s$ for $n=0$, or for any $n$ as $b \rightarrow \infty$, is around 2.303 (2.996).
On the other hand, the upper limit given by the FC method decreases as a function of $b$ and approaches a constant value at large $b$. For example, for $n=0$, the 90\% (95\%) CL upper limit given by the FC method, after requiring it to be non-increasing as a function of $b$,
is approximately 0.8 (1.34) at large $b$.

It is important for the following that the result for CL$_s(n,b,s)$ in the case of a single Poisson channel in eq.~(\ref{eq:CLs_singlechannel}) can also be obtained \cite{Helene:1982pb} as a Bayesian credible interval, using a flat prior for the signal and likelihoods ${\cal L}(s|n,b) \propto P(n|s+b)$:
\beq 
\CLexcl
(n, b, s) &=& 
\frac{\displaystyle\scaleobj{0.85}{\int_{\scaleobj{1.4}{s}}^{\scaleobj{1.2}{\infty}}} ds'\, {\cal L}(s'|n,b)}{
\displaystyle\scaleobj{0.85}{\int_{\scaleobj{1.2}{0}}^{\scaleobj{1.2}{\infty}}} ds'\, {\cal L}(s'|n,b)}
\>=\>
\frac{
\displaystyle\scaleobj{0.85}{\int_{\scaleobj{1.4}{s}}^{\scaleobj{1.2}{\infty}}}
ds^\prime \, e^{-(s^\prime + b)} \, (s^\prime + b)^n
}
{
\displaystyle\scaleobj{0.85}{\int_{\scaleobj{1.2}{0}}^{\scaleobj{1.2}{\infty}}}
ds^\prime \, e^{-(s^\prime + b)} \, (s^\prime + b)^n
}
.
\label{eq:CLsBayes_excl}
\eeq
Performing the integrations, $\CLexcl(n,b,s)$ as defined by eq.~(\ref{eq:CLsBayes_excl}) is precisely equal to ${\rm CL}_s(n,b,s)$ as defined by eq.~(\ref{eq:CLs_singlechannel}).\footnote{If the signal mean is instead allowed to be negative with $s + b \ge 0$ (see previous footnote), then $\CLexcl(n, b, s)$ can be defined as
$\CLexcl(n, b, s)=\frac{
\displaystyle\scaleobj{0.85}{\int_{\scaleobj{1.4}{s}}^{\scaleobj{1.2}{\infty}}}
ds^\prime \, e^{-(s^\prime + b)} \, (s^\prime + b)^n
}
{
\displaystyle\scaleobj{0.85}{\int_{\scaleobj{1.2}{-b}}^{\scaleobj{1.2}{\infty}}}
ds^\prime \, e^{-(s^\prime + b)} \, (s^\prime + b)^n
}$. After performing the integrations, $\CLexcl(n,b,s)$ is now precisely equal to
$p_{\rm excl} (n,b,s)$ as defined in eq.~(\ref{eq:pexcl_singlechannel}).}
However, despite the numerical equivalence, the interpretation is quite different, since the ratio of frequentist $p$-values is not directly a Bayesian confidence interval. Moreover, the equivalence between CL$_s$ and $\CLexcl$ is only approximate in more complicated generalizations. Looking ahead to the case of experiments which collect counts in multiple independent channels governed by Poisson statistics, and which may have nuisance parameters including uncertainties in the backgrounds, we will argue for a generalization based 
straightforwardly on the Bayesian version $\CLexcl$ as given in eq.~(\ref{eq:CLsBayes_excl}) rather than CL$_s$ given in eq.~(\ref{eq:CLs_general}) or its specialization eq.~(\ref{eq:CLs_singlechannel}).

For a single-channel counting experiment, the discovery confidence level statistic defined in eq.~(\ref{eq:CLdisc_general}) becomes
\beq
\CLdisc(n,b,s) &=& 
\frac{P(n|b)}{P(n|b+s)} \>=\>
\frac{e^s}{(1 + s/b)^n},
\label{eq:CLdiscsinglechannel}
\eeq
which can be used in place of $p$ in eq.~(\ref{eq:Zfromp}) to obtain a discovery significance.
(If the result is greater than 1, then clearly no discovery claim should be contemplated.)
Note that unlike $p_{\rm disc}(n,b)$, the result for $\CLdisc(n,b,s)$ depends on the strength of the signal whose discovery is under investigation.
It is always more conservative than $p_{\rm disc}(n,b)$ in claiming discovery, 
just as CL$_{\rm s}$ is more conservative than $p_{\rm excl}$ in claiming exclusion.
For example, in the extreme case $s=0$, one has $\CLdisc(n,b,s=0) = 1$ for any $b$ and $n$, so one would never claim discovery using that criteria. In contrast, the frequentist statistic $p_{\rm disc}(n,b)$
can be arbitrarily small, implying an arbitrarily large discovery significance $Z$, even in situations where the physics provides absolutely no possible source for a signal.\footnote{For example, imagine a search for a 
new fundamental particle of mass 1 TeV,
conducted by dropping a bag of hammers from the top of a tall building, with a somewhat noisy detector surrounding the impact point on the sidewalk. For this experiment, theoretical modeling confidently predicts $s=0$, so one should reasonably refrain from announcing discovery even if one estimated $b = 0.01$ and observed $n=3$.}
As we will see below, $\CLdisc$ also generalizes more straightforwardly to cases that have multiple independent channels governed by Poisson statistics, and which may have nuisance parameters including uncertainties in the backgrounds. 

\begin{figure}
  \begin{minipage}[]{0.495\linewidth}
    \includegraphics[width=8cm]{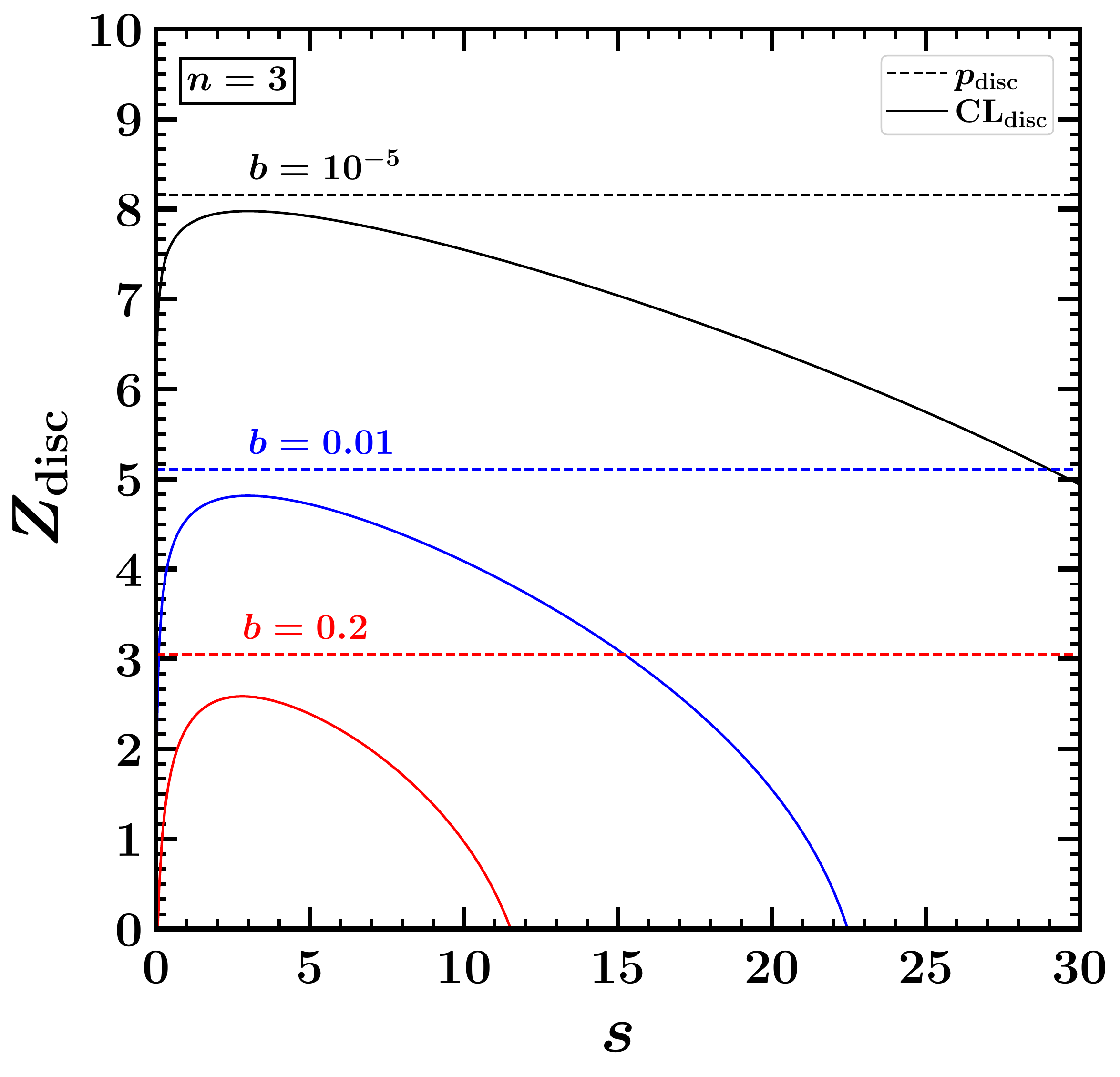}
  \end{minipage}
  \begin{minipage}[]{0.495\linewidth}
    \includegraphics[width=8cm]{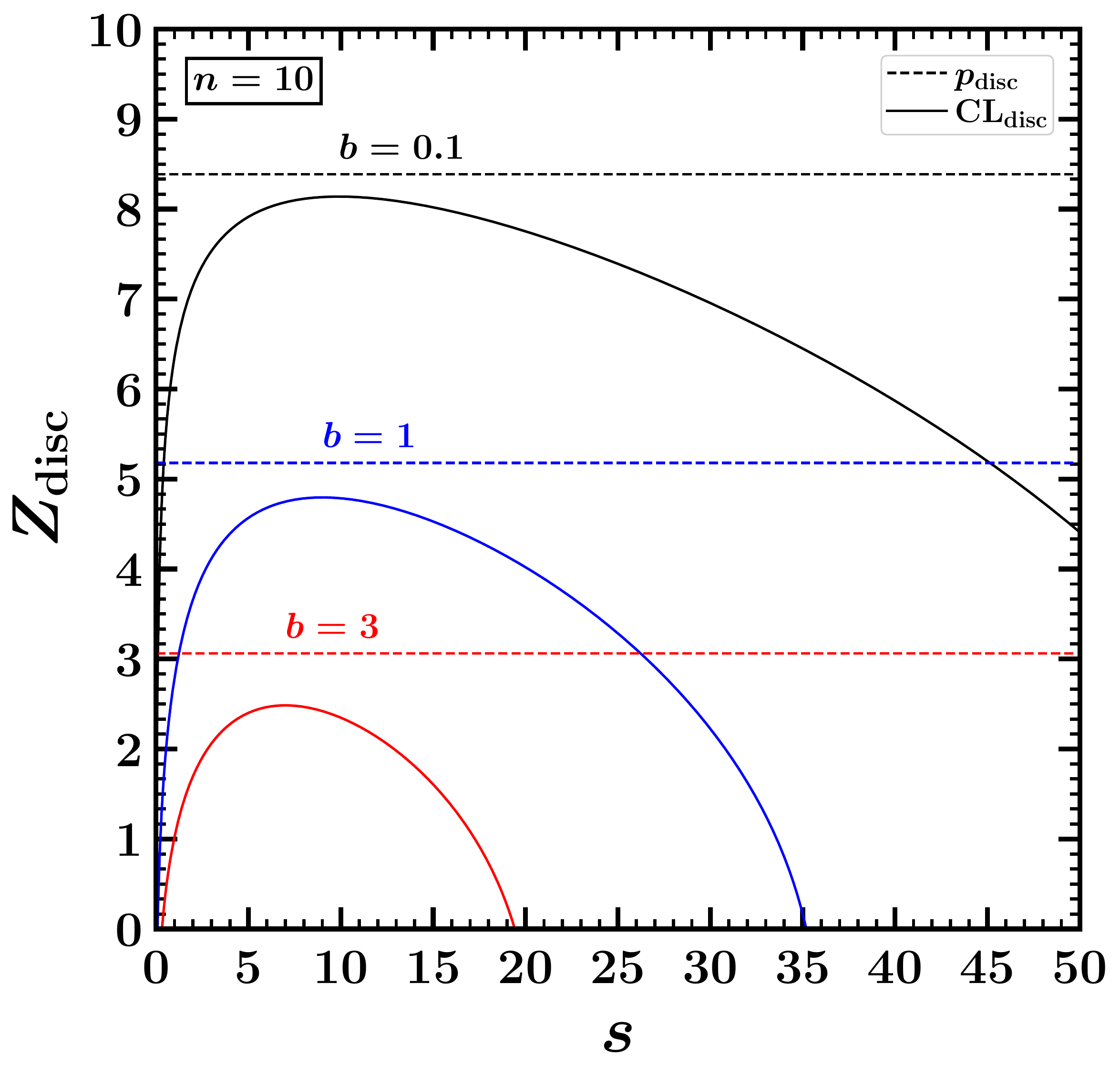}
  \end{minipage}
\begin{minipage}[]{0.95\linewidth}
\caption{Comparison of discovery significances obtained using eq.~(\ref{eq:Zfromp}) from $p_{\rm disc}$
[dashed lines, from eq.~(\ref{eq:pdisc_singlechannel})]
and ${\rm CL}_{\rm disc}$ [solid lines, from eq.~(\ref{eq:CLdiscsinglechannel})]
as a function of $s$ for $n=3$ (left panel) and $n=10$ (right panel), for various choices of $b$.
\label{fig:pdisc_vs_CLdisc}}
\end{minipage}
\end{figure}

Figure~\ref{fig:pdisc_vs_CLdisc} compares the discovery significance obtained from $p_{\rm disc}$
and $\CLdisc$ as a function of $s$ for fixed $n$, with different curves for different values of $b$.
Note that the discovery significance obtained from $\CLdisc$, which is always more
conservative than that of $p_{\rm disc}$, is maximized at $s = n - b$.

\begin{figure}
  \begin{minipage}[]{0.495\linewidth}
    \includegraphics[width=8cm]{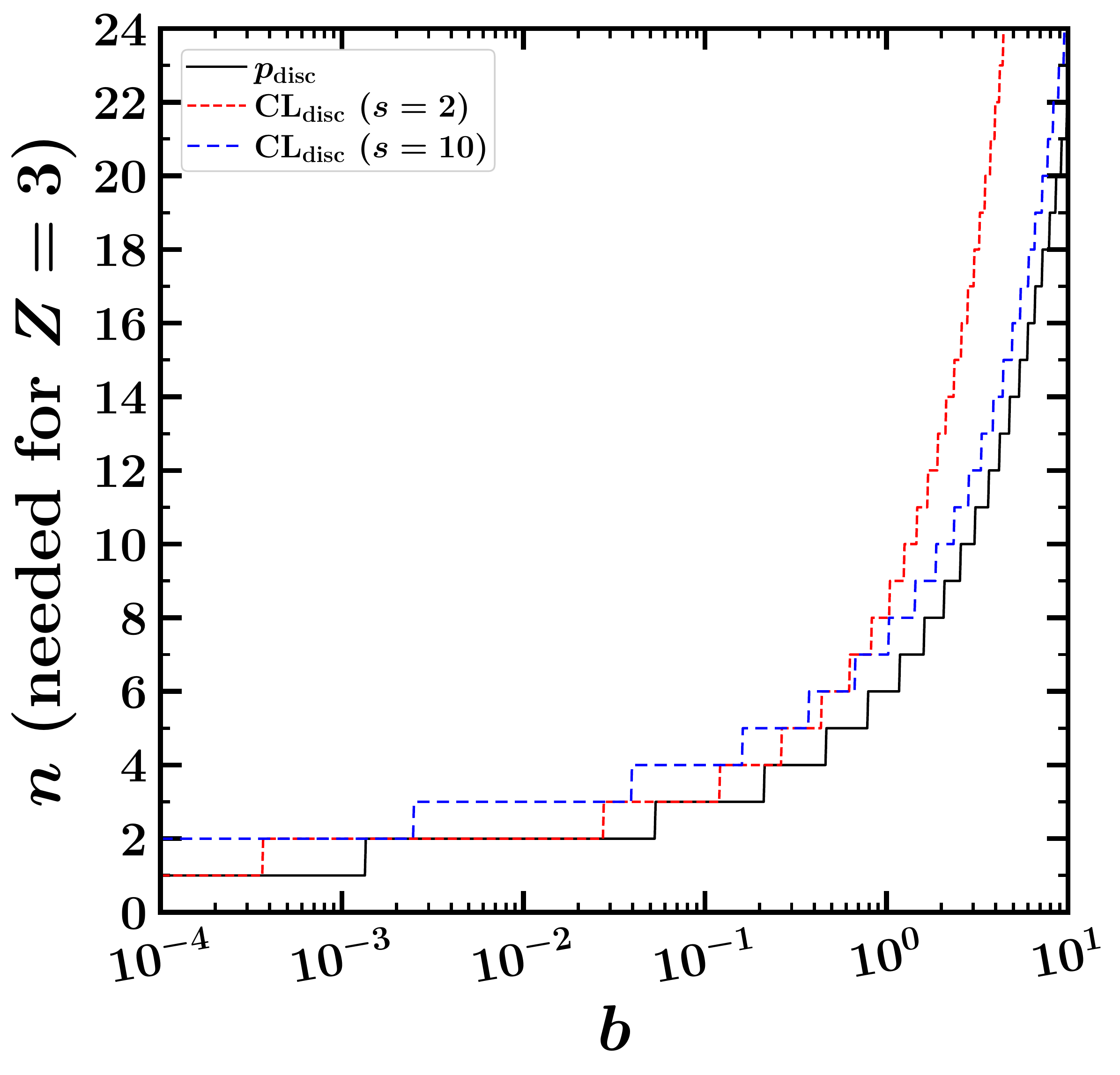}
  \end{minipage}
  \begin{minipage}[]{0.495\linewidth}
    \includegraphics[width=8cm]{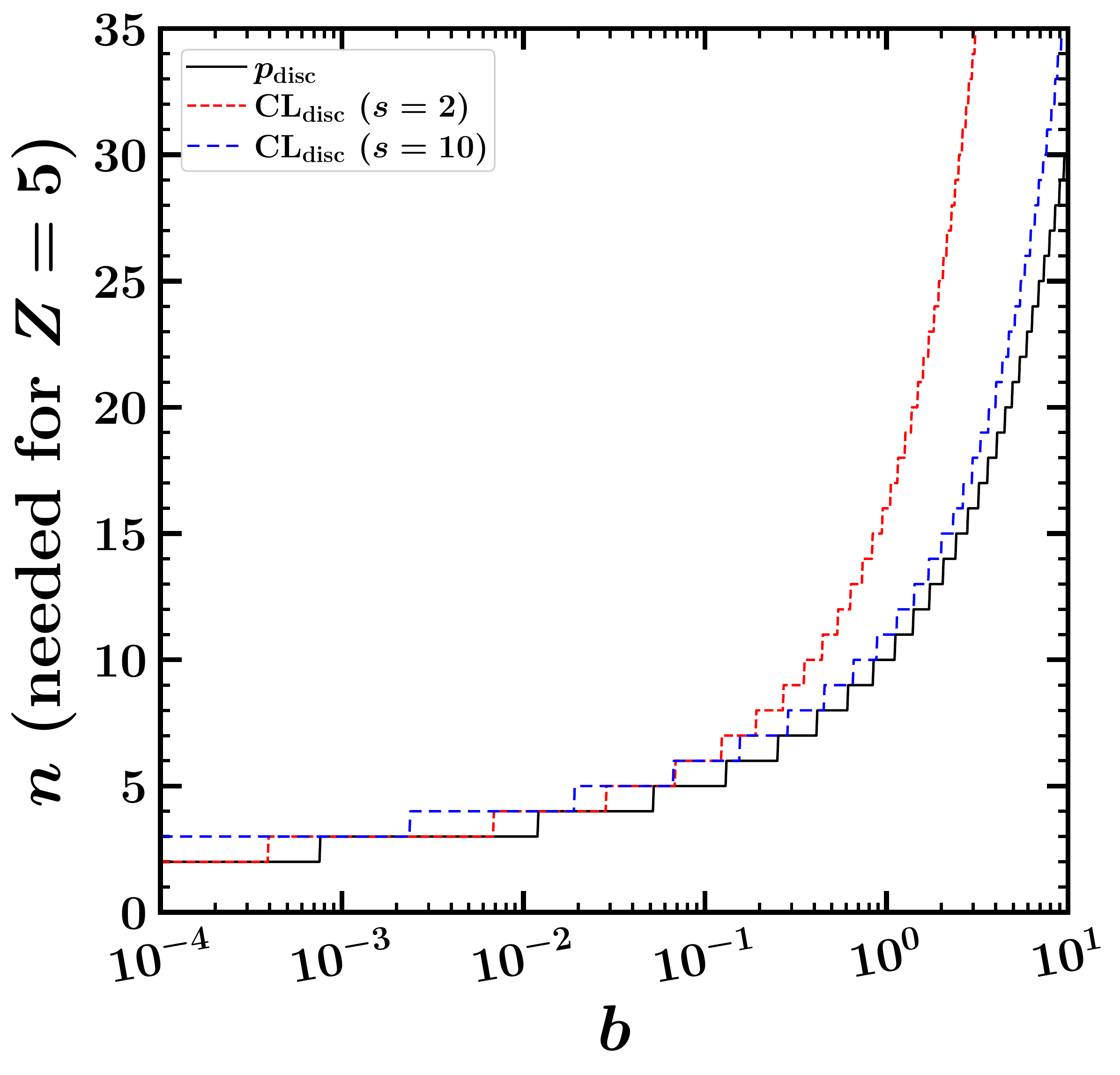}
  \end{minipage}
 \caption{The observed number of events $n$ needed for $Z=3$ evidence (left panel) and
 $Z=5$ discovery (right panel) as functions of the expected background mean $b$.
The solid black lines show the result obtained from eq.~(\ref{eq:n_pdisc}) using the standard frequentist approach based on $p_{\rm disc}$, which is independent of the signal mean $s$. The dashed red and blue lines show the results obtained from eq.~(\ref{eq:n_CLdisc}) using $\CLdisc$ for the cases of signal mean $s =2$ and $10$, respectively.
\label{fig:ndisc_vs_b}}
\end{figure}

Given the number of observed events $n$ and an expected background mean,
the standard $p$-value for discovery $p_{\rm disc}$ does not depend on the signal.
So, for a perfectly known background mean $b$,
we can compute the number of events needed for discovery
at a significance $Z$ by solving for $n$ from
[see eqs.~(\ref{eq:Zfromp}) and (\ref{eq:pdisc_singlechannel})]
\beq
\frac{1}{2} {\rm erfc}\left(\frac{Z}{\sqrt 2}\right) &=& \frac{\gamma(n, b)}{\Gamma(n)}
\qquad (p_{\rm disc}\>\text{method})
.
\label{eq:n_pdisc}
\eeq
On the other hand, ${\rm CL}_{\rm disc}$ depends also on the signal,
in which case the number of events needed for discovery for a known background $b$ and signal mean $s$ at a given significance $Z$ can be obtained by solving for $n$ from
[see eqs.~(\ref{eq:Zfromp}) and (\ref{eq:CLdiscsinglechannel})]
\beq
\frac{1}{2} {\rm erfc}\left(\frac{Z}{\sqrt 2}\right) &=& \frac{e^s}{(1 + s/b)^{n}}
\qquad (\CLdisc \> \text{method})
.
\label{eq:n_CLdisc}
\eeq
Figure~\ref{fig:ndisc_vs_b} shows the observed number of events
required for $Z=3$ evidence (left panel) and $Z=5$ discovery (right panel)
given by the $p_{\rm disc}$ approach (solid black lines), and the ${\rm CL}_{\rm disc}$ approach
for two choices of the signal mean $s=2$ (dashed red lines) and $10$ (dashed blue lines) as functions of $b$. It is clear from the figure that, for a given background mean,
the observed number of events needed for discovery given by the $\CLdisc$ approach
are at least as large as the result given by the $p_{\rm disc}$ criterion, and often much larger when the background is not very small.

We now turn to the question of projecting expectations for exclusion and discovery at ongoing and future experiments. In simulations or assessments of a proposed experiment, one considers the statistics of pseudo-data generated under an alternative hypothesis $H_1$. For assessments of prospects for exclusion the alternative hypothesis is that the signal source is absent, $H_1 = H_b$, while for discovery, the pseudo-data is generated assuming that both signal and background are present, $H_1 = H_{s+b}$.

\begin{figure}
  \begin{minipage}[]{0.495\linewidth}
    \includegraphics[width=8cm]{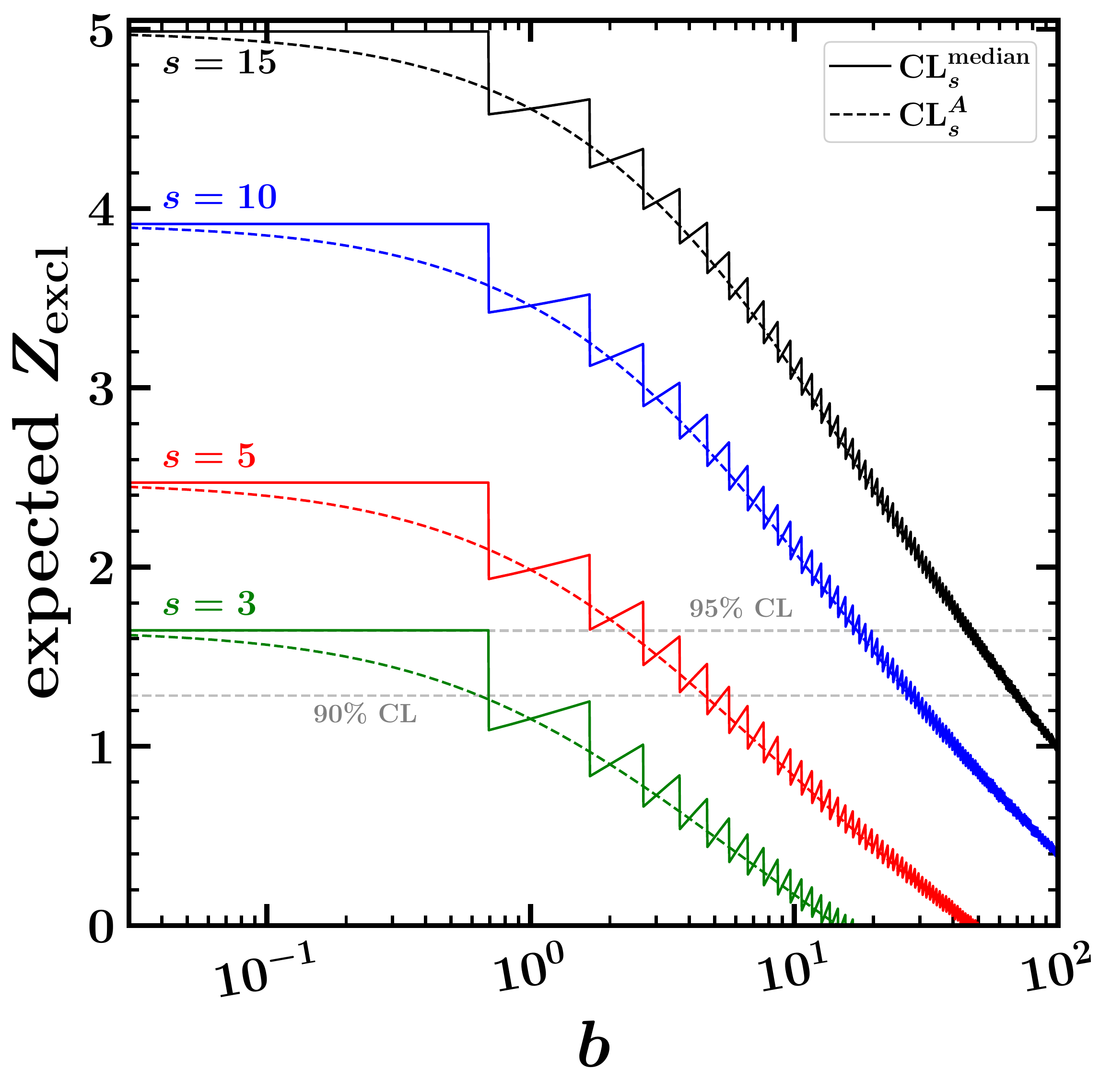}
  \end{minipage}
  \begin{minipage}[]{0.495\linewidth}
    \includegraphics[width=8cm]{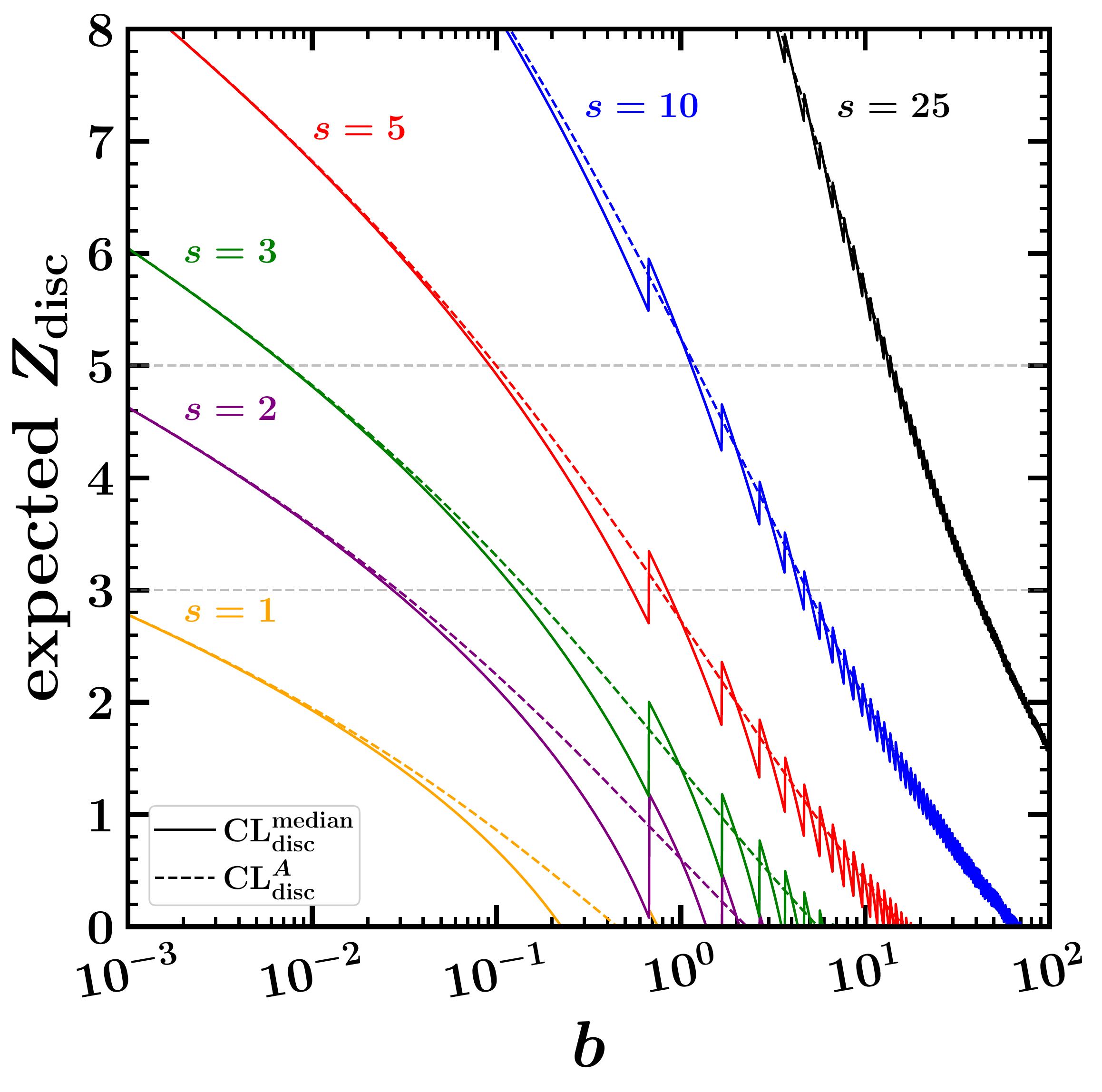}
  \end{minipage}
\begin{minipage}[]{0.95\linewidth}
\caption{Median (solid lines) and exact Asimov (dashed lines) expected significances obtained
using eq.~(\ref{eq:Zfromp})
from ${\rm CL}_s = \CLexcl$ [from eqs.~(\ref{eq:CLs_singlechannel}) and (\ref{eq:CLsAsimov})] for exclusion (left panel) and $\CLdisc$ [from eqs.~(\ref{eq:CLdiscsinglechannel}) and (\ref{eq:CLdiscAsimov})] for discovery (right panel),
as a function of the background mean $b$ for various values of signal mean $s$,
for a single-channel Poisson counting experiment. Due to the discrete nature of Poisson statistics,
the median expected significances suffer from a sawtooth behavior.
On the other hand, the exact Asimov expected significances behave sensibly as they decrease monotonically with $b$.
\label{fig:sawtoothCLS}}
\end{minipage}
\end{figure}

A common way to project an expected result is to set the number of events $n$
equal to the median expected value under the hypothesis $H_1$. However,
due to the discrete nature of Poisson statistics events, the median expected outcome has the striking flaw that it can predict smaller significances if an experiment takes more data or reduces its background. 
This counterintuitive feature of the median expected significance
was pointed out and studied in detail in refs.~\cite{Cowan:2010js,Cowan}, and in \cite{Bhattiprolu:2020mwi} where it was referred to as the ``sawtooth problem". It occurs for the median expected CL$_s$ and $\CLdisc$ as well.
The sawtooth behavior of the median expected CL$_s$ and $\CLdisc$
as a function of the background mean $b$, for various values of signal mean $s$, is evident from
Figure~\ref{fig:sawtoothCLS}.
For comparison, Figure~\ref{fig:sawtoothCLS} also show the significances obtained from
the exact Asimov expected CL$_s$ and $\CLdisc$ (dashed lines),
detailed below, that are smooth and sensible.

\begin{figure}
  \begin{minipage}[]{0.495\linewidth}
    \includegraphics[width=8cm]{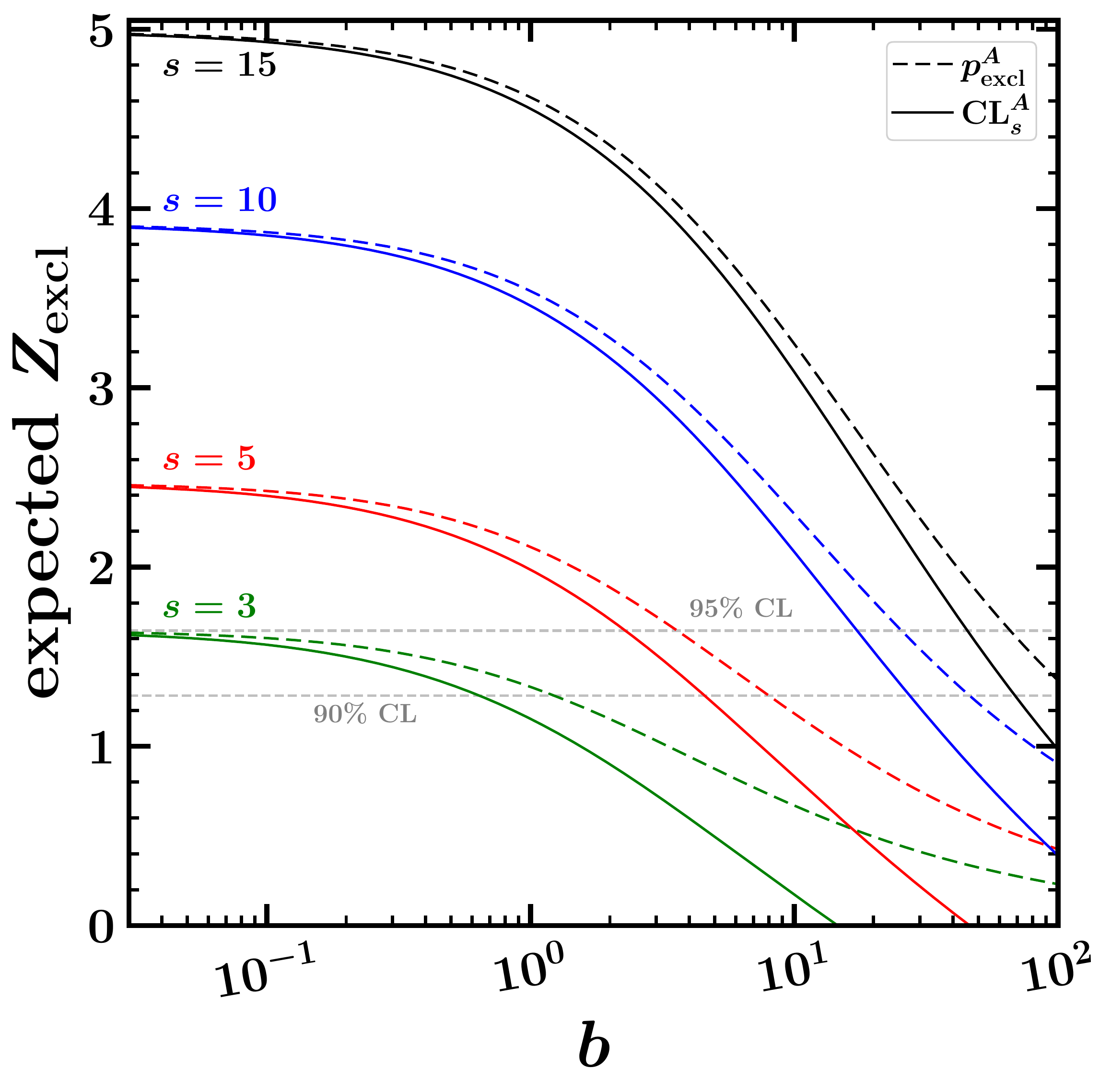}
  \end{minipage}
  \begin{minipage}[]{0.495\linewidth}
    \includegraphics[width=8cm]{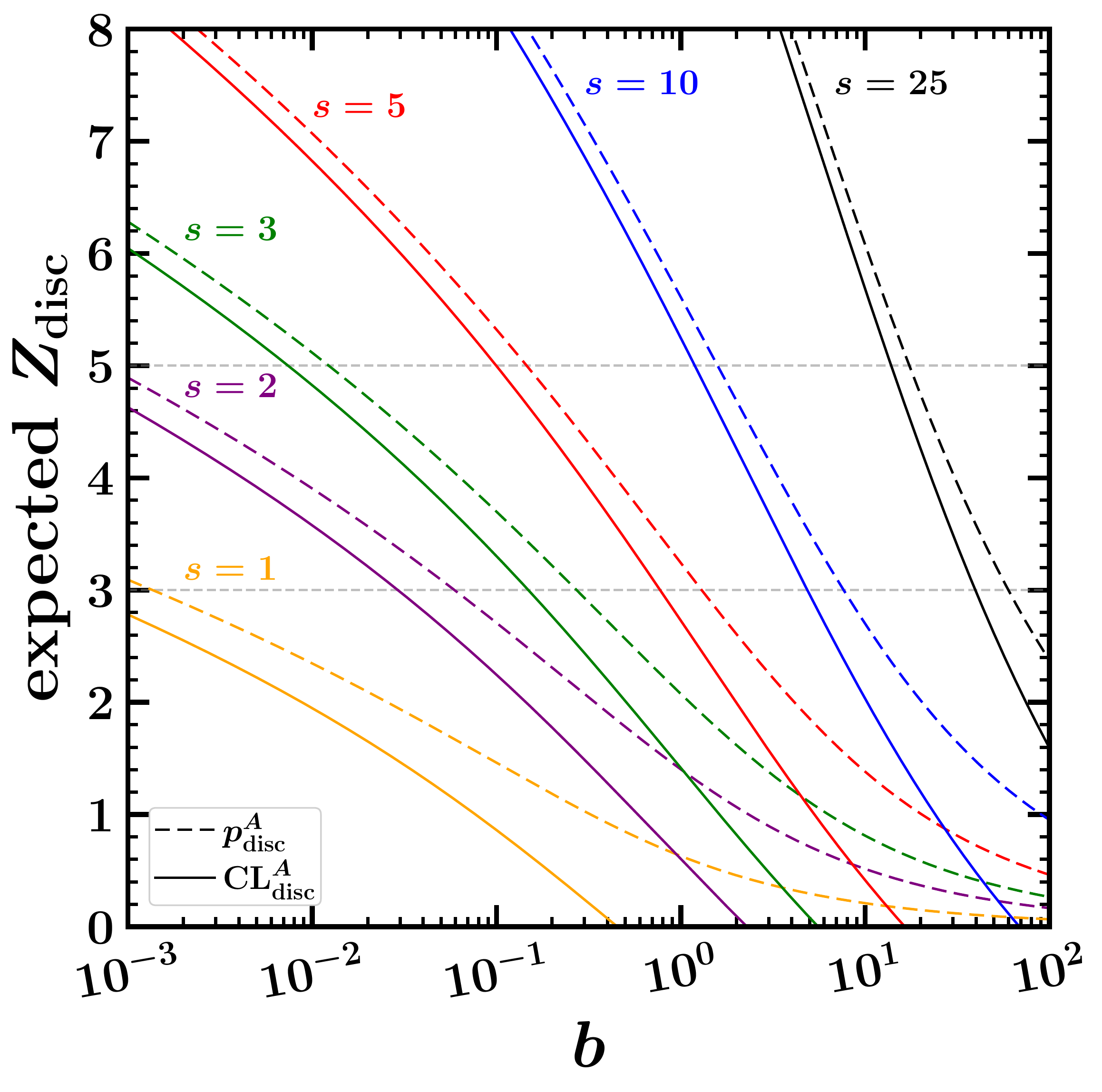}
  \end{minipage}
\begin{minipage}[]{0.95\linewidth}
\caption{The exact Asimov expected significances obtained from frequentist $p$-values (dashed lines) and modified frequentist ${\rm CL}_s$/Bayesian ${\rm CL}_{\rm disc}$ confidence levels (solid lines), converted to significances $Z$ using eq.~(\ref{eq:Zfromp}),
for a single-channel Poisson counting experiment. Results are presented as functions of the background mean $b$ for various values of signal mean $s$. The term ``exact Asimov" means that we set the number of events equal to the mean expected according to the hypothesis $H_1$, so $n=b$ for exclusion and $n=s+b$ for discovery.
The left panel compares $p_{\rm excl}^A$ to ${\rm CL}_s^A$ for exclusion, from eqs.~(\ref{eq:pexcl_singlechannelAsimov}) and (\ref{eq:CLsAsimov}).
The right panel compares $p_{\rm disc}^A$ to $\CLdiscA$ for discovery, from eqs.~(\ref{eq:pdisc_singlechannelAsimov}) and (\ref{eq:CLdiscAsimov}).
\label{fig:Z_vs_b}}
\end{minipage}
\end{figure}

Therefore, in ref.~\cite{Bhattiprolu:2020mwi}, we proposed instead to use
an exact Asimov approach for projecting sensitivities of planned experiments, where the observed number of events $n$ is replaced by its mean expected value $\langle n_{\rm excl} \rangle = b$ for exclusion and $\langle n_{\rm disc} \rangle = s + b$ for discovery. From 
eqs.~(\ref{eq:pexcl_singlechannel}) and (\ref{eq:CLs_singlechannel}) we thus obtain for the expected exclusion
in the case of a single-channel counting experiment with signal and background means
$s$ and $b$:
\beq
p_{\rm excl}^A &=& \frac{\Gamma(b+1, s+b)}{\Gamma(b+1)},
\label{eq:pexcl_singlechannelAsimov}
\\
{\rm CL}_s^A \,=\, \CLexclA &=& \frac{\Gamma(b+1, s+b)}{\Gamma(b+1, b)},
\label{eq:CLsAsimov}
\eeq
Similarly, for the expected discovery significance, we obtain from eqs.~(\ref{eq:pdisc_singlechannel}) and
(\ref{eq:CLdiscsinglechannel}):
\beq
p_{\rm disc}^A &=& \frac{\gamma(s+b, b)}{\Gamma(s+b)},
\label{eq:pdisc_singlechannelAsimov}
\\
\CLdiscA &=& \frac{e^s}{(1 + s/b)^{s+b}}.
\label{eq:CLdiscAsimov}
\eeq
Figure~\ref{fig:Z_vs_b} compares the exact Asimov expected significances obtained from frequentist (dashed lines) and modified frequentist ${\rm CL}_s$/Bayesian $\CLdisc$ (solid lines) confidence levels, for both exclusion (left panel) and discovery (right panel) cases.
This illustrates the more general fact that CL$_s$ and $\CLdisc$ are 
more conservative than $p_{\rm excl}$ and $p_{\rm disc}$, respectively.

In order to project expected exclusions based on the $p_{\rm excl}$ or CL$_s$ approaches,
we set eq.~(\ref{eq:pexcl_singlechannelAsimov}) or (\ref{eq:CLsAsimov}) equal to the desired 
$\alpha = 0.10$ or $0.05$, and then solve for $s$.
We also consider projections based on the FC method, in two different ways.
One is the Feldman-Cousins experimental sensitivity, advocated within ref.~\cite{Feldman:1997qc},
that is defined as the arithmetic mean of the upper limits obtained by the FC method
at a chosen confidence level\footnote{These upper limits on signal
are defined in ref.~\cite{Feldman:1997qc}, and are shown as a function of background $b$ for
$n = 0, 1, 2$ with solid black lines in Figure~\ref{fig:sULexcl_vs_b_fix_n}.}
$s^{\rm UL}_{\rm FC} (n, b)$ in a large number of pseudo-experiments with data generated under background-only hypothesis:\footnote{An implementation of the Feldman-Cousins
method to evaluate the upper limits and the experimental sensitivity, advocated within
ref.~\cite{Feldman:1997qc}, is made available with the {\sc Zstats v2.0} package \cite{Zstats:v2}.}
\beq
\text{FC sensitivity} &=& \sum_{n=0}^{\infty} P(n | b) \, s^{\rm UL}_{\rm FC} (n, b).
\label{eq:FCsensitivity}
\eeq
The other way is to simply compute the upper limit on signal given by the
FC method with the observed number of events taken to be the nearest integer
to the expected background mean $n = {\rm round}(b)$\footnote{When rounding half-integral values of
$b=0.5, 1.5, 2.5, 3.5, \ldots$, we follow the IEEE 754 standard of taking the nearest even integer such that $\text{round}(b) = 0, 2, 2, 4, \ldots$.}.
We consider the latter for future reference,
as it was alluded to in ref.~\cite{Alt:2020blf} while projecting exclusion sensitivity for proton decay in
$p \rightarrow \overline{\nu} K^+$ channel at DUNE.
\begin{figure}
  \begin{minipage}[]{0.495\linewidth}
    \includegraphics[width=8cm]{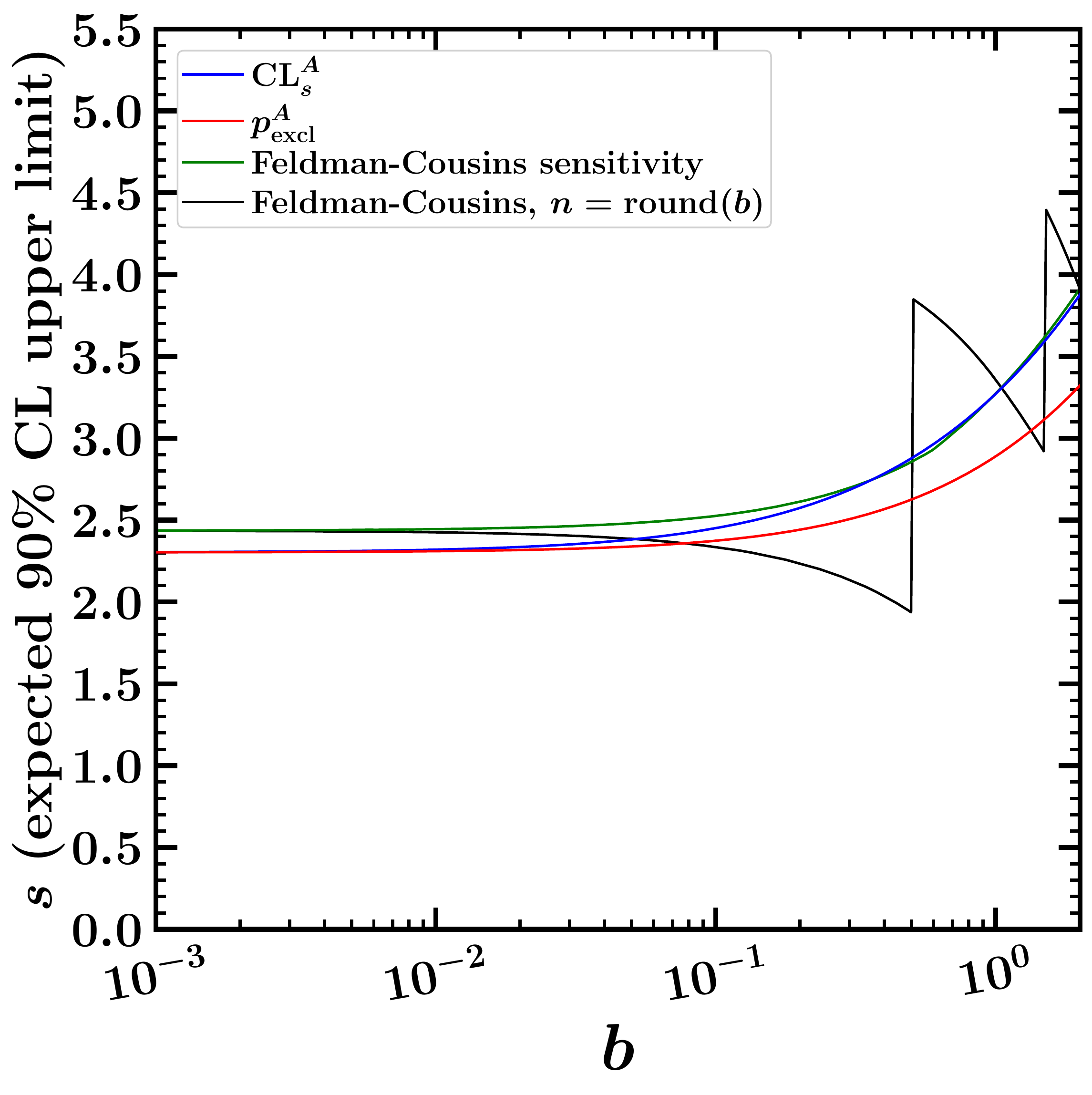}
  \end{minipage}
  \begin{minipage}[]{0.495\linewidth}
    \includegraphics[width=8cm]{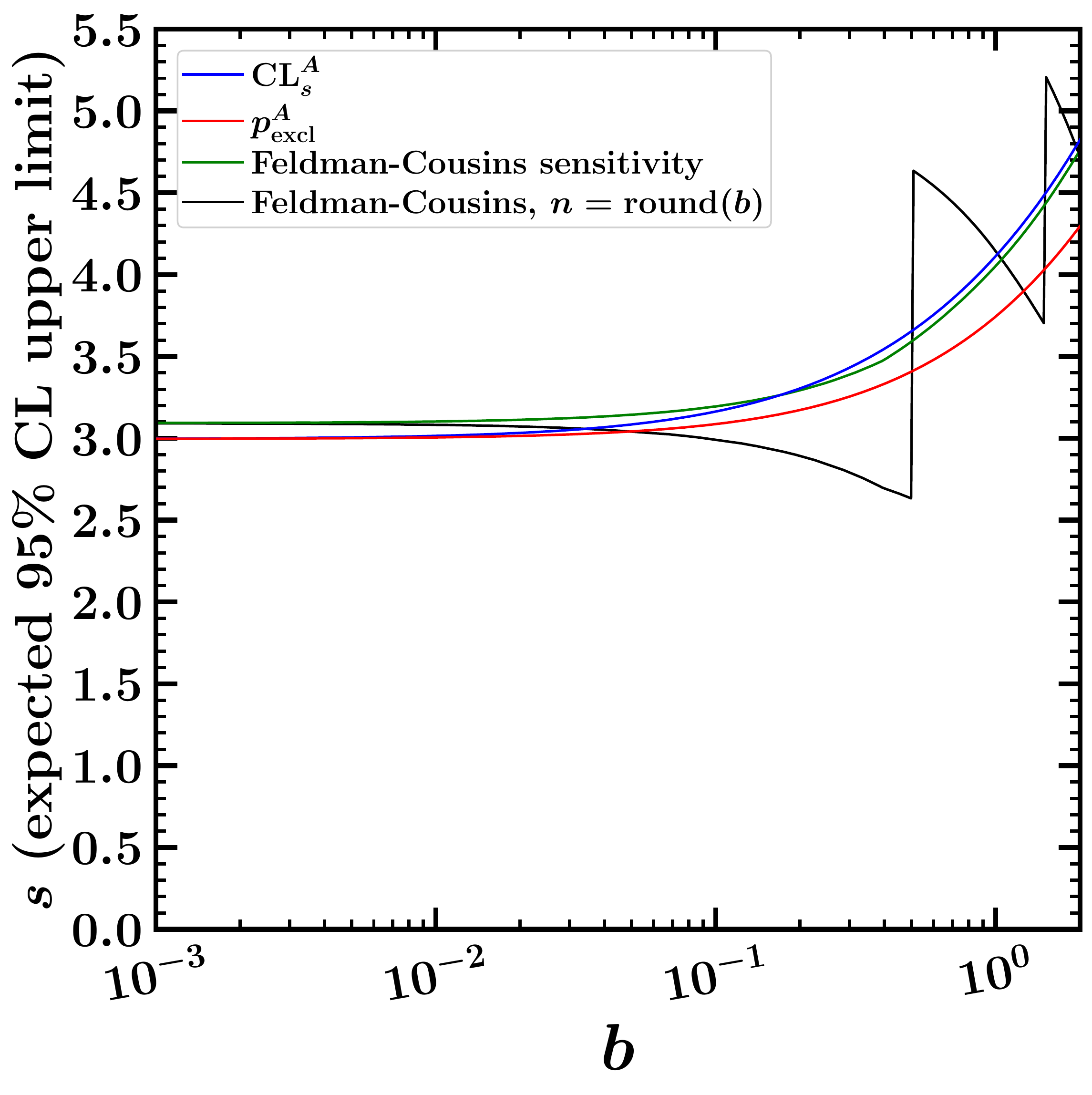}
  \end{minipage}
  \begin{minipage}[]{0.495\linewidth}
    \includegraphics[width=8cm]{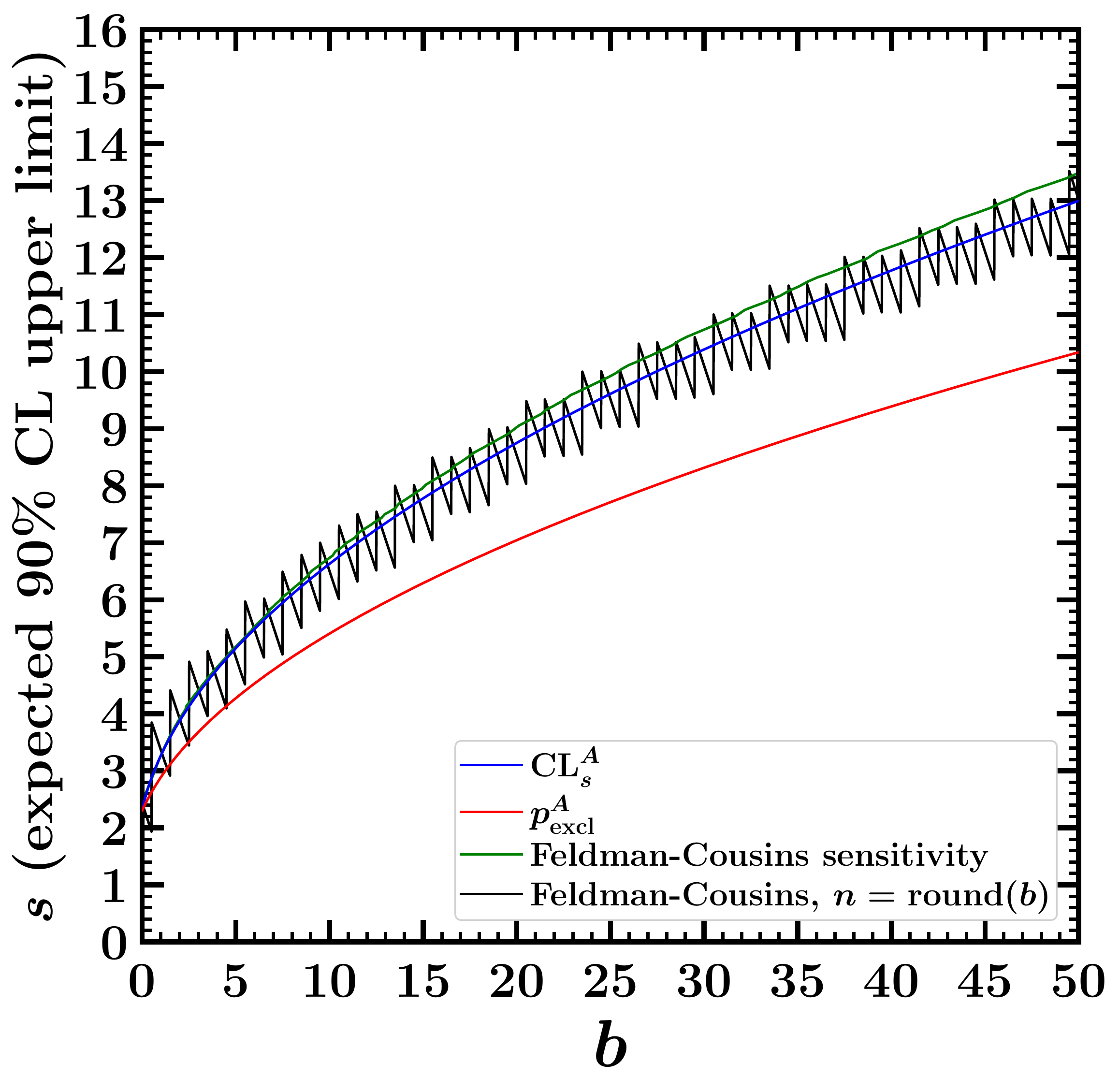}
  \end{minipage}
  \begin{minipage}[]{0.495\linewidth}
    \includegraphics[width=8cm]{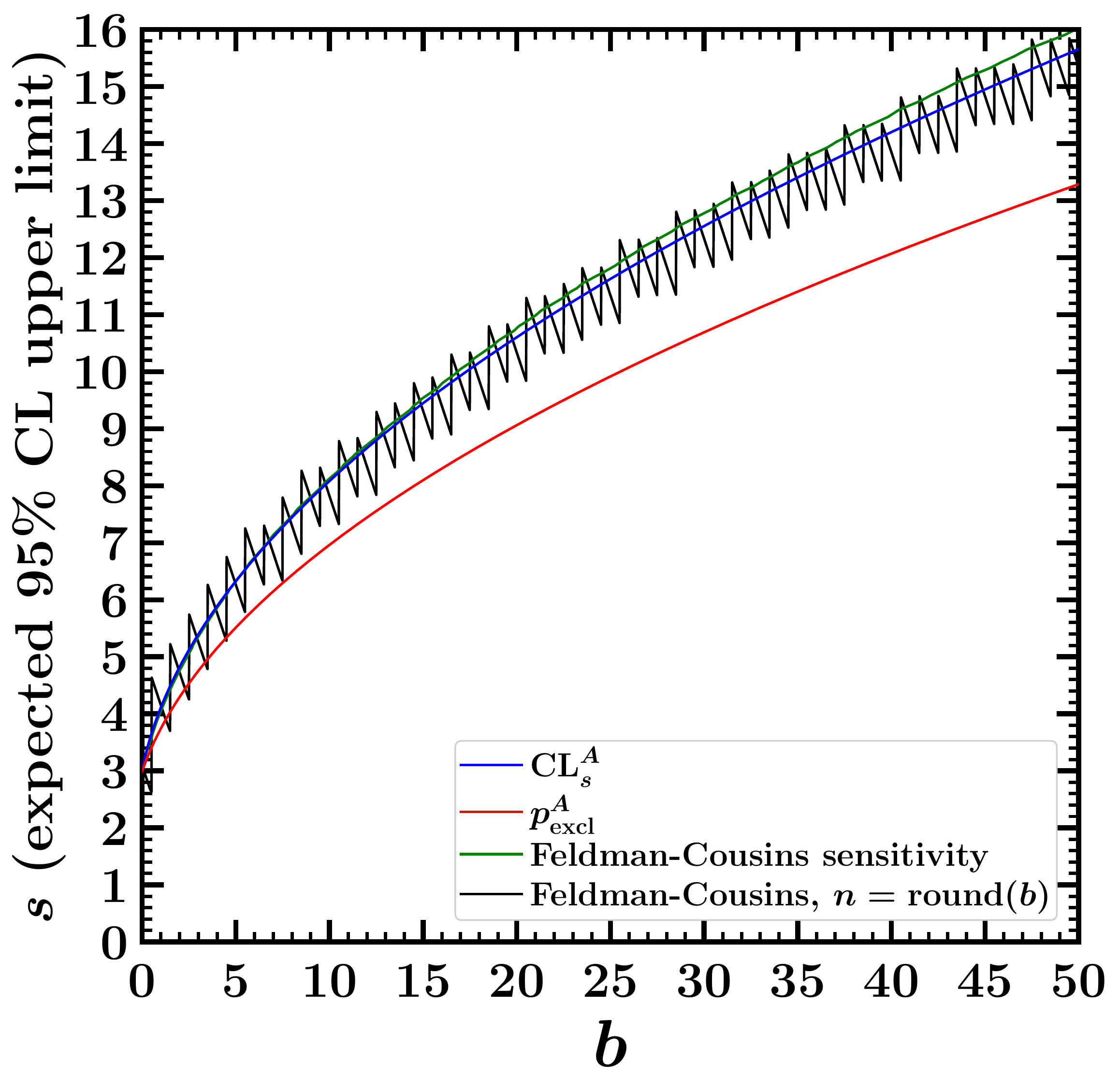}
  \end{minipage}
\begin{minipage}[]{0.95\linewidth}
 \caption{
  The expected 90\% CL (left panels) and 95\% CL (right panels) upper limit on signal
  as a function of the background mean, using the exact Asimov modified frequentist
  ${\rm CL}_s$ [blue lines, from eq.~(\ref{eq:CLsAsimov})] and standard frequentist
  $p$-value [red lines, from eq.~(\ref{eq:pexcl_singlechannelAsimov})],
  the Feldman-Cousins experimental sensitivity
  [green lines, from ref.~\cite{Feldman:1997qc} and eq.~(\ref{eq:FCsensitivity})],
  and the Feldman-Cousins method from ref.~\cite{Feldman:1997qc} with $n=\textrm{round}(b)$ [black lines].
  The top and bottom panels show the same information but with logarithmic and linear scales,
  respectively, for $b$.
\label{fig:sULexcl_vs_b}}
\end{minipage}
\end{figure}

In Figure~\ref{fig:sULexcl_vs_b}, we compare the expected 90\% CL (left panels)
and 95\% CL (right panels) upper limits on the signal mean $s$, obtained using the
exact Asimov ${\rm CL}_s$ (blue lines) and $p_{\rm excl}$ (red lines),
FC experimental sensitivity\footnote{In evaluating the FC sensitivity, we used the upper limits obtained by the Feldman-Cousins method for a fixed $n$ before requiring them to be non-increasing as a function of background mean. This does not make much difference as the FC upper limit differs from its non-increasing (with $b$) version only when the number of observed events are few compared to the expected background mean $b$, for which the probability of occurrence is small and will rapidly fall off for even smaller $n$.} (green lines), and FC upper limit with $n={\rm round}(b)$ (black lines). We note the following from the figure.
First, unlike the case with the observed upper limits (i.e. fixed $n$), the $p_{\rm excl}$
method gives sensible positive expected upper limits with the exact Asimov approach for all $b$,
but still is less conservative than the CL$_s$ and FC sensitivity results.
Second, the upper limit given by the FC method with $n={\rm round}(b)$ suffers from a
sawtooth problem and is therefore counter-intuitive and flawed as a method of comparing experimental prospects for different scenarios, as it implies that
an experiment could become more sensitive if it had larger background.
Finally, the FC sensitivity
and the upper limits given by
exact Asimov ${\rm CL}_s$ are both sensible as they increase monotonically with $b$,
and are also comparable at small backgrounds.
At large backgrounds, however, the FC sensitivity is slightly more conservative.
We also note that ${\rm CL}_s$ upper limits are much easier to evaluate than the FC upper limits.

\begin{figure}
  \begin{minipage}[]{0.495\linewidth}
    \includegraphics[width=8cm]{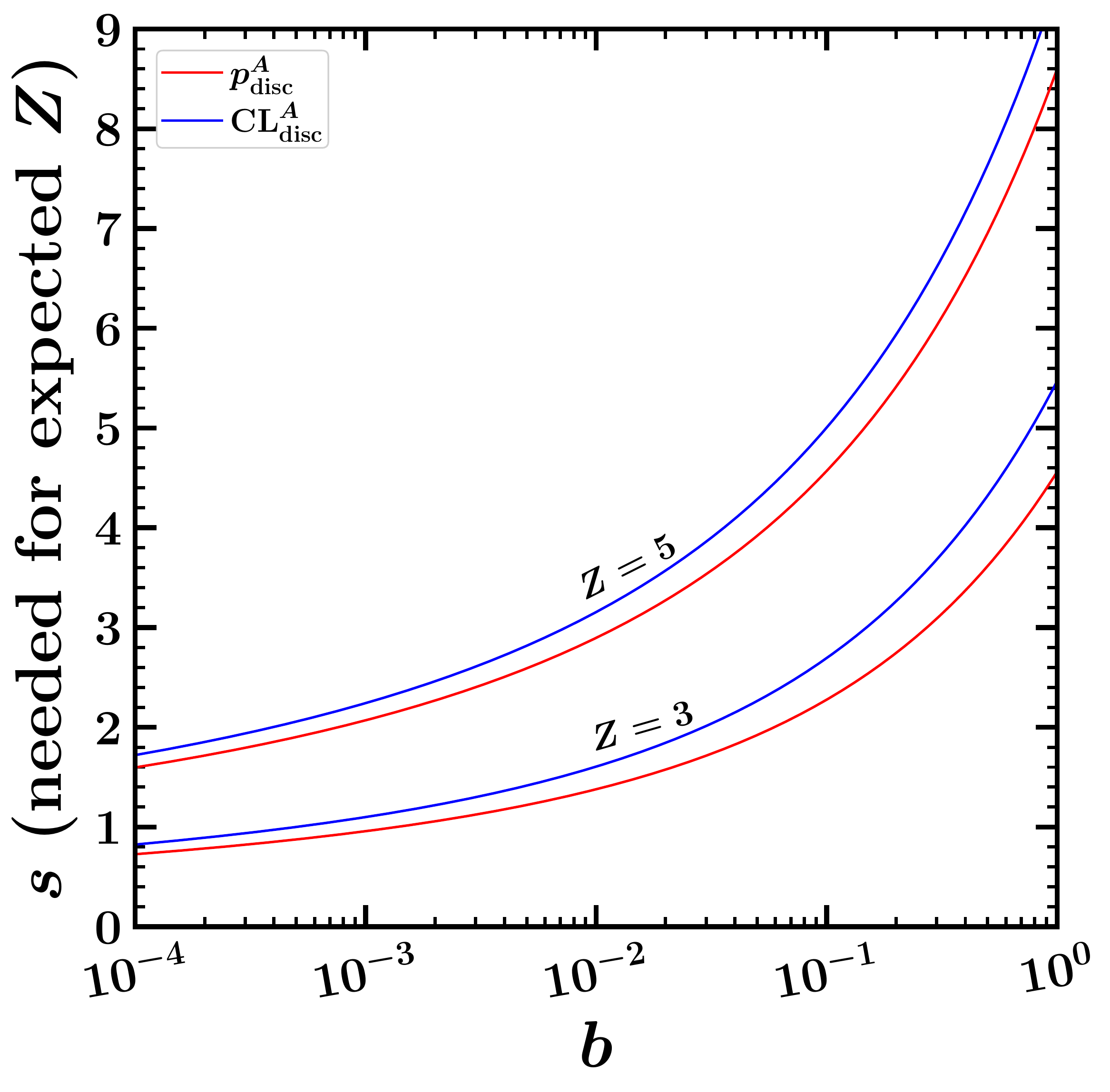}
  \end{minipage}
  \begin{minipage}[]{0.495\linewidth}
    \includegraphics[width=8cm]{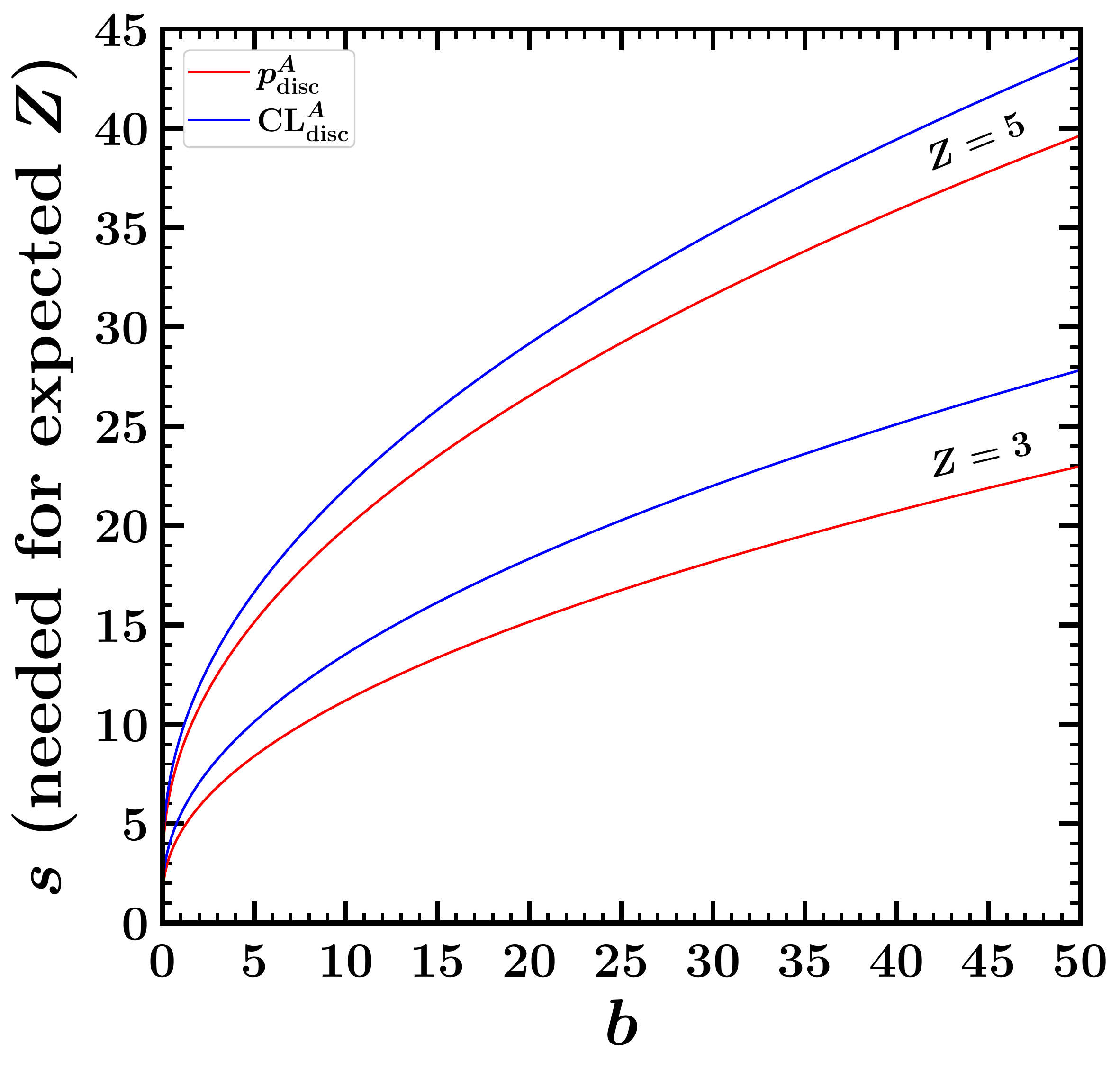}
  \end{minipage}
 \caption{The signal needed for an expected $Z=3$ evidence (lower curves) or $Z=5$ discovery (higher curves), as a function of background mean $b$,
using the exact Asimov $p_{\rm disc}$ [red lines, from eq.~(\ref{eq:pdisc_singlechannelAsimov})]
and ${\rm CL}_{\rm disc}$ [blue lines, from eq.~(\ref{eq:CLdiscAsimov})].
\label{fig:sdisc_vs_b}}
\end{figure}

We now turn to the issue of prospects for discovery, using the exact Asimov criterion. The signal mean needed for an expected discovery at a significance $Z$ is given by the solution for $s$ in
setting eq.~(\ref{eq:pdisc_singlechannelAsimov}) for $p_{\rm disc}$, or (\ref{eq:CLdiscAsimov}) for $\CLdisc$, equal to $\frac{1}{2} {\rm erfc}\left(\frac{Z}{\sqrt 2}\right)$ for the desired $Z$.
Figure~\ref{fig:sdisc_vs_b} compares the signals $s$ needed for an expected $Z=3$ evidence
or $Z=5$ discovery, as a function of background mean $b$, based on $p_{\rm disc}^A$ and $\CLdiscA$.
We note that as expected the results from $\CLdiscA$
are more conservative than those obtained from $p_{\rm disc}^A$.

\begin{figure}[!h]
  \begin{minipage}[]{0.55\linewidth}
    \includegraphics[width=0.94\linewidth]{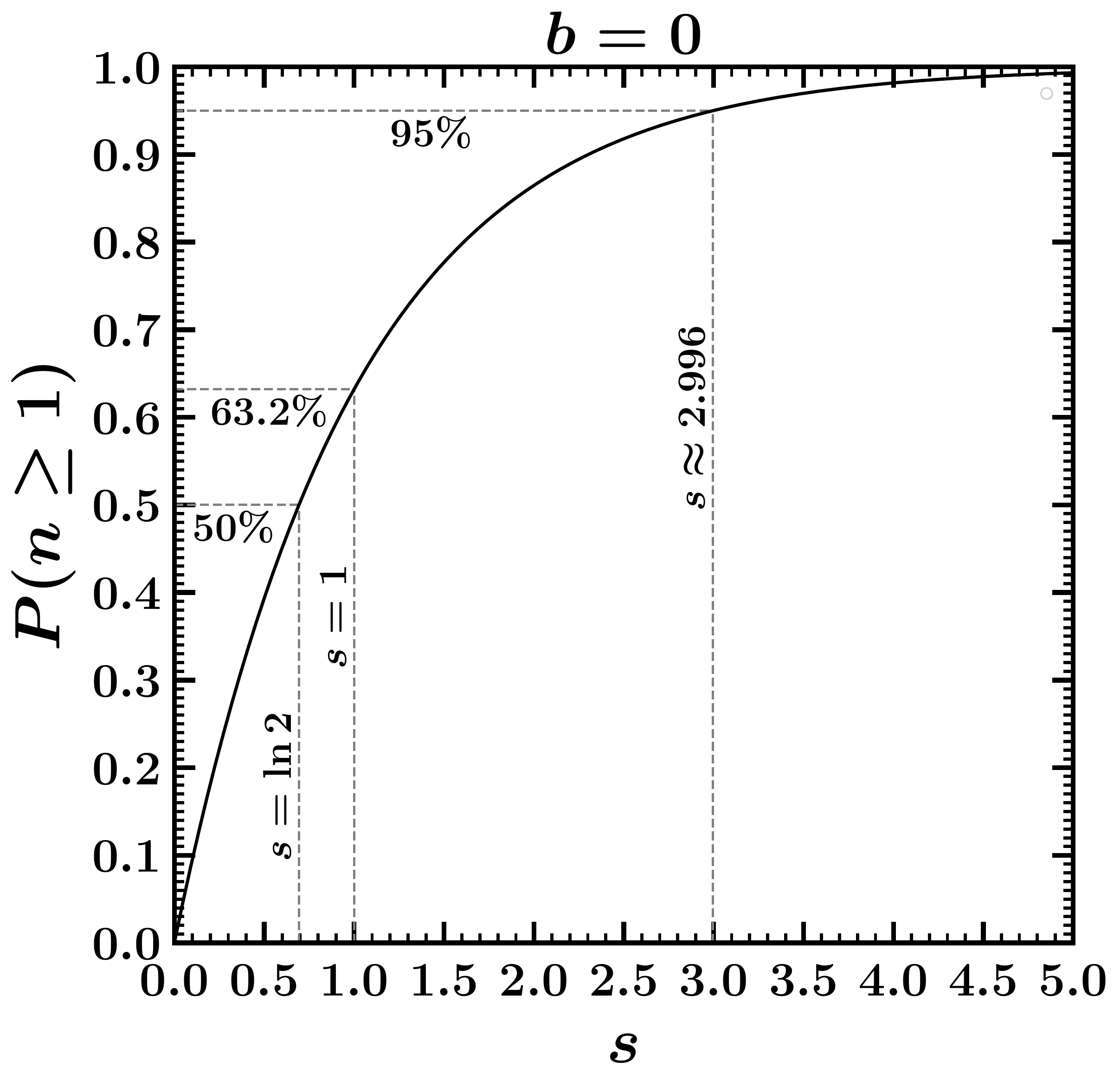}
  \end{minipage}
\begin{minipage}[]{0.44\linewidth}
 \caption{Probability of observing at least one event obtained from eq.~(\ref{eq:P_n_ge_1_disc}),
 as a function of the signal mean $s$, in the case with no background $b=0$.
\label{fig:P_n_ge_1_disc}}
\end{minipage}
\end{figure}

For very small $b$, note that for $Z=3$ the $s$ needed in Figure \ref{fig:sdisc_vs_b} is actually less than 1. Here, it is important to note that the discovery statistics $p_{\rm disc}$ and $\CLdisc$ are not well-defined in the strict background-free limit $b \rightarrow 0$. Specifically,
\beq
p_{\rm disc} (n, 0) &=&
\begin{cases}
0 \textrm{ if } n \ne 0\\[-5pt]
1 \textrm{ if } n = 0,
\end{cases}\\
{\rm CL}_{\rm disc} (n, 0, s) &=&
\begin{cases}
0 \textrm{ if } n \ne 0, s \ne 0\\[-5pt]
1 \textrm{ otherwise.}
\end{cases}
\eeq
Since $\langle n_{\rm disc} \rangle = s$ for $b=0$, the above implies that the exact Asimov expected discovery significances are both infinite, $Z(p^{\rm A}_{\rm disc}) = Z({\rm CL}^A_{\rm disc}) = \infty$, for any non-zero $s$ (however small). However, as a practical matter, it is clearly unreasonable to suggest an expectation of a discovery if the mean expected number of signal events is much less than 1.
Therefore, in order to be conservative, in cases with an extremely small background 
we can impose an additional requirement that $P(n \ge 1)$ should be greater than
some fixed value in order to claim an expected discovery. 
Figure~\ref{fig:P_n_ge_1_disc} shows the probability of observing at least one event,
\beq
P(n \ge 1) = \sum_{n = 1}^\infty P(n | s) = 1 - e^{-s},
\label{eq:P_n_ge_1_disc}
\eeq
as a function of signal mean $s$.
For example, if we require $P(n \ge 1) > (50\%, 63.2\%, 95\%)$ then the signal mean has to be $s > (0.693,\, 1.0,\, 2.996)$ respectively. Requiring $s > \ln 2 \approx 0.693$ guarantees the median number of events is at least 1, and $s > 1$ guarantees the expected mean number of events $\langle n_{\rm disc} \rangle > 1$.

\subsection{Exclusion for multi-channel counting experiments\label{subsec:exclusionmultichannel}}

Consider a counting experiment with $N$ independent channels. For each channel $i=1,\ldots,N$, the background and possible signal are assumed to be governed by Poisson distributions 
with means $b_i$ and $s_i$. For future convenience, define 
\beq
s &=& \sum_{i=1}^N s_i,
\\
r_i &=& s_i/s,
\eeq 
so that $s$ is the total mean expected signal in all channels, and the $r_i$ are the expected fractions of the total signal events for each channel.

Given an observation $\{n_i\}$, the $p$-value for exclusion is\footnote{In the following we use $\vec{n}$ as the argument 
of a function to denote the dependence on the full set $\{n_i\}$. 
This applies similarly for $\vec{b}$ and $\vec{s}$ 
to represent the dependences on $\{ b_i \}$ and $\{s_i\}$.} 
\beq
p_{\rm excl}(\vec{n}, \vec{b}, \vec{s}) 
&=&
\sum_{\{k_i\}}
\, 
\prod_{i=1}^N 
\,
P (k_i|s_i+b_i)
,
\label{eq:pexclmultichannel}
\eeq
where the sums over non-negative integer numbers of events $\{k_i\}$ are restricted according to the condition that 
\beq
Q(\vec{k}) \leq Q(\vec{n}). 
\label{eq:QkilessQni}
\eeq
where $Q$ is an appropriately chosen test-statistic with the property that larger $Q$ is more signal-like. 

We can also compute:
\beq
p_{\rm excl}(\vec{n}, \vec{b}, 0) 
&=&
\sum_{\{k_i\}}
\, 
\prod_{i=1}^N 
\,
P (k_i|b_i)
,
\label{eq:pexclbmultichannel}
\eeq
with the same restrictions on $k_i$ as in eq.~(\ref{eq:QkilessQni}).
Then we have
\beq
{\rm CL}_s(\vec{n}, \vec{b}, \vec{s}) &=& \frac{p_{\rm excl}(\vec{n}, \vec{b}, \vec{s})}{p_{\rm excl}(\vec{n}, \vec{b}, 0)} ,
\label{eq:CLsmultichannelPoisson}
\eeq
which is interpreted as the confidence level in the hypothesis that the signal is present.

For the single channel case, the obvious choice for $Q$ is the observed number of events, but in the multi-channel case one can consider different choices for $Q$. A simple and good choice\footnote{There are other choices, including the profile likelihood ratio, but these are more complicated and end up giving very similar (and often identical) results.} 
of test-statistic $Q$ is the likelihood ratio, 
\beq
q(\vec{n}, \vec{b}, \vec{s}) &=& \prod_{i=1}^N \frac{P(n_i|s_i+b_i)}{P(n_i|b_i)},
\eeq
which simplifies to
\beq
q &=& \prod_{i=1}^N e^{-s_i} \left (1 + \frac{s_i}{b_i}\right )^{n_i}
.
\label{eq:defineQmultichannelPoisson}
\eeq
It is more convenient to use instead
\beq
Q = \ln(q) = -s + \sum_{i=1}^N  n_i \ln(1 + s_i/b_i),
\label{eq:deflnQformultichannel}
\eeq 
which gives exactly the same results for $p_{\rm excl}$ and ${\rm CL}_s$ as $Q=q$, since $\ln(q)$ increases monotonically with $q$.
The contribution $-s$ is an irrelevant constant (independent of the data $\{n_i\}$),
so the use of $Q = \ln(q)$ amounts to taking the sum of the individual $n_i$'s, but weighting each of the channels by the factor 
$
w_i = \ln(1 + s_i/b_i).
$ 
This means that, using eq.~(\ref{eq:deflnQformultichannel}) in eq.~(\ref{eq:QkilessQni}), 
the restriction on the $\{k_i\}$
appearing in the sums in eqs.~(\ref{eq:pexclmultichannel}) and (\ref{eq:pexclbmultichannel}) becomes:
\beq
\sum_{i=1}^N (n_i - k_i) \ln(1 + s_i/b_i) &\geq& 0
.
\label{eq:restrictionkiLR}
\eeq

In contrast, the Bayesian way is to define, as a generalization of eq.~(\ref{eq:CLsBayes_excl}):
\beq
\CLexcl
(\vec{n},\vec{b},\vec{s}) &=&
\frac{
\displaystyle\scaleobj{0.85}{\int_{\scaleobj{1.4}{s}}^{\scaleobj{1.2}{\infty}}} ds' \, 
\scaleobj{0.85}{\prod_{\scaleobj{1.2}{i=1}}^{\scaleobj{1.2}{\hspace{1.9pt}N}}}\, P(n_i | r_i s'+ b_i)}{
\displaystyle\scaleobj{0.85}{\int_{\scaleobj{1.2}{0}}^{\scaleobj{1.2}{\infty}}} ds' \, 
\scaleobj{0.85}{\prod_{\scaleobj{1.2}{i=1}}^{\scaleobj{1.2}{\hspace{1.9pt}N}}}\, P(n_i | r_i s'+ b_i)}
.
\label{eq:multichannelCLsBayesian}
\eeq
Unlike in the special case of a single channel, $\CLexcl (\vec{n},\vec{b},\vec{s})$ defined in this way is not exactly equal to  ${\rm CL}_s(\vec{n},\vec{b},\vec{s})$
defined by eq.~(\ref{eq:CLsmultichannelPoisson}). Therefore, we will now study some simple test cases to illustrate the differences.

\begin{figure}
 \begin{minipage}[]{0.495\linewidth}
    \includegraphics[width=8cm]{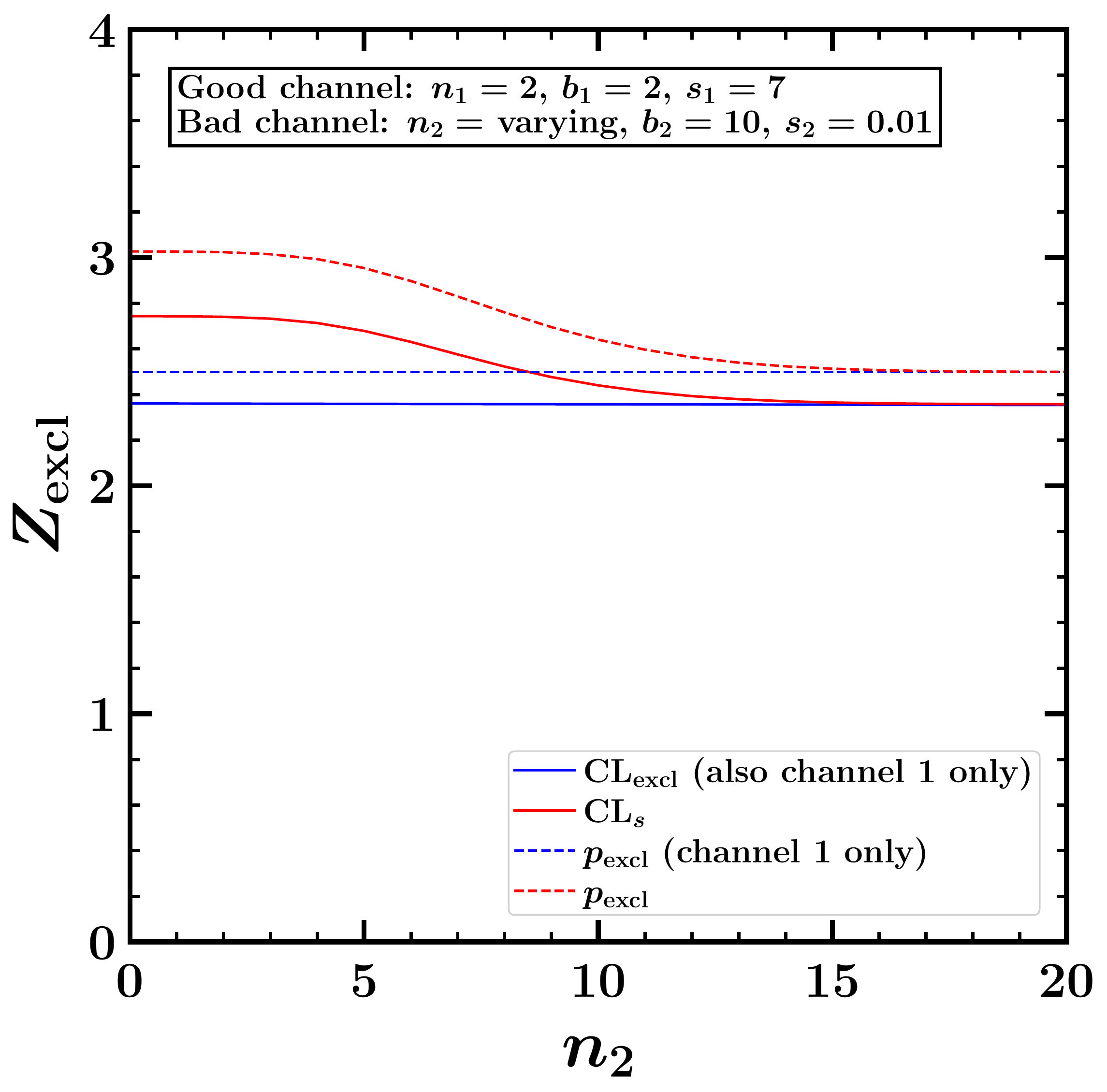}
 \end{minipage}
 \begin{minipage}[]{0.495\linewidth}
    \includegraphics[width=8cm]{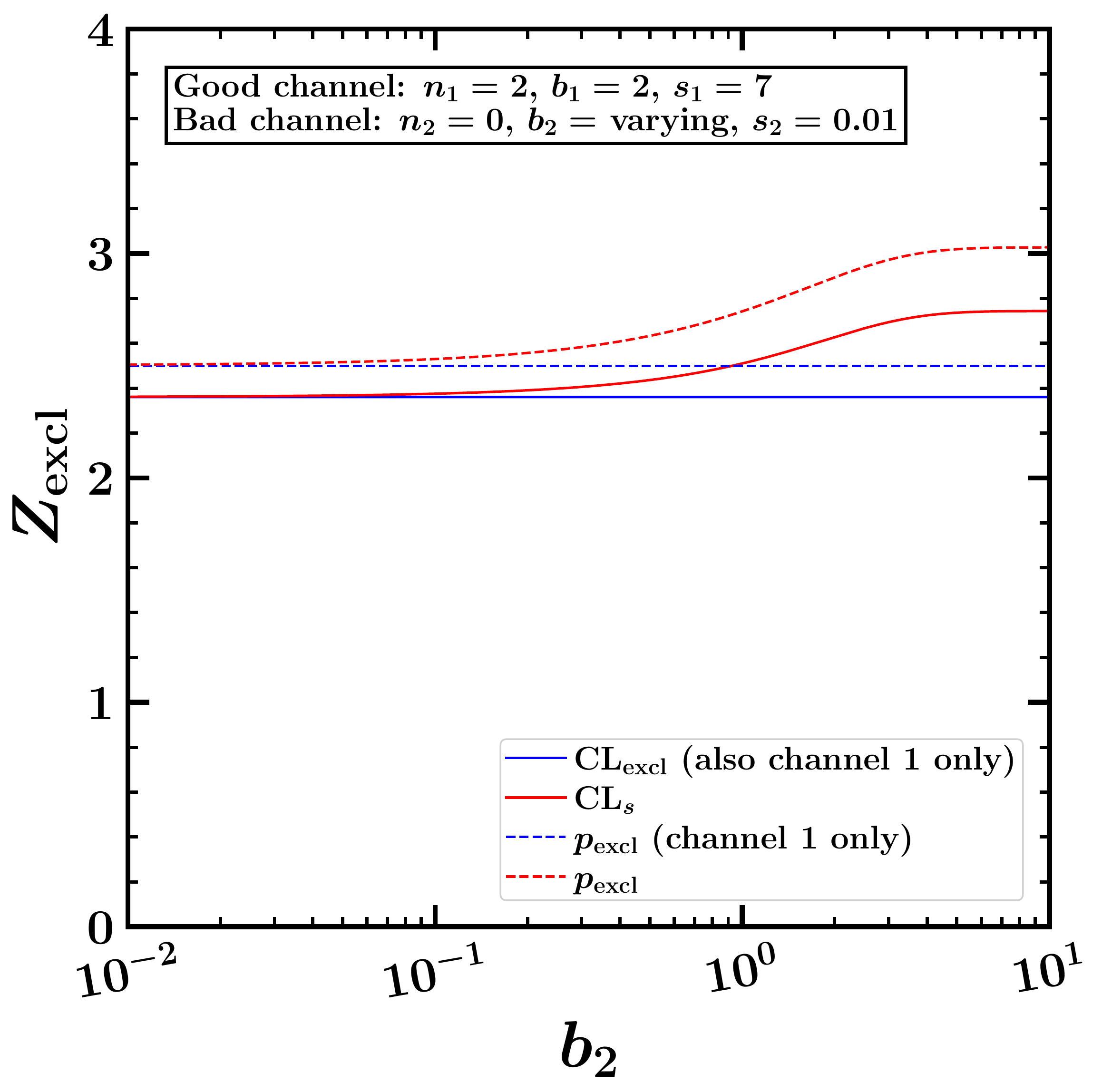}
 \end{minipage}
\begin{minipage}[]{0.95\linewidth} 
\caption{Comparison of exclusion significances $Z$ in the case of a counting experiment with a good channel and a bad channel. The solid lines are the modified frequentist ${\rm CL}_s$ [solid red line, from eqs.~(\ref{eq:pexclmultichannel})-(\ref{eq:CLsmultichannelPoisson}) and (\ref{eq:restrictionkiLR})] and $\CLexcl$ [solid blue line, from eq.~(\ref{eq:multichannelCLsBayesian})]. In this example, $\CLexcl$ is visually indistinguishable from the result  obtained from channel 1 only, conforming with the fact that channel 2 contains essentially no information about the signal. Also shown are the results for $p_{\rm excl}$ obtained from considering channel 1 only [dashed blue line, from eq.~(\ref{eq:pexcl_singlechannel})] and from both channels [dashed red line, from eqs.~(\ref{eq:pexclmultichannel}) and (\ref{eq:restrictionkiLR})]. 
\label{fig:CLsFvsB}}
\end{minipage}
\end{figure}

First, let us consider what happens when there are two channels, one of which (the ``bad", or non-informative channel) has a much lower signal and higher background than the other (the ``good channel"). As a specific numerical case, suppose:
\beq
&b_1 = 2,\qquad &s_1 = 7,\quad\qquad\> n_1 = 2,\qquad\qquad\>\>\mbox{(good channel)}
,
\\
&b_2 = 10,\qquad &s_2 = 0.01,\qquad n_2 = \text{varying},\qquad\mbox{(bad channel)}
.
\eeq
In this case, because the bad channel 2 has a tiny expected signal $s_2$ and a large background $b_2$, one intuitively expects it to provide essentially no information about the correctness of the signal hypothesis, no matter what $n_2$ is observed. Considering only the good channel 1, we obtain
\beq
p_{\rm excl} &=&  0.006232, \qquad Z_{\rm excl} = 2.4987\qquad \mbox{(channel 1 alone)},
\\
{\rm CL}_s \>=\> \CLexcl &=&  0.009210,\qquad Z_{\rm excl} =2.3571,\qquad
\mbox{(channel 1 alone)}
.
\eeq
However, combining both channels using the formulas (\ref{eq:pexclmultichannel}), (\ref{eq:CLsmultichannelPoisson}), and (\ref{eq:multichannelCLsBayesian}) above, we 
have the results shown in the left panel of Figure \ref{fig:CLsFvsB}.
Counterintuitively, adding another channel with a larger background and almost no expected signal has increased our confidence in the exclusion as measured by either the frequentist 
$p_{\rm excl}$ or the modified frequentist
${\rm CL}_s$ measures, when $n_2$ is small. 
In contrast, $\CLexcl$ behaves as intuitively expected; the result obtained including both channels is numerically almost independent of $n_2$ and almost identical to the result obtained only from channel 1.

To understand the origin of this counterintuitive effect for $p_{\rm excl}$ and CL$_{s}$, let us consider which integers $k_1, k_2$ contribute to the sums in 
eqs.~(\ref{eq:pexclmultichannel}) and (\ref{eq:pexclbmultichannel}). In general, $k_1=0$
and $1$ each contribute for a very large range of $k_2$, so that very nearly we have a factor
$\sum_{k_2 = 0}^\infty P(k_2|s_2 + b_2) \approx 1$ for channel 2 in eq.~(\ref{eq:pexclmultichannel}). However, for $k_1 = n_1 = 2$, we only get a factor of $\sum_{k_2 = 0}^{n_2} P(k_2|s_2 + b_2) < 1$
contributing to the $p$-values. The problem boils down to this fact: for the contributions with $k_1 = n_1$, only a subset of the $k_2$ values contribute, even though any result for $k_2$ should give us essentially no information about the presence of the (tiny) signal. This explains why the counterintuitive problem disappears for reasonably large $n_2$, where we see from the left panel of Figure \ref{fig:CLsFvsB} that $\CLexcl \approx {\rm CL}_s$ and $p_{\rm excl}$ agree with their counterparts from channel 1 only. 

To show another facet of this disturbing effect, in the right panel of Figure \ref{fig:CLsFvsB} we use the same data except that $n_2=0$ is fixed and $b_2$ is varying. Again, we see that despite channel 2 containing essentially no information about the signal, the modified frequentist ${\rm CL}_s$ including both channels depends on $b_2$, while $\CLexcl$ is almost exactly flat, conforming to intuitive expectation.

\begin{figure}
\begin{minipage}[]{0.495\linewidth}
    \includegraphics[width=\linewidth]{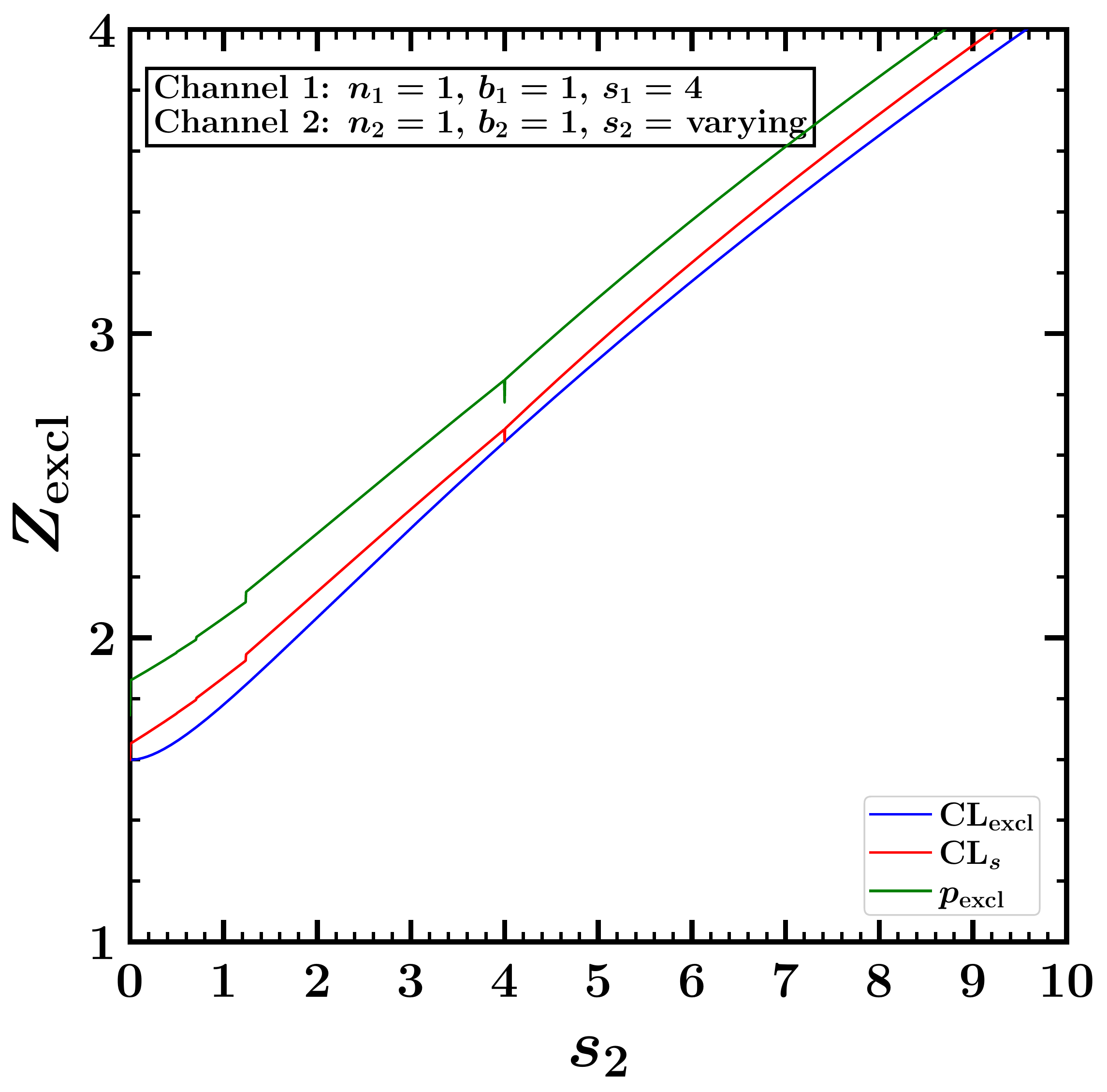}
\end{minipage}
\begin{minipage}[]{0.495\linewidth}
    \includegraphics[width=\linewidth]{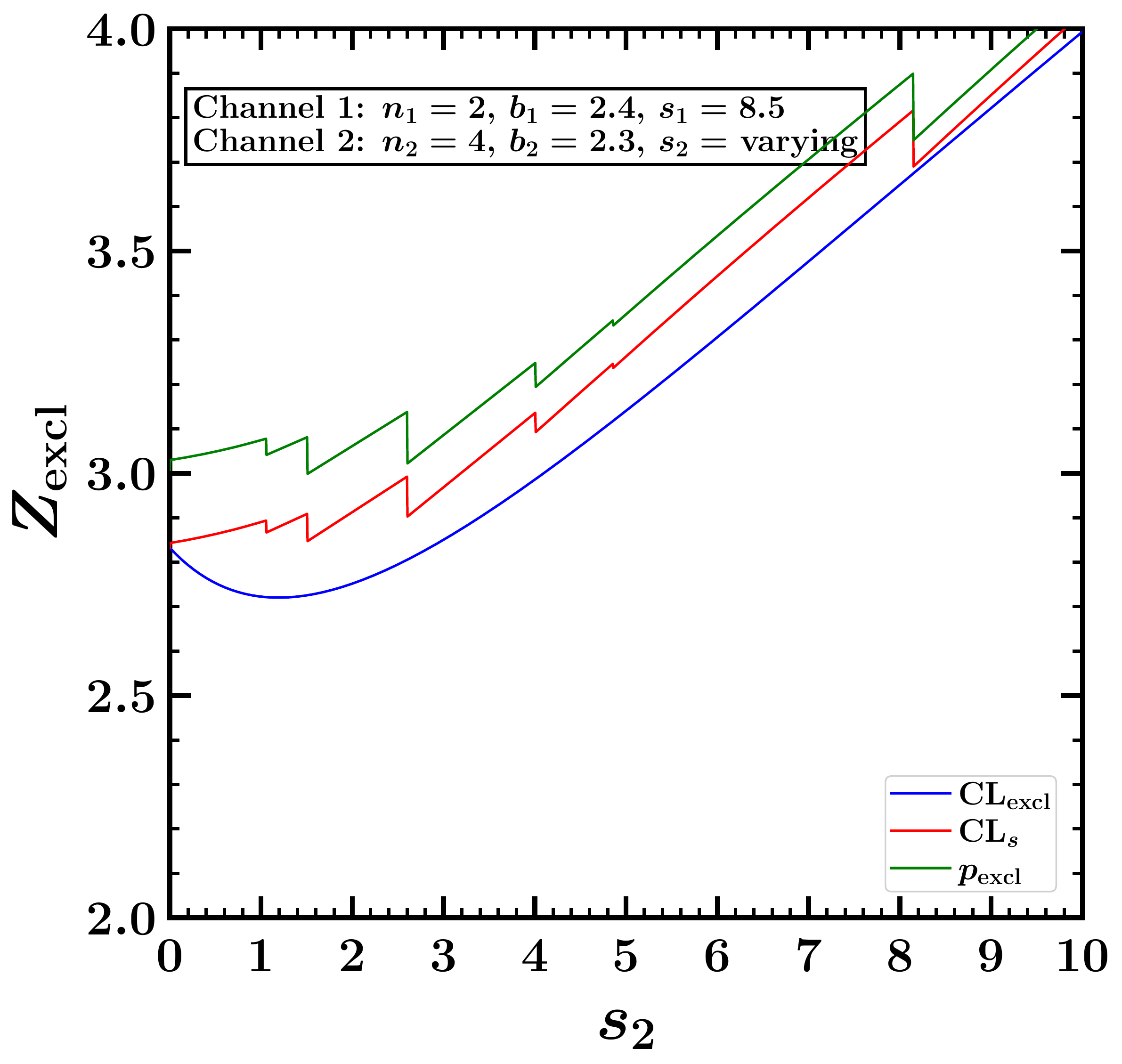}
\end{minipage}
\begin{minipage}[]{0.95\linewidth}  
   \caption{Comparison of exclusion significances $Z$ obtained from 
${\rm CL}_s$ [red line, from eqs.~(\ref{eq:pexclmultichannel})-(\ref{eq:CLsmultichannelPoisson}) and (\ref{eq:restrictionkiLR})] 
and $\CLexcl$ [blue line, from eq.~(\ref{eq:multichannelCLsBayesian})]
and $p_{\rm excl}$ [green line, from eqs.~(\ref{eq:pexclmultichannel}) and (\ref{eq:restrictionkiLR})],
for the test cases of eqs.~(\ref{eq:casestudy11411var1})-(\ref{eq:casestudy11411var2}) [left panel] and (\ref{eq:casestudy3})-(\ref{eq:casestudy4}) [right panel].
The results for $p_{\rm excl}$ and CL$_s$ exhibit discontinuities as $s_2$ is varied, due to abrupt changes in which outcomes $k_1$ and $k_2$ are summed over. The
Bayesian version $\CLexcl$ does not have such discontinuities.
\label{fig:CLsFvsBagain}}
\end{minipage}
\end{figure}

Another study case is shown in the first panel of Figure \ref{fig:CLsFvsBagain}, with:
\beq
&&n_1 = 1,\qquad b_1 = 1,\qquad s_1 = 4,
\label{eq:casestudy11411var1}
\\
&&n_2 = 1,\qquad b_2 = 1,\qquad s_2 = {\rm varying}.
\label{eq:casestudy11411var2}
\eeq
The variation of exclusion significances as a function of $s_2$ is shown in the first panel of Figure \ref{fig:CLsFvsBagain}. For $s_2 = s_1 = 4$ exactly, the results satisfy ${\rm CL}_s = \CLexcl$ and agree precisely with the result that would be obtained for a single combined channel with $n=2$, $b=2$, $s=8$. 
However, the $Z$ value for CL$_s$ has a small discontinuity at exactly $s_2 = 4$, such that
for all other values of $s_2$, ${\rm CL}_s$ has a higher exclusion significance $Z$ than 
$\CLexcl$.
Numerically:
\beq
{\rm CL}_s &=& 0.004093,\quad(Z = 2.644),\qquad \mbox{(for $s_2 = 4$)}
\label{eq:casestudy111144}
\\
{\rm CL}_s &=& 0.003616,\quad(Z = 2.686),\qquad \mbox{(for $s_2 = 4 \pm \epsilon$)},
\eeq
for $\epsilon$ arbitrarily small but non-zero.
This discontinuity can be traced to the fact that for $s_2=4$ exactly, the weights satisfy $w_1 = w_2$ exactly for the two channels, which affects which integers are summed over due to eq.~(\ref{eq:restrictionkiLR}). There are also discontinuities in CL$_s$ at $s_2=\sqrt{5}-1 \approx 1.23607$, where $w_1 = 2 w_2$, and at $s_2 = 5^{1/3} - 1 \approx 0.709976$, where $w_1 = 3 w_2$, etc. 

For another case study, consider:
\beq
&&n_1 = 2,\qquad b_1 = 2.4,\qquad s_1 = 8.5,
\label{eq:casestudy3}
\\
&&n_2 = 4,\qquad b_2 = 2.3,\qquad s_2 = {\rm varying}.
\label{eq:casestudy4}
\eeq
The results are depicted in the second panel of Figure \ref{fig:CLsFvsBagain}, and show more pronounced discontinuities in both frequentist $p_{\rm excl}$ and the modified frequentist ${\rm CL}_s$. In contrast, the Bayesian result $\CLexcl$ is smooth as we vary $s_2$, and gives more
conservative exclusion significances. 

\begin{figure}
\begin{minipage}[]{0.495\linewidth}
    \includegraphics[width=\linewidth]{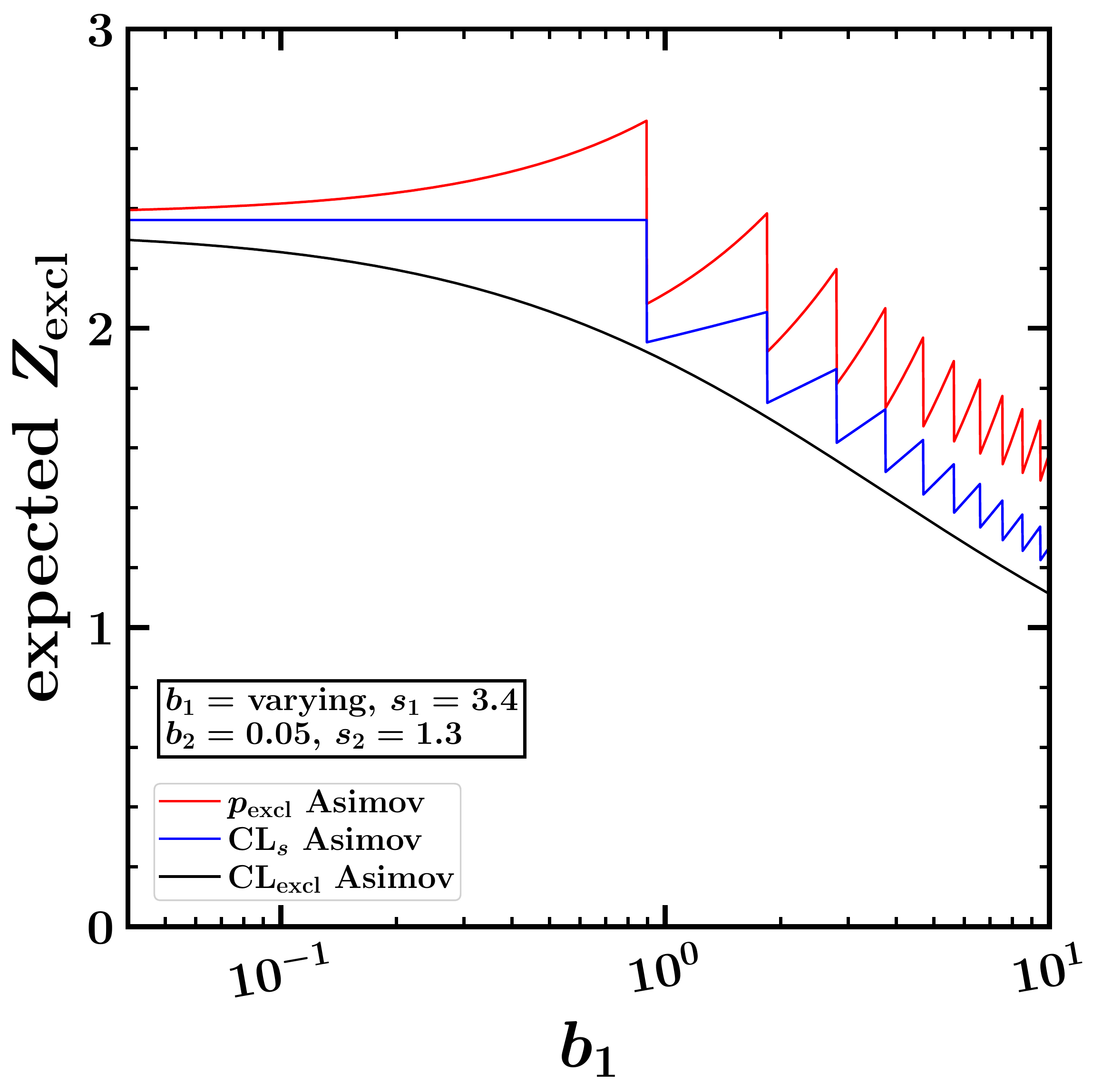}
\end{minipage}
\begin{minipage}[]{0.495\linewidth}
    \includegraphics[width=\linewidth]{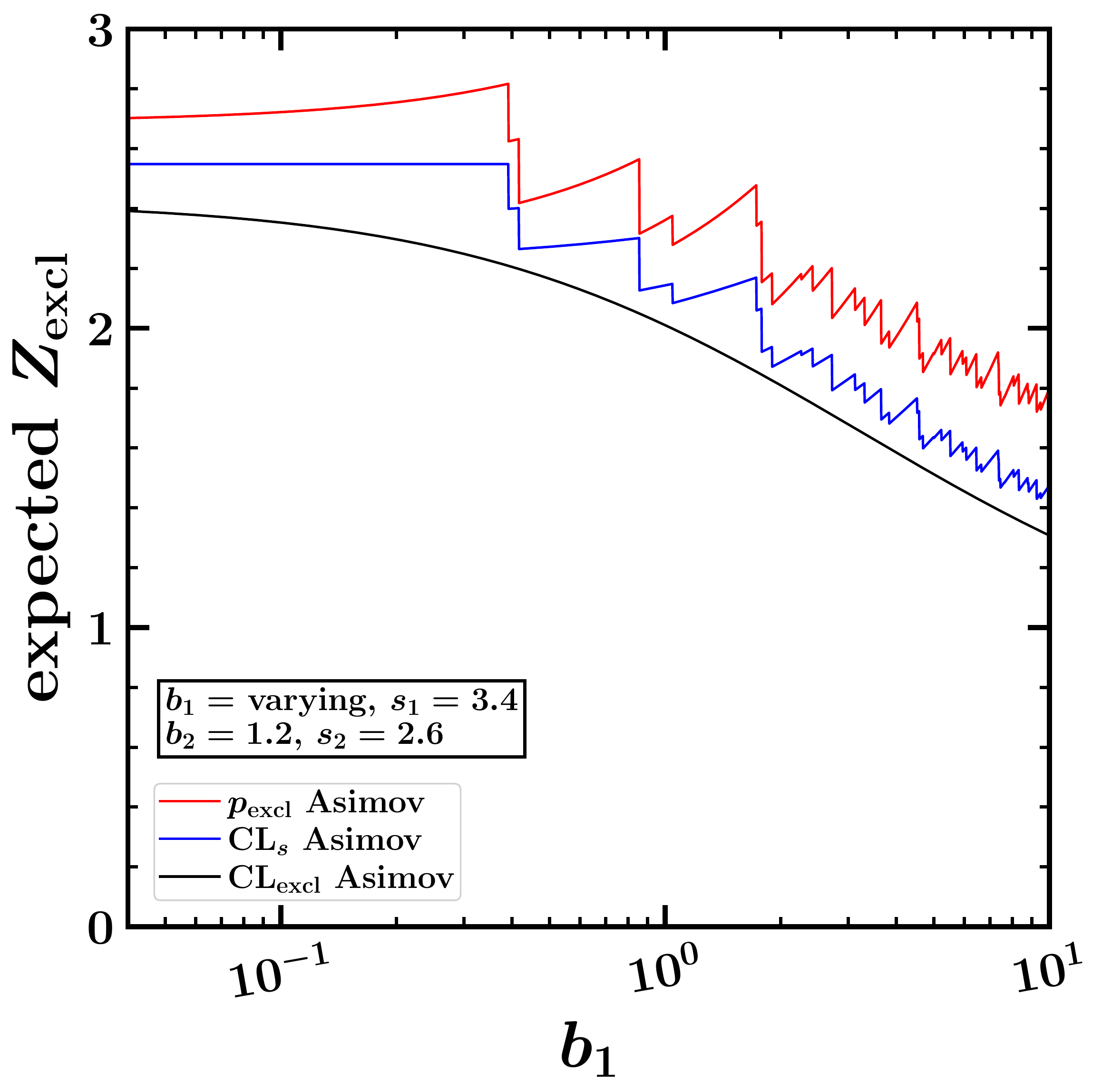}
\end{minipage}
\begin{minipage}[]{0.95\linewidth}  
\caption{Comparison of the Asimov expected exclusion significances $Z$ obtained from 
$p_{\rm excl}$ [red lines, from eq.~(\ref{eq:pexclmultichannel})],
${\rm CL}_s$ [blue lines, from eq.~(\ref{eq:CLsmultichannelPoisson})],
and 
$\CLexcl$ [black lines, from eq.~(\ref{eq:multichannelCLsBayesian})],
for two test cases with two independent channels, as labeled. The Asimov results are obtained by setting $n_i = b_i$ for each channel.
Due to the non-continuous effect of the restriction of eq.~(\ref{eq:restrictionkiLR}), the Asimov $p_{\rm excl}$ and ${\rm CL}_s$ have a counterintuitive non-monotonic behavior as the first channel background mean $b_1$ is varied, while the Asimov $\CLexcl$ is monotonic in the expected way.  
\label{fig:ZAsimovexcl}}
\end{minipage}
\end{figure}

Let us now consider the question of projecting expected exclusion significances for future experiments. In the multi-channel case, one can define Asimov results
for $p_{\rm excl}$ and ${\rm CL}_s$
by replacing each $n_i$ in eqs.~(\ref{eq:pexclmultichannel}) and (\ref{eq:CLsmultichannelPoisson}) by the mean expected result $b_i$ in the restriction eq.~(\ref{eq:restrictionkiLR}). However,
in the multi-channel case, the resulting sets of $\{k_i\}$ that contribute to the sums will depend discontinuously on the $\{s_i\}$ and $\{b_i\}$, leading to the same sort of sawtooth problems that occurs in the median expected significance. In particular, an increase in the backgrounds often leads, counterintuitively, to a larger expected significance. (This problem did not occur in the single-channel case, because the sum $\sum_{k=0}^n$ was evaluated in closed form in terms of 
incomplete $\Gamma$ functions, after which the argument $n$ could be interpreted as a continuous real number rather than an integer.) 
In contrast, if one uses $\CLexcl(\vec{n},\vec{b},\vec{s})$, then the exact Asimov method is perfectly straightforward and continuous, since it does not involve sums over integers subject to restrictions. Thus one can simply replace $n_i$ by $b_i$ in eq.~(\ref{eq:multichannelCLsBayesian}) to obtain the exact Asimov result. The Asimov results for $p_{\rm excl}$, ${\rm CL}_s$, and $\CLexcl$ are compared in Figure \ref{fig:ZAsimovexcl} for two test cases, showing the sawtooth behavior of the first two and the smooth, monotonic (and more conservative) behavior of the latter.

In view of the preceding discussion, we propose $\CLexcl$ in eq.~(\ref{eq:multichannelCLsBayesian}) as the preferred
statistic for exclusion for multi-channel counting experiments. Unlike $p_{\rm excl}$ and CL$_s$ 
(with which it coincides in the single-channel case), it does not suffer from the problem of being affected significantly by the presence of a bad channel, and does not have discontinuities when signal and background means are changed infinitesimally. The exact Asimov result is straightforward to obtain and behaves continuously and monotonically in the expected way with respect to changes in the background. Furthermore, the introduction of 
background uncertainties and probability distributions for nuisance parameters is more straightforward, avoiding discontinuities in the integrand, as we will see below.

\subsection{Discovery for multi-channel counting experiments\label{subsec:discoverymultichannel}}

For the discovery case, the frequentist $p$-value is defined by 
\beq
p_{\rm disc}(\vec{n}, \vec{b}, \vec{s}) &=& \sum_{\{k_i\}} \prod_{i=1}^N P(k_i| b_i).
\label{eq:pdiscmultichannel}
\eeq
The sum over $\{k_i\}$ is restricted by the condition that the test-statistic $\ln(q)$
defined by eq.~(\ref{eq:deflnQformultichannel}) is not smaller for $\{k_i\}$ than for the observed data $\{n_i\}$, so:
\beq
\sum_{i=1}^N (n_i - k_i) \ln (1 + s_i/b_i) &\leq& 0.
\label{eq:pdiscmultichannelkirestriction}
\eeq
Unlike the single-channel special case, $p_{\rm disc}$ depends on the signal strengths $s_i$ when there is more than one channel because of this restriction.
Note that the inequality has the opposite sense compared to the exclusion case, eq.~(\ref{eq:restrictionkiLR}). 

\begin{figure}
\mbox{
\includegraphics[width=8.0cm]{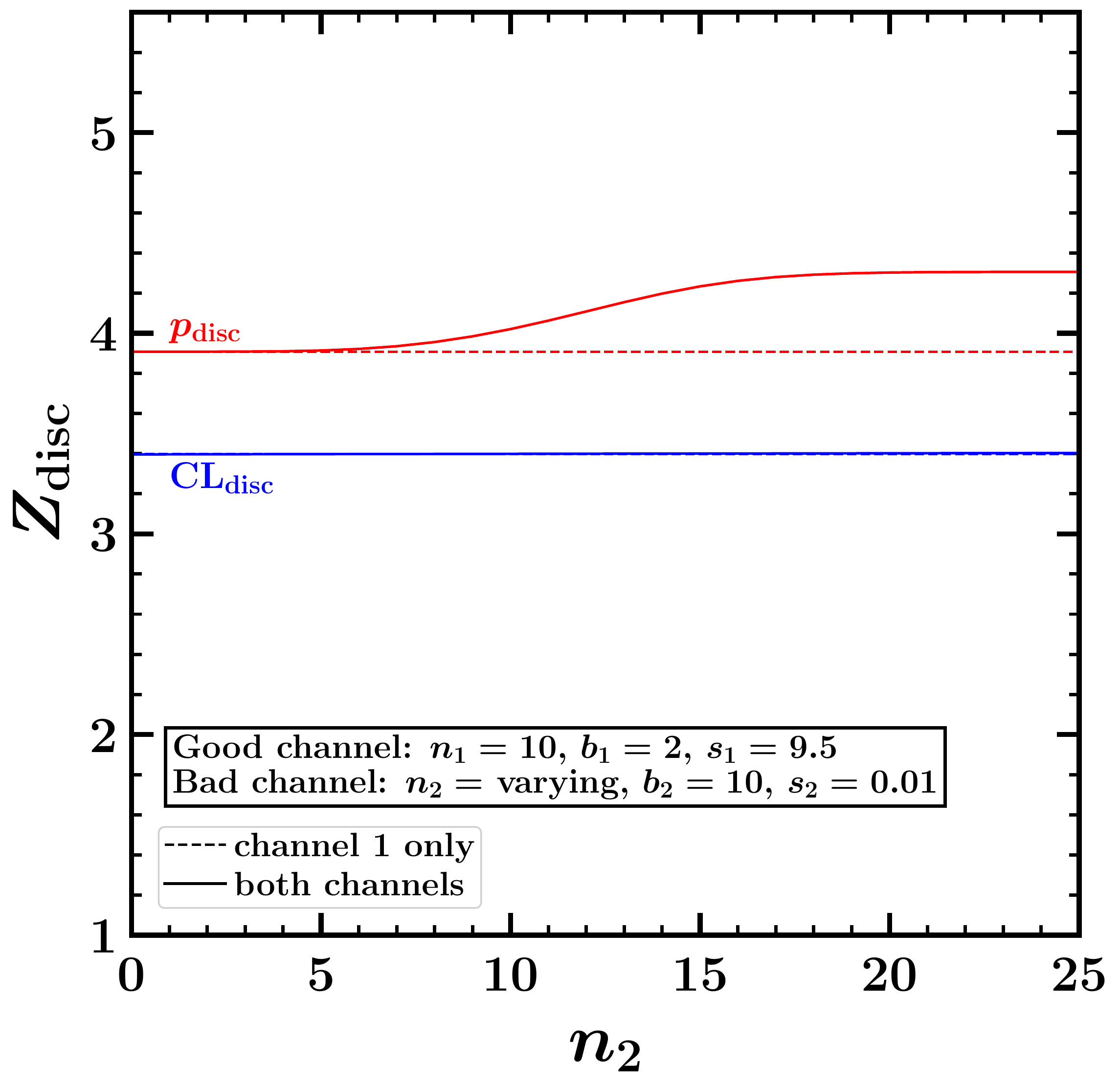}
\includegraphics[width=8.0cm]{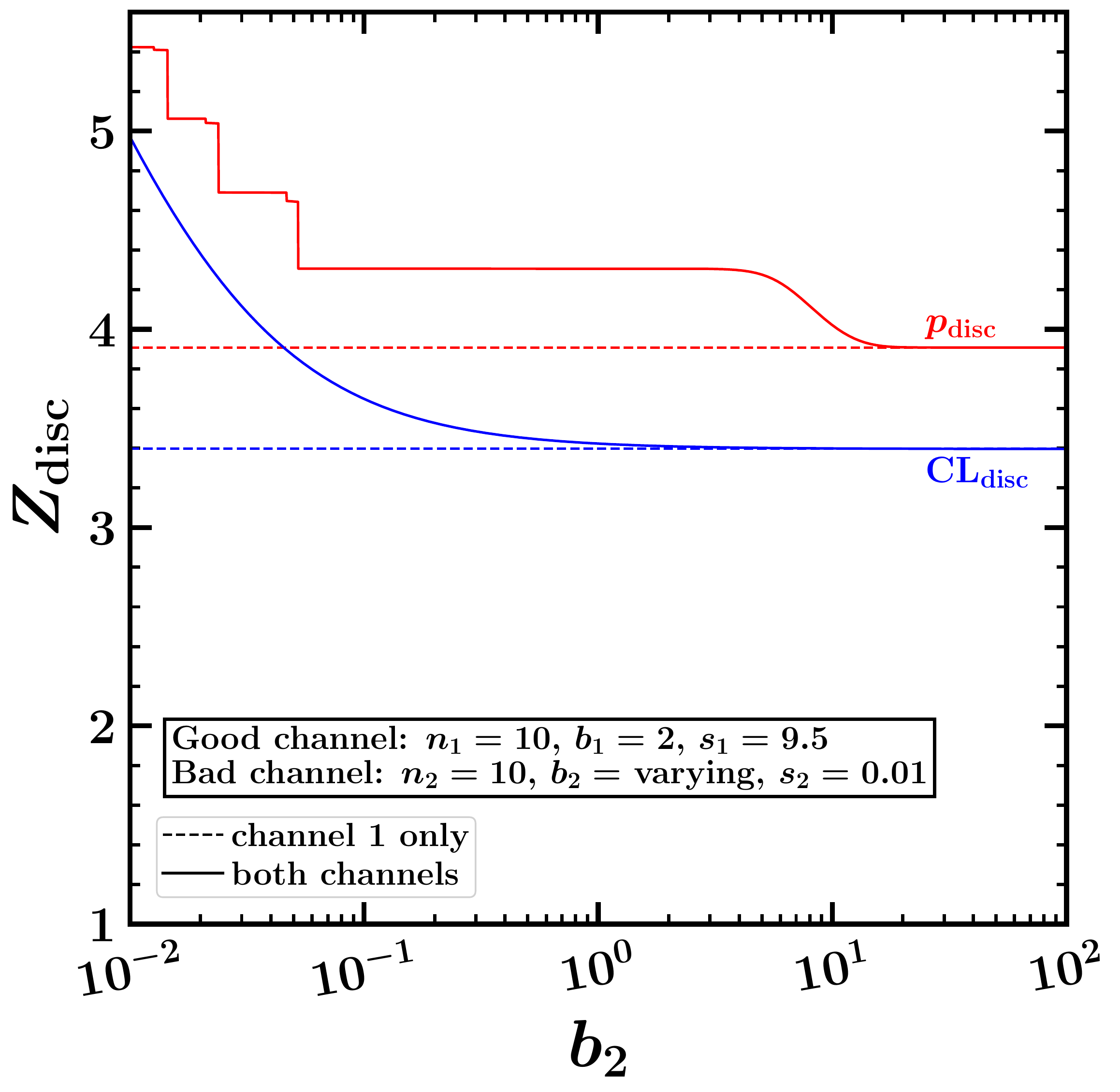}
}
\begin{minipage}[]{0.95\linewidth}
\caption{Comparison of discovery significance $Z$ in the case of a counting experiment with a good channel and a bad channel. The solid  lines are obtained from 
$p_{\rm disc}$ [red lines, from eqs.~(\ref{eq:pdiscmultichannel}) and (\ref{eq:pdiscmultichannelkirestriction})] and $\CLdisc$ [blue lines, from eq.~(\ref{eq:CLBdiscoverymultichannel})]. The dashed lines are obtained in the same way, but considering only the data from channel 1. In this example, $\CLdisc$ is more resistant to the effects of the non-informative channel, except in the case that $b_2$ is very small.
The step function discontinuities in $p_{\rm disc}$ in the right panel 
are not numerical artifacts, but
occur at values of $b_2$ such that the ratio of weights $w_1/w_2 = \ln(1+s_1/b_1)/\ln(1 + s_2/b_2)$ is rational.
\label{fig:casestudydiscgoodbad}}
\end{minipage}
\end{figure}

A more conservative, and simpler, alternative to $p_{\rm disc}(\vec{n}, \vec{b}, \vec{s})$ is the generalization of eq.~(\ref{eq:CLdiscsinglechannel}),
\beq
\CLdisc(\vec{n}, \vec{b}, \vec{s}) 
&=&
\prod_{i=1}^N\frac{ P(n_i|b_i)}{P(n_i| s_i + b_i)}
.
\label{eq:CLBdiscoverymultichannel}
\eeq
In order to compare these criteria for discovery, we first consider a case with one good channel and one bad channel, starting from the following values:
\beq
&b_1 = 2,\qquad &s_1 = 9.5,\qquad\>\> n_1 = 10,\qquad\mbox{(good channel)}
,
\\
&b_2 = 10,\qquad &s_2 = 0.01,\qquad n_2 = 10,\qquad\mbox{(bad channel)}
.
\eeq
In Figure \ref{fig:casestudydiscgoodbad}, we show the results for the discovery significance $Z$
obtained from $p_{\rm disc}$ and $\CLdisc$, considering variations in both $n_2$ and $b_2$ as the other quantities are held fixed, and compare to the same results using only channel 1.
As in the exclusion case, we note that $p_{\rm disc}$ is affected in a non-trivial way by the presence of the bad channel, contrary to intuitive expectations.
The step function discontinuities in $p_{\rm disc}$ are not a numerical artifact, but
occur at values of $b_2$ such that the ratio of weights $w_1/w_2 = \ln(1+s_1/b_1)/\ln(1 + s_2/b_2)$ is a rational number, so that the integer 
number of terms appearing in the $\sum_{\{k_i\}}$ in 
eq.~(\ref{eq:pdiscmultichannel}) changes discontinuously.

In contrast, $\CLdisc$ is seen to be much less affected by the presence of the bad channel. 
The reason for this is that for any channel $i$ with very small $s_i$, the numerator and denominator factors for that channel will cancel in the limit $s_i/b_i \rightarrow 0$ in eq.~(\ref{eq:CLBdiscoverymultichannel}).
The exception (in the right panel of Figure \ref{fig:casestudydiscgoodbad}) occurs in the
case that $b_2$ is also small, in which case $n_2=10$ is a surprising outcome for both
the background-only and background+signal hypotheses. 

\begin{figure}
\mbox{
\includegraphics[width=8.0cm]{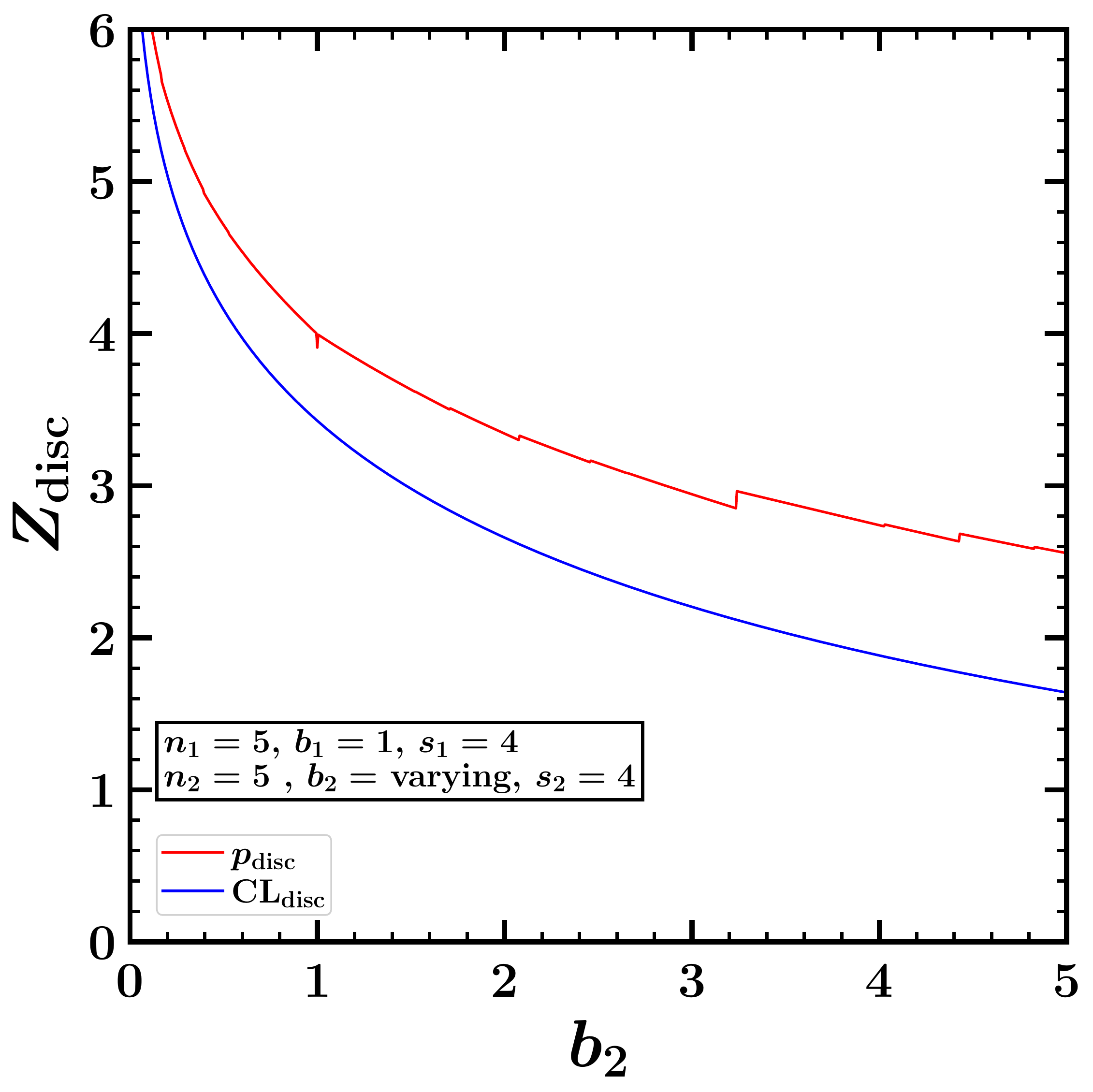}
\includegraphics[width=8.0cm]{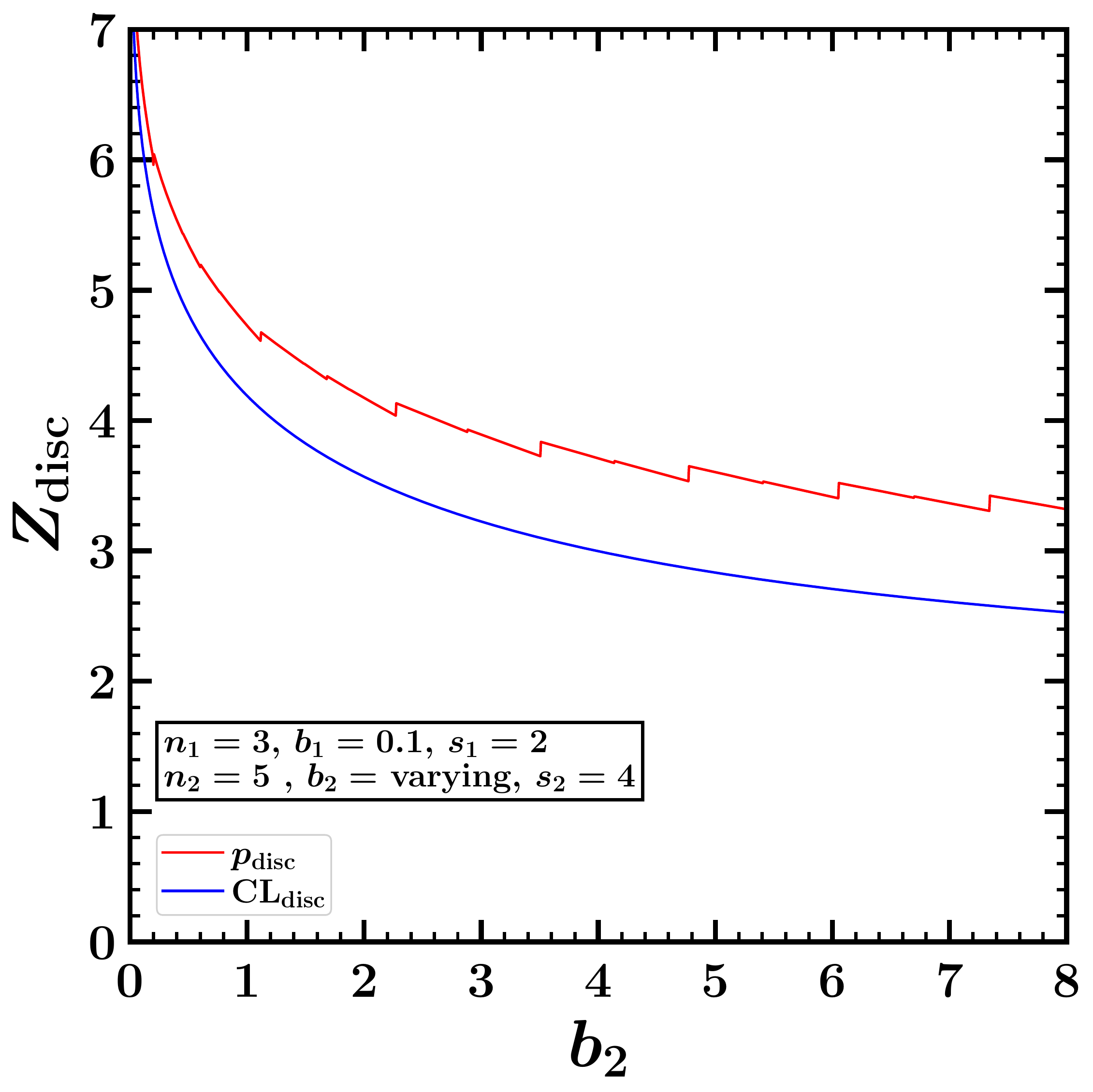}}
\caption{Comparison of significances $Z$ for discovery, obtained using $p_{\rm disc}$ [red lines, from eqs.~(\ref{eq:pdiscmultichannel}) and (\ref{eq:pdiscmultichannelkirestriction})] and $\CLdisc$ [blue lines, from eq.~(\ref{eq:CLBdiscoverymultichannel})], for two 2-channel test cases with data as labeled.
\label{fig:casestudydisctwo}}
\end{figure}

Further comparisons between the significances obtained from $p_{\rm disc}$ and $\CLdisc$ for two test cases are shown in Figure \ref{fig:casestudydisctwo}.
The results obtained from $p_{\rm disc}$ have numerous discontinuities, which are small numerically but have the disturbing property of being non-monotonic as the background $b_2$ is varied. The results from $\CLdisc$ are reliably more conservative, as we have already noted, and do not suffer discontinuities because there is no restricted sum over integers in its definition.

\begin{figure}
\begin{minipage}[]{0.495\linewidth}
    \includegraphics[width=\linewidth]{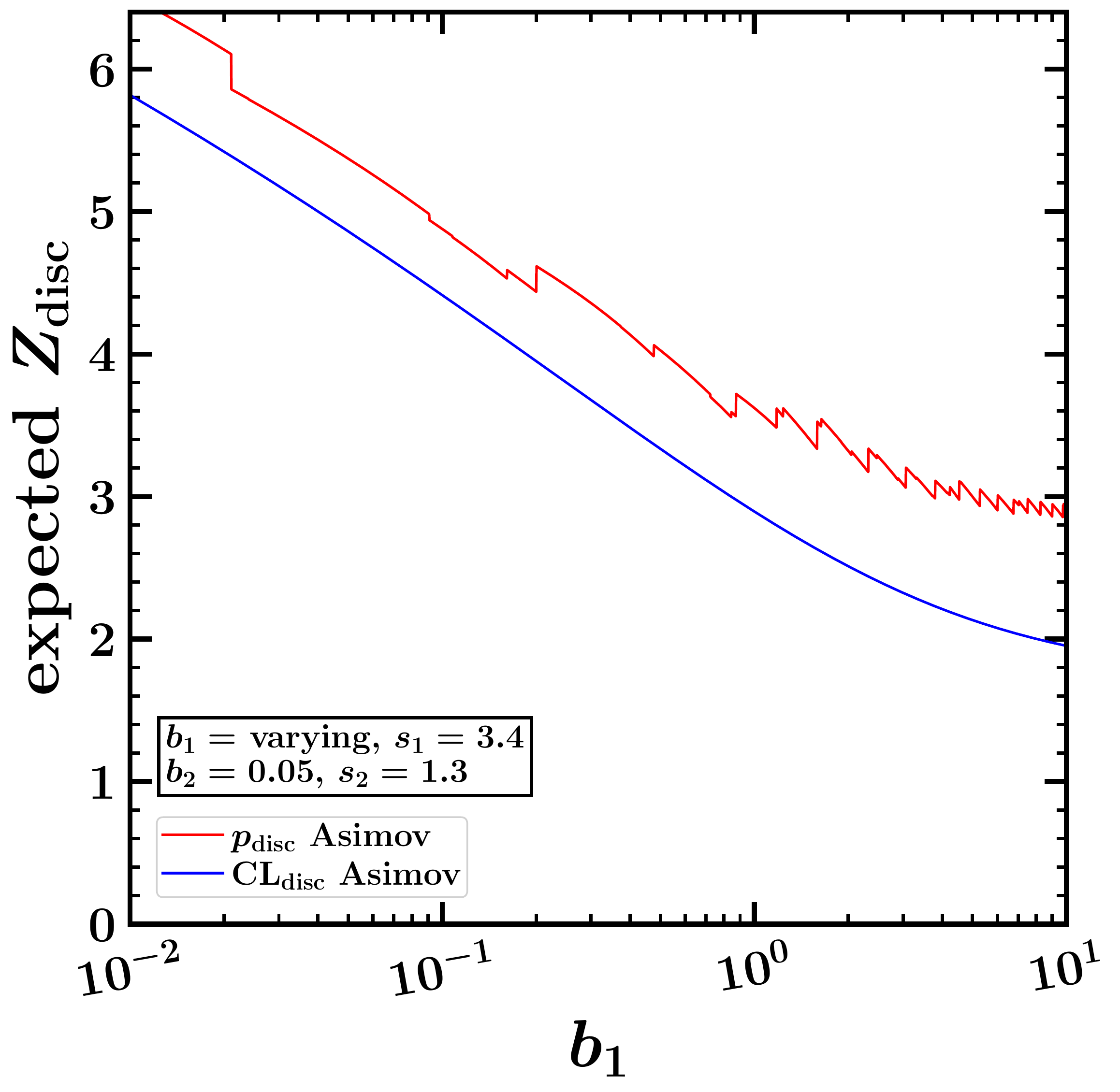}
\end{minipage}
\begin{minipage}[]{0.495\linewidth}
    \includegraphics[width=\linewidth]{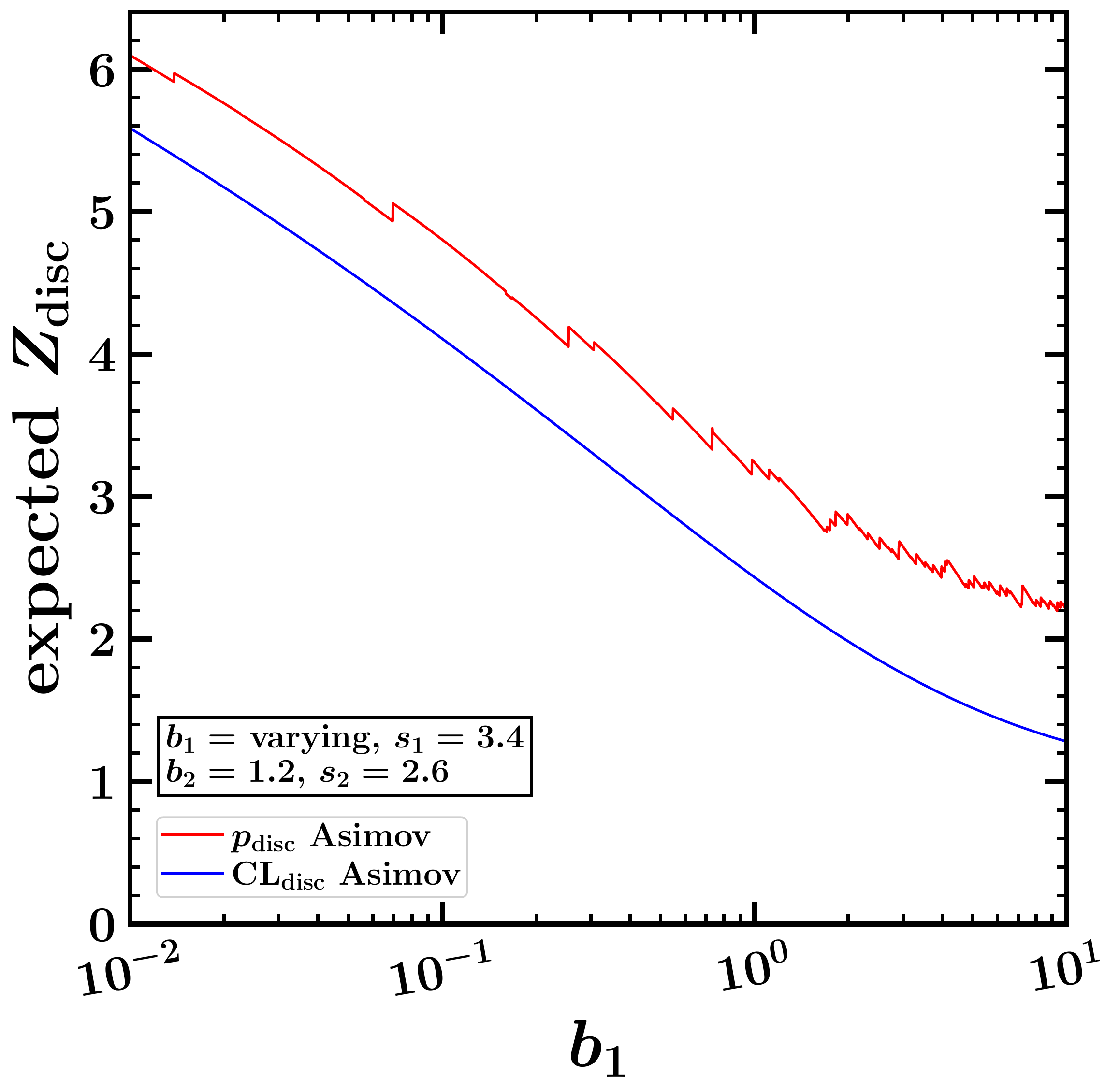}
\end{minipage}
\begin{minipage}[]{0.95\linewidth}  
\caption{Comparison of the Asimov expected discovery significances $Z$ obtained from 
$p_{\rm disc}$ [red lines, from eqs.~(\ref{eq:pdiscmultichannel}) and (\ref{eq:pdiscmultichannelkirestriction})] and
$\CLdisc$ [blue lines, from eq.~(\ref{eq:CLBdiscoverymultichannel})],
for two test cases with two independent channels, as labeled. The Asimov results are obtained by setting $n_i = b_i + s_i$ for each channel.
Due to the non-continuous effect of the restriction of eq.~(\ref{eq:pdiscmultichannelkirestriction}), the Asimov $p_{\rm disc}$ has a counterintuitive non-monotonic behavior as the first channel background mean $b_1$ is varied, while the Asimov $\CLdisc$ is monotonic in the expected way, and more conservative.  
\label{fig:ZAsimovdisc}}
\end{minipage}
\end{figure}

For the purpose of projecting discovery prospects in future experiments, one can again define
the Asimov values of $p_{\rm disc}$ and $\CLdisc$ by replacing $n_i$ with $b_i + s_i$ in eqs.~(\ref{eq:pdiscmultichannel}) and (\ref{eq:CLBdiscoverymultichannel}) respectively.
These are compared for two test cases in Figure \ref{fig:ZAsimovdisc}.
In the case of $p_{\rm disc}$, the constraint put on the sum by eq.~(\ref{eq:pdiscmultichannelkirestriction}) leads to a non-monotonic sawtooth behavior,
although much less pronounced than in the exclusion case in Figure \ref{fig:ZAsimovexcl}.

For the reasons just discussed, and because of the ease of generalization to the case of background uncertainties as discussed in the next section, we propose to use $\CLdisc$ as the figure of merit for the significance of a possible discovery, and for projecting the discovery reach of future experiments.

 
\subsection{Background uncertainty and other nuisance parameters\label{subsec:nuisance}}

In the real world, the background level is never perfectly known. Furthermore, the background and signal may depend on other nuisance parameter(s), to be called $\nu$ below. These can be dealt with in a Bayesian approach by assuming probability densities $f(b)$ and $g(\nu)$, subject to the normalization conditions $\int_0^\infty db\, f(b) = 1$ and $\int d\nu\, g(\nu)= 1$.

For example, following \cite{Bhattiprolu:2020mwi},  we can model the background 
uncertainty in terms of an on-off problem \cite{Li:1983fv,Gehrels:1986mj,ZhangRamsden:1990,Alexandreas:1993,Linnemann:2003vw,Cousins:2008zz}, 
where $m$ is the number of Poisson events in a signal-off (background-only) region, and the ratio of background means in the signal-off and signal-on regions is called $\tau$. In terms of $m$ and $\tau$, the point estimate for the background and its variance are
\beq
\hat b = m/\tau, \qquad \Delta_{b} = \sqrt{m}/\tau,
\label{eq:on-off_relations}
\eeq
or equivalently
\beq
\tau =  \hat b/\Delta_b^2,\qquad m = \hat b^2/\Delta_b^2,
\eeq
so that the probability density of $b$ is
\beq
f(b) &=& f(b \hspace{1pt}|\hspace{1pt} \hat b, \Delta_b) \>=\> \tau^{m+1} b^m e^{-\tau b}/m!,
\label{eq:Pbonoff}
\eeq
the posterior probability distribution for $b$ obtained by using Bayes' theorem with Poisson likelihood for background in the signal-off region $P(m | \tau b)$ and flat prior for $b$.
Note that this probability distribution can be used as a model even in situations where the estimates of the background and its uncertainty come partly or completely from theory rather than some signal-off region data.

In the case of eq.~(\ref{eq:Pbonoff}),
the probability for observing $n$ events in the signal-on region is obtained by averaging over $b$ \cite{Haines:1986ii,Alexandreas:1993,Cousins:1991qz,Linnemann:2003vw,Cousins:2008zz} to obtain
\beq
\Delta P (n,\hat b, \Delta_b, s) &=& \int_0^\infty db \, f(b | \hat b, \Delta_b) P(n | s + b),
\eeq
We can then extend the definitions of frequentist $p$-values and 
to the uncertain background case by simply replacing the Poisson probability $P(n | s + b)$ with $\Delta P(n, \hat b, \Delta_b, s)$ \cite{Zech:1988un}:
\beq
p_{\rm excl}(n, 
\hat b, \Delta_b,
s) &=& \sum_{k=0}^n \Delta P(k, \hat b, \Delta_b, s).
\label{eq:pexcl_db}
\\
p_{\rm disc}(n, 
\hat b, \Delta_b
) &=& \sum_{k=n}^\infty \Delta P(k, \hat b, \Delta_b, 0),
\label{eq:pdisc_db}
\eeq
Explicit formulas for 
$\Delta P(n, \hat b, \Delta_b, s)$, 
$p_{\rm excl}(n, 
\hat b, \Delta_b,
s)$, and 
$p_{\rm disc}(n, 
\hat b, \Delta_b
)$ can be found in eqs.~(12)-(15) of ref.~\cite{Bhattiprolu:2020mwi}. Besides these, we note the simple formula:
\beq
p_{\rm excl}(n, 
\hat b, \Delta_b,
0) &=& \frac{B(
1/(1 + \Delta_b^2/\hat b)
, m+1, n+1)}{B(m+1, n+1)}.
\eeq
Similarly, the confidence levels discussed in the previous sections can be obtained in the uncertain background case as
\beq
{\rm CL}_s(n, 
\hat b, \Delta_b,
s) &=& \frac{p_{\rm excl}(n, \hat b, \Delta_b, s)}{p_{\rm excl}(n, \hat b, \Delta_b, 0)}\label{eq:CLs_db}
,
\\
\CLexcl(n, 
\hat b, \Delta_b,
s) = \frac{\displaystyle\scaleobj{0.85}{\int_{\scaleobj{1.4}{s}}^{\scaleobj{1.2}{\infty}}} ds'\, \Delta P(n, \hat b, \Delta_b, s')}{
\displaystyle\scaleobj{0.85}{\int_{\scaleobj{1.2}{0}}^{\scaleobj{1.2}{\infty}}} ds'\, \Delta P(n, \hat b, \Delta_b, s')}
&=& \frac{p_{\rm excl}(n, \hat b, \Delta_b, s)}{p_{\rm excl}(n, \hat b, \Delta_b, 0)}\label{eq:CLexclunc}
,
\\
\CLdisc(n, 
\hat b, \Delta_b
,s) &=& \frac{\Delta P(n,\hat b, \Delta_b,0)}{\Delta P(n,\hat b, \Delta_b,s)}
.
\label{eq:CLdiscunc}
\eeq
Note that we retain the property $\CLexcl = {\rm CL}_s$ in the single-channel case
with non-zero background uncertainty.

The exact Asimov expectations for $p_{\rm excl}$, ${\rm CL}_s = \CLexcl$ and  $p_{\rm disc}$, $\CLdisc$ in the uncertain background case are obtained by replacing $n$ in 
the preceding equations
by its expected mean in each case:
\beq
\langle n_{\rm excl} \rangle &=& \sum_{n=0}^\infty n\, \Delta P(n,\hat b, \Delta_b,0)
\>=\>
\hat b + \Delta^2_{b}/\hat b,
\label{eq:nexcl_mean}
\\
\langle n_{\rm disc} \rangle &=& \sum_{n=0}^\infty n\, \Delta P(n,\hat b, \Delta_b,s)
\>=\>
s + \hat b + \Delta^2_{b}/\hat b,
\label{eq:ndisc_mean}
\eeq
for exclusion and discovery, respectively.

More generally, for any probability distributions $f(\vec{b})$ and $g(\nu)$ for the background and other nuisance parameters, one can marginalize (integrate) over $b_i$ and $\nu$.
In the case of exclusion, eq.~(\ref{eq:pexclmultichannel}) generalizes to
\beq
p_{\rm excl} &=& 
\int d \nu\, g(\nu)
\int d\vec{b}\, f(\vec{b})
\>
\sum_{\{k_i\}}
\,
\prod_{i=1}^N 
\,
P (k_i|s_i+b_i)
,
\label{eq:pexclmultichanneluncertain}
\eeq
and similarly for eq.~(\ref{eq:pexclbmultichannel}), which then gives CL$_s$.
However, note that the sum $\sum_{\{k_i\}}$ is subject to the restriction eq.~(\ref{eq:restrictionkiLR}), 
so that even the numbers of terms in the sum depends in a discontinuous way on $\nu$ and $b_i$ as we integrate over them in the multichannel case.
Ref.~\cite{ATLAS:2011tau} contains a discussion of various ways to account for the uncertainties
in the background and nuisance parameters in the frequentist methods.
As argued above, we prefer instead to generalize eq.~(\ref{eq:multichannelCLsBayesian}), resulting in:
\beq
\CLexcl
&=&
\frac{1}{D}  \int d \nu\, g(\nu)\> \int d\vec{b}\, f(\vec{b})\, 
\int_{s}^\infty ds' \,\prod_{i=1}^N P(n_i | r_i s' + b_i).
\label{eq:CLsBuncertain}
\eeq
Here we have used a short-hand notation to be used several times below, such that the normalization factor $D$ is equal to the expression that follows it with $s=0$. 

Similarly, in the case of discovery in the presence of background uncertainties and nuisance parameters, we can generalize eq.~(\ref{eq:pdiscmultichannel}) to obtain
\beq
p_{\rm disc} &=& 
\int d \nu\, g(\nu)
\int d\vec{b}\, f(\vec{b})
\>
\sum_{\{k_i\}}
\,
\prod_{i=1}^N 
\,
P (k_i|b_i)
,
\label{eq:pdiscmultichanneluncertain}
\eeq
this time subject to the constraint eq.~(\ref{eq:pdiscmultichannelkirestriction}) on the terms in the sum. However, as argued above, we prefer to use the more conservative
\beq
\CLdisc &=& 
\frac{\displaystyle\int d \nu\, g(\nu) \int d\vec{b}\, f(\vec{b}) \> \prod_{i=1}^N P(n_i|b_i)}
{\displaystyle\int d \nu\, g(\nu) \int d\vec{b}\, f(\vec{b}) \> \prod_{i=1}^N P(n_i| s_i + b_i)}.
\label{eq:CLdiscmultichanneluncertain}
\eeq
To obtain the Asimov results, one can substitute in the mean expected values for $n_i$, namely
\beq
\langle n_{i, \rm excl} \rangle &=& 
\int d\nu\, g(\nu) \int_0^\infty d\vec{b}\, f(\vec{b})\, \sum_{n_i=0}^\infty n_i P(n_i|b_i),
\\
\langle n_{i, \rm disc} \rangle &=& 
\int d\nu\, g(\nu) \int_0^\infty d\vec{b}\, f(\vec{b})\, \sum_{n_i=0}^\infty n_i P(n_i|s_i+b_i)
.
\eeq

\section{Application to proton decay\label{sec:protondecay}}
\setcounter{equation}{0}
\setcounter{figure}{0}
\setcounter{table}{0}
\setcounter{footnote}{1}

In this section, we will first consider the application of the Bayesian statistic $\CLexcl$ to estimate the current lower limits on proton partial lifetimes in $p \rightarrow \overline{\nu} K^+$ and $p \rightarrow e^+ \pi^0$ modes, based on Super-Kamiokande's data, at various confidence levels generalizing  the 90\% CL published limits. We will then consider the prospects for exclusion or discovery of these proton decay modes for several planned future neutrino experiments: 
DUNE \cite{DUNE:2015lol}, JUNO \cite{JUNO:2015zny},
Hyper-Kamiokande \cite{Hyper-Kamiokande:2018ofw}, and THEIA \cite{Theia:2019non}. We do this by applying the Bayesian approach of using $\CLexcl$ and $\CLdisc$ with the exact Asimov criterion of replacing the observed counts by their respective expected means.

As discussed above, the Bayesian approaches $\CLexcl$ for exclusion and $\CLdisc$ for discovery are ideal methods to obtain these limits and projections, as they:
1) guard against claiming exclusion (or discovery) when an experiment is actually
not sensitive to the signal model, and therefore are more conservative
than the frequentist $p_{\rm excl}$  and $p_{\rm disc}$;
2) are well-behaved in multi-channel counting experiments in the sense that, unlike the
(modified) frequentist approach, $\CLexcl$ and $\CLdisc$
are not overly affected by the presence of non-informative channels and do not have any discontinuities as the
signal and background means are varied;
3) are easily able to include uncertainties in the backgrounds and the signal selection
efficiencies, especially for multi-channel counting experiments.

The estimates for the backgrounds and the signal selection efficiencies in a specific proton decay mode have been obtained by the DUNE, JUNO, and THEIA collaborations by modeling the experiments as single-channel counting experiments, whereas Hyper-Kamiokande searches for proton decay
are modeled as multi-channel counting experiments
based on the signal regions and search strategies used at Super-Kamiokande.
Before we present our results,
we first review the methods we employ to obtain the limits/projections for
proton partial lifetimes at single-channel and multi-channel counting experiments,
based on the methods elucidated in section~\ref{sec:statistics}.

The number of decays in a specific decay channel at an experiment
with $N_0$ initial number of protons for a runtime of $\Delta t$
is given by
\beq
\Delta N = N_0 \Gamma \Delta t,
\eeq
where the proton partial width $\Gamma$ is extremely small.
(More generally, $\Delta N = N_0 (1 - e^{-\Gamma \Delta t})$.)
Therefore the signal can be computed as
\beq
s = \epsilon ( \Delta N ) = \epsilon N_0 \Gamma \Delta t,
\label{eq:signal_for_proton_decay}
\eeq
where $0 \leq \epsilon \leq 1$ is the signal selection efficiency. In terms of the number of protons
per kiloton of detector material $N_p$ and the exposure $\lambda$
($=$ runtime $\times$ number of kilotons of detector material)
of the experiment in units of kiloton-years, we can reexpress
eq.~(\ref{eq:signal_for_proton_decay}) as
\beq
s = \Gamma N_p \epsilon \lambda.
\label{eq:s_protondecay}
\eeq
The present exclusion limit at confidence level $1-\alpha$
for the proton partial lifetime is then provided by \cite{Bueno:2007um}
\beq
\tau_p = 1/\Gamma = N_p \epsilon \lambda/s,
\label{eq:lifetime reach/limit for one search channel}
\eeq
where $s$ is the number of signal events that gives $\CLexcl$ equal to $\alpha$.
For a future experiment, the exclusion reach for the proton partial lifetime at confidence level $1-\alpha$ 
is given by the same formula
eq.~(\ref{eq:lifetime reach/limit for one search channel}),
where $s$ is now the signal that makes the exact Asimov ${\rm CL}^A_{\rm excl}$ equal to $\alpha$.
The discovery reach for a given significance $Z$ is likewise obtained from 
eq.~(\ref{eq:lifetime reach/limit for one search channel}) 
using the $s$ that provides for ${\rm CL}^A_{\rm disc} = \frac{1}{2}{\rm erfc}(Z/\sqrt{2})$.

Eq.~(\ref{eq:lifetime reach/limit for one search channel}) holds for an experiment with a single search channel with known background $b$ and signal selection efficiency $\epsilon$.
For the more general case of an experiment with one or more independent search channels with
possibly uncertain backgrounds and signal efficiencies,
we employ a Bayesian approach to obtain the limit/reach for proton partial lifetime,
as discussed above.
First, for the exclusion case, given the number of observed events $n_i$ in each 
search channel labeled $i$, the upper limit on proton partial width at a confidence level $1-\alpha$
is obtained by solving for $\Gamma$ in (see eq.~(\ref{eq:CLsBuncertain}), and ref.~\cite{Super-Kamiokande:2005lev}):
\beq
\alpha &=&
\frac{1}{D}
\int_\Gamma^\infty d\Gamma^\prime \, \prod_{i=1}^N \, \int_0^1
d\epsilon_{i} \, 
g(\epsilon_i)\,
\int_0^\infty db_i \, 
f(b_i)\,
P(n_i | s_i' + b_i)
.
\label{eq:Bayesian for exclusion for independent search channels}
\eeq
Here, $D$ is a normalization factor, defined to equal the expression that follows it evaluated at $\Gamma=0$, and in each search channel labeled by $i$, the signal rate is
\beq
s_i' = N_p \epsilon_{i} \lambda_i \Gamma^\prime 
,
\eeq
and
$g(\epsilon_i)$
and 
$f(b_i)$
are
the probability distributions for the signal efficiency $\epsilon_i$ and the background $b_i$. These distributions can take different forms to parameterize our lack of perfect knowledge of the efficiency and background,
such that
$\int_0^1 d\epsilon_{i} \, 
g(\epsilon_i)
= 1$
and
$\int_0^\infty db_i \, 
f(b_i) 
= 1$.
For example, 
the probability distribution of true signal selection efficiency $\epsilon_{i}$
might be taken to be a truncated Gaussian distribution
with central value $\hat \epsilon_i$ and variance $\Delta_{\epsilon_i}$, as in the Super-Kamiokande search analyses in refs.~\cite{Super-Kamiokande:2014otb,Super-Kamiokande:2020wjk}:
\beq
g(\epsilon_{i} | \hat \epsilon_{i}, \Delta_{\epsilon_{i}}) &=& \sqrt{\frac{2}{\pi}}
\frac{\exp\left [{- \frac{(\epsilon_{i} - \hat \epsilon_{i})^2}{2 \Delta^2_{\epsilon_{i}}}}\right ]}
{\Delta_{\epsilon_{i}}
\left (
    {\rm erf} \left (\frac{1 - \hat \epsilon_{i}}{\sqrt{2} \Delta_{\epsilon_{i}}} \right )
    +
    {\rm erf} \left (\frac{\hat \epsilon_{i}}{\sqrt{2} \Delta_{\epsilon_{i}}} \right )
\right )}
.
\label{eq:Probability distribution of true signal efficiency}
\eeq
The probability distribution of true background $b_i$
in the $i^{\rm th}$ search channel
$f(b_i | \hat b_i, \Delta_{b_i})$ can be taken to be given by eq.~(\ref{eq:Pbonoff}) as in the on-off problem, in terms of quantities $m_i$ and $\tau_i$,
related to the central value $\hat b_i$ and variance 
$\Delta_{b_i}$ by eq.~(\ref{eq:on-off_relations}).
Eq.~(\ref{eq:Bayesian for exclusion for independent search channels}) assumes that the
search channels are independent.

If the background and the signal selection efficiencies are perfectly
known, i.e. 
$f(b_i | \hat b_i, \Delta_{b_i}) = \delta(b_i - \hat b_i)$ and
$g(\epsilon_i | \hat \epsilon_i, \Delta_{\epsilon_i}) = \delta(\epsilon_i - \hat \epsilon_i)$,
then we get
\beq
\alpha &=&
\frac{1}{D} \int_{\Gamma}^\infty \,d\Gamma' \, \prod_i P(n_i| s_i' + b_i),
\eeq
with $s_i' = N_p \epsilon_i \lambda_i \Gamma'$  after identifying $\hat b_i = b_i$ and $\hat \epsilon_i = \epsilon_i$. This corresponds to eq.~(\ref{eq:multichannelCLsBayesian}).
Specializing further to a single search channel (dropping the subscript $i$),
this reduces
to eq.~(\ref{eq:CLsBayes_excl}) with
$s^\prime = N_p \epsilon \lambda \Gamma^\prime$.

For projecting the exclusion reach for partial lifetime at future experiments, we make use of the
exact Asimov method by replacing the number of events $n_i$ in each search channel by
their respective expected means, 
\beq
\langle b_i \rangle &=& \int_0^\infty db_i\, 
f(b_i)
\, b_i  ,
\label{eq:defineexpectedbi}
\eeq
for example $\langle b_i \rangle = (m_i + 1)/\tau_i = 
\hat b_i + \Delta_{b_i}^2/b_i$ if the on-off problem treatment is used for the background.
The expected confidence level $1-\alpha$ upper limit  
on partial width $\Gamma$ is then solved from
eq.~(\ref{eq:Bayesian for exclusion for independent search channels}) 
with $n_i$ replaced by $\langle b_i \rangle$:
\beq
\alpha &=&
\frac{1}{D}
\int_\Gamma^\infty d\Gamma^\prime \, \prod_i \, \int_0^1
d\epsilon_{i} \, 
g(\epsilon_i)
\,
\int_0^\infty db_i \, 
f(b_i)
\,
P(\langle b_i \rangle | s_i' + b_i)
.
\label{eq:Bayesian projections for exclusion}
\eeq
Equation (\ref{eq:Bayesian projections for exclusion}) gives the Asimov expected lower limit on the partial lifetime via $\tau_p = 1/\Gamma$.

For the expected discovery reach for proton partial widths at future experiments, we use a method based on the exact Asimov evaluation of the statistic $\CLdisc$. In particular, we solve for $\Gamma$ from
(see eq.~(\ref{eq:CLdiscmultichanneluncertain}))
\beq
\frac{1}{2} {\rm erfc}\left(\frac{Z}{\sqrt 2}\right) 
=
\frac{
\displaystyle \prod_i \, \int_0^\infty db_i \, 
f(b_i)
\, P(\langle n_i \rangle | b_i)
}{
\displaystyle \prod_i \, \int_0^1 d\epsilon_i \, 
g(\epsilon_i)
\, \int_0^\infty db_i \, 
f(b_i) 
\, P(\langle n_i \rangle | s_i + b_i)
}
,
\label{eq:Bayesian projections for discovery}
\eeq
where $s_i = N_p \lambda_i \epsilon_i \Gamma$ and
$\langle n_i \rangle = \langle s_i \rangle + \langle b_i \rangle$, with
$\langle b_i \rangle$ as given in eq.~(\ref{eq:defineexpectedbi}), and
\beq
\langle s_i \rangle 
&=& 
\Gamma N_p \lambda_i \, \int_0^1 d\epsilon_{i} \, 
g(\epsilon_i)
\, \epsilon_{i}
.
\label{eq:exp_signal}
\eeq
This gives the expected discovery reach for partial lifetime using $\tau_p = 1/\Gamma$
corresponding to a chosen significance $Z$.

\begin{figure}[!tb]
  \begin{minipage}[]{0.495\linewidth}
    \includegraphics[width=8cm]{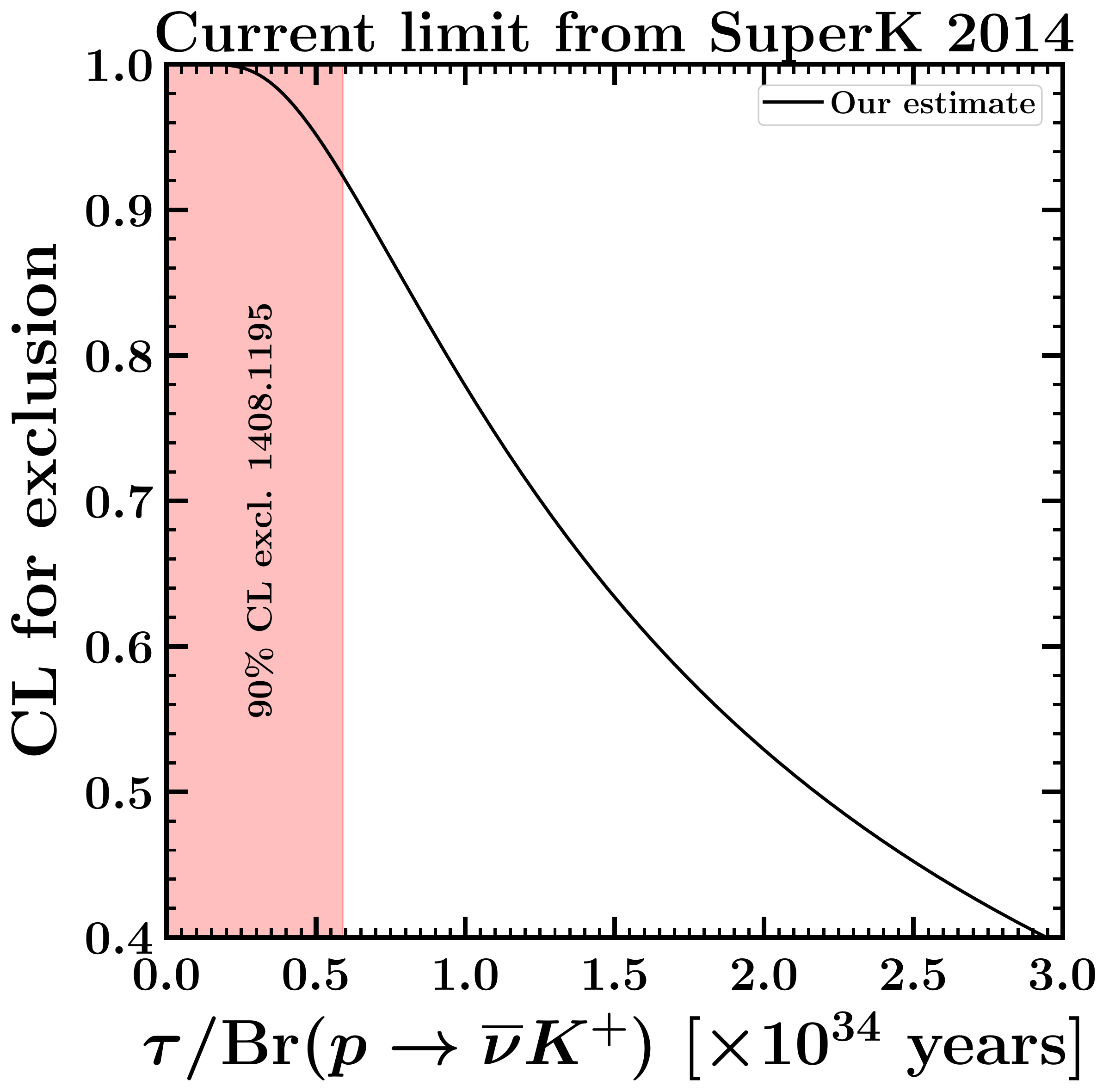}
  \end{minipage}
  \begin{minipage}[]{0.495\linewidth}
    \includegraphics[width=8cm]{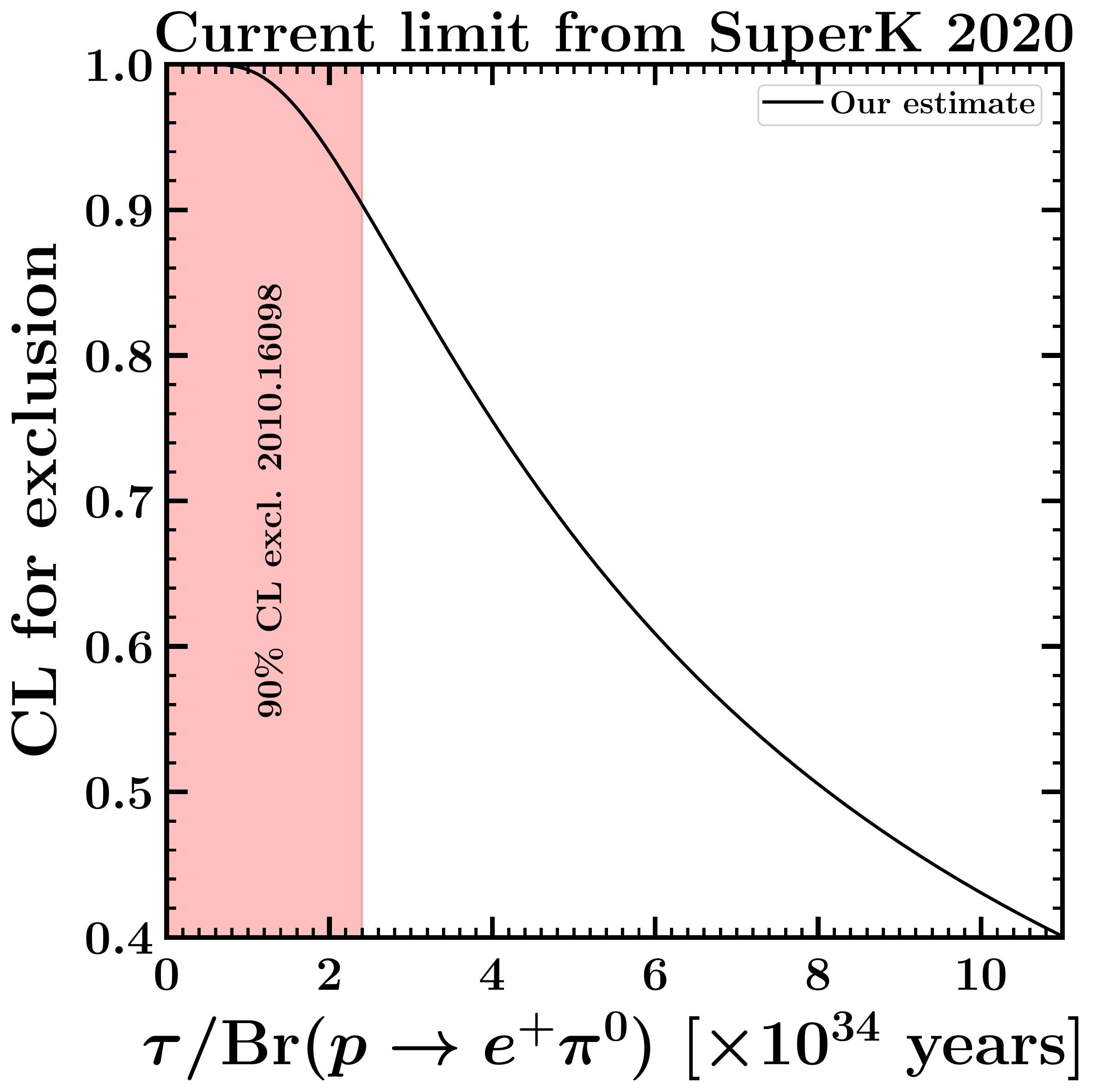}
  \end{minipage}
 \caption{The current confidence level for the exclusion of proton decay in
 $p \rightarrow \overline{\nu} K^+$ (left panel) and
 $p \rightarrow e^+ \pi^0$ (right panel) channels, as a function of the respective
 proton partial lifetimes.
 Our confidence level estimates (solid black lines) are obtained using
 eq.~(\ref{eq:Bayesian for exclusion for independent search channels})
 based on Super-Kamiokande's data through
 2014 \cite{Super-Kamiokande:2014otb} (left panel) and
 2020 \cite{Super-Kamiokande:2020wjk} (right panel),
 summarized in Table~\ref{tab:SuperK_data}.
 The red shaded regions correspond to Super-Kamiokande's published exclusions
 on proton partial lifetimes at $90\%$ CL, from \cite{Super-Kamiokande:2014otb}
 and \cite{Super-Kamiokande:2020wjk}.
\label{fig:SuperK_CL}}
\end{figure}

\begin{table}
\caption{Super-Kamiokande's data for $p \rightarrow \overline{\nu} K^+$ and
$p \rightarrow e^+ \pi^0$ decay modes, taken from refs.~\cite{Super-Kamiokande:2014otb} and
\cite{Super-Kamiokande:2020wjk}, respectively. In each decay mode, the exposures $\lambda_i$ in
kton-years, total backgrounds $\hat b_i \pm \Delta_{b_i}$, signal efficiencies
$\hat \epsilon_i \pm \Delta_{\epsilon_i}$, and the observed number of counts $n_i$ are listed.
$\langle s_i^{90\text{CL}} \rangle$ are the expected signal events, defined in eq.~(\ref{eq:exp_signal}),
for proton partial lifetime set equal to its 90\% CL lower limit.
The last column gives a brief description of each of the channels referring
to the detector period (SK I-IV) and the name of the search method used in refs.~\cite{Super-Kamiokande:2014otb,Super-Kamiokande:2020wjk}.
\label{tab:SuperK_data}}
\begin{center}
\begin{tabular}{|c || c | c | c | c || c || c |}
\hline
~Decay mode~ & ~$\lambda_i$~ & ~$\hat b_i \pm \Delta_{b_i}$~  &  ~$\hat \epsilon_i \pm \Delta_{\epsilon_i}$ [\%]~  & ~$n_i$~ & ~$\langle s_i^{90\text{CL}} \rangle$~ & ~Comment~\\
\hline
\hline
    ~$p \rightarrow \overline{\nu} K^+$~ & ~91.7~ & ~$
    0.08
 \pm 0.02$~ & ~$7.9 \pm 0.1$~ & ~0~ & ~0.37~ & ~SK-I, prompt $\gamma$~
    \\[1pt]
    \cline{3-7}
    ~~ & ~~ & ~$0.18 \pm 0.04$~ & ~$7.8 \pm 0.1$~ & ~0~ & ~0.36~ & ~SK-I, $\pi^+ \pi^0$~
    \\[1pt]
    \cline{3-7}
    ~~ & ~~ & ~$193.21 \pm 3.58$~ & ~$33.9 \pm 0.3$~ & ~177~ & ~1.57~ & ~SK-I, $p_\mu$ spectrum~
    \\[1pt]
\cline{2-7}
    ~~ & ~49.2~ & ~$0.14 \pm 0.03$~ & ~$6.3 \pm 0.1$~ & ~0~ & ~0.16~ & ~SK-II, prompt $\gamma$~
    \\[1pt]
    \cline{3-7}
    ~~ & ~~ & ~$0.17 \pm 0.03$~ & ~$6.7 \pm 0.1$~ & ~0~ & ~0.17~ & ~SK-II, $\pi^+ \pi^0$~
    \\[1pt]
    \cline{3-7}
    ~~ & ~~ & ~$94.27 \pm  1.72$~ & ~$30.6 \pm 0.3$~ & ~78~ & ~0.76~ & ~SK-II, $p_\mu$ spectrum~
    \\[1pt]
\cline{2-7}
    ~~ & ~31.9~ & ~$0.03 \pm 0.01$~ & ~$7.7 \pm 0.1$~ & ~0~ & ~0.12~ & ~SK-III, prompt $\gamma$~
    \\[1pt]
    \cline{3-7}
    ~~ & ~~ & ~$
    0.09
    \pm 0.01$~ & ~$7.9 \pm 0.1$~ & ~0~ & ~0.13~ & ~SK-III, $\pi^+ \pi^0$~
    \\[1pt]
    \cline{3-7}
    ~~ & ~~ & ~$69.00 \pm 1.28$~ & ~$32.6 \pm 0.3$~ & ~85~ & ~0.53~ & ~SK-III, $p_\mu$ spectrum~
    \\[1pt]
\cline{2-7}
    ~~ & ~87.3~ & ~$0.13 \pm 0.03$~ & ~$9.1 \pm 0.1$~ & ~0~ & ~0.4~ & ~SK-IV, prompt $\gamma$~
    \\[1pt]
    \cline{3-7}
    ~~ & ~~ & ~$
    0.18
    \pm 0.03$~ & ~$10.0 \pm 0.1$~ & ~0~ & ~0.44~ & ~SK-IV, $\pi^+ \pi^0$~
    \\[1pt]
    \cline{3-7}
    ~~ & ~~ & ~$223.14 \pm 4.10$~ & ~$37.6 \pm 0.3$~ & ~226~ & ~1.66~ & ~SK-IV, $p_\mu$ spec.~
    \\[1pt]
\hline
\hline
~$p \rightarrow e^+ \pi^0$~ & ~111.4~ & ~$0.01 \pm 0.01$~ & ~$18.3 \pm 1.7$~ & ~0~ & ~0.28~ & ~SK-I, lower~
\\[1pt]
\cline{3-7}
~~ & ~~ & ~$0.15 \pm 0.06$~ & ~$ 20.0\pm 3.3$~ & ~0~ & ~0.3~ & ~SK-I, upper~
\\[1pt]
\cline{2-7}
~~ & ~59.4~ & ~$0.01 \pm 0.01$~ & ~$16.6 \pm 1.7$~ & ~0~ & ~0.13~ & ~SK-II, lower~
\\[1pt]
\cline{3-7}
~~ & ~~ & ~$0.11 \pm 0.04$~ & ~$19.4 \pm 3.0$~ & ~0~ & ~0.16~ & ~SK-II, upper~
\\[1pt]
\cline{2-7}
~~ & ~38.6~ & ~$0.01$~ & ~$18.7 \pm 1.7$~ & ~0~ & ~0.1~ & ~SK-III, lower~
\\[1pt]
\cline{3-7}
~~ & ~~ & ~$0.07 \pm 0.03$~ & ~$20.3 \pm 3.3$~ & ~0~ & ~0.11~ & ~SK-III, upper~
\\[1pt]
\cline{2-7}
~~ & ~241.3~ & ~$0.01$~ & ~$18.2 \pm 1.5$~ & ~0~ & ~0.6~ & ~SK-IV, lower~
\\[1pt]
\cline{3-7}
~~ & ~~ & ~$0.25 \pm 0.11$~ & ~$19.2 \pm 3.1$~ & ~0~ & ~0.63~ & ~SK-IV, upper~
\\[1pt]
\hline
\end{tabular}
\end{center}
\end{table}
Based on Super-Kamiokande's data, taken from
refs.~\cite{Super-Kamiokande:2014otb,Super-Kamiokande:2020wjk}, which we quote for completeness in Table~\ref{tab:SuperK_data}, we now compute the upper limit on proton partial widths
in the $p \rightarrow \overline{\nu} K^+$ and $p \rightarrow e^+ \pi^0$
decay modes that are excluded at various confidence levels
(e.g.~95\%, 90\%, 68\%, 50\% CL) using
eq.~(\ref{eq:Bayesian for exclusion for independent search channels}),
which can then be translated
into corresponding lower limits on the proton partial lifetime.
Super-Kamiokande uses a water Cerenkov detector with a fiducial mass of 22.5 ktons,
and the analysis for $p \rightarrow e^+ \pi^0$ in ref.~\cite{Super-Kamiokande:2020wjk}
also includes data from an enlarged fiducial mass of 27.2 ktons.
While Super-Kamiokande can probe for proton decay in both $p \rightarrow \overline{\nu} K^+$
and $p \rightarrow e^+ \pi^0$ decay modes, it is less sensitive to the former decay mode,
because the $K^+$ is produced below its Cerenkov threshold in water and
is only identified from its decay constituents.
Figure~\ref{fig:SuperK_CL} shows our own computed estimates of the current confidence levels
for the exclusion of proton decay at Super-Kamiokande in
$p \rightarrow \overline{\nu} K^+$ (left panel) and
$p \rightarrow e^+ \pi^0$ (right panel) channels as a function of
proton partial lifetime in the respective decay channels. This generalizes the results presented by the Super-Kamiokande collaboration, which gave results only for 90\% CL exclusions.
From the data in Table \ref{tab:SuperK_data},
we estimated the current lower limits on proton partial lifetimes to be 
\beq
\tau_p/{\rm Br}(p \rightarrow \overline{\nu} K^+)
&>&
\begin{cases}
5.1 \times 10^{33} \text{ years at 95\% CL},\\[-7pt]
6.6 \times 10^{33} \text{ years at 90\% CL},\\[-7pt]
1.3
\times 10^{34} \text{ years at 68\% CL},\\[-7pt]
2.2 \times 10^{34} \text{ years at 50\% CL},
\end{cases}
\eeq
and
\beq
\tau_p/{\rm Br}(p \rightarrow e^+ \pi^0)
&>&
\begin{cases}
1.9 \times 10^{34} \text{ years at 95\% CL},\\[-7pt]
2.4 \times 10^{34} \text{ years at 90\% CL},\\[-7pt]
4.9 \times 10^{34} \text{ years at 68\% CL},\\[-7pt]
8.1 \times 10^{34} \text{ years at 50\% CL}.
\end{cases}
\eeq
In comparison, the published $90\%$ CL
exclusion limit on proton partial lifetimes from the Super-Kamiokande collaboration are 
\beq
\tau_p/{\rm Br}(p \rightarrow \overline{\nu} K^+)
&>& 5.9 \times 10^{33} \text{ years at 90\% CL (SuperK 2014 \cite{Super-Kamiokande:2014otb})},\\
\tau_p/{\rm Br}(p \rightarrow e^+ \pi^0)
&>& 2.4 \times 10^{34} \text{ years at 90\% CL (SuperK 2020 \cite{Super-Kamiokande:2020wjk})}.
\eeq
shown as the shaded red regions in Figure \ref{fig:SuperK_CL}. 
We see that in the case of $p \rightarrow \overline\nu K^+$, our\footnote{Besides using the probability distribution for true
background as in the on-off problem (eq.~(\ref{eq:Pbonoff})), we have
considered various other distributions such as a Gaussian,
and a convolution of Gaussian and Poisson (only for search channels with extremely low backgrounds)
as done in refs.~\cite{Super-Kamiokande:2014otb,Super-Kamiokande:2020wjk},
but there was no noticeable change in our results.
In Super-Kamiokande's analysis for $p \rightarrow \overline{\nu} K^+$ decay mode in
ref.~\cite{Super-Kamiokande:2014otb},
the search channels with large backgrounds that are referred to as
``$p_\mu$ spectrum" in Table~\ref{tab:SuperK_data} were 
further divided into sub-channels, but due to insufficient data made available, we are not able to include that subdivision.} estimate for the 90\% CL limit is slightly stronger ($6.6 \times 10^{33}$ years rather than $5.9 \times 10^{33}$ years) than the journal published limit in ref.~\cite{Super-Kamiokande:2014otb}.
In this paper, we only consider the limits from data published in journal articles. In the case of $p\rightarrow \overline\nu K^+$, there is more data from the continuation of run SK-IV, which was not used for the published limit in 
ref.~\cite{Super-Kamiokande:2014otb}. It is therefore quite possible that a future limit, based on data already taken, will be stronger.
In the case of $p \rightarrow e^+ \pi^0$, our estimate for the 90\% CL limit agrees perfectly
with the Super-Kamiokande published limit in ref.~\cite{Super-Kamiokande:2020wjk}. 

We now discuss projections for exclusion and discovery of proton decay at possible future neutrino
detectors DUNE, JUNO, Hyper-Kamiokande, and THEIA.
Both DUNE and JUNO will be primarily searching for proton decay in $p \rightarrow \overline{\nu} K^+$ decay mode.
For these searches, DUNE uses its far detector with a total of 40 kiloton (kton) fiducial mass of liquid argon
\cite{DUNE:2015lol} and can track and reconstruct charged kaons with high efficiency,
and JUNO uses its central detector with a 20 kton fiducial mass of a liquid scintillator \cite{JUNO:2015zny}.
On the other hand, Hyper-Kamiokande \cite{Hyper-Kamiokande:2018ofw} uses a water Cerenkov detector
with 186 ktons of fiducial mass and is sensitive to
both $p \rightarrow \overline{\nu} K^+$ and  $p \rightarrow e^+ \pi^0$ decay modes among others.
As was the case with Super-Kamiokande, Hyper-Kamiokande will be more sensitive to the
$p \rightarrow e^+ \pi^0$ mode, compared to the $p \rightarrow \overline{\nu} K^+$ mode, due to much better reconstruction of the Cerenkov rings of the positron and the electromagnetic showers emanating from $\pi^0 \rightarrow \gamma \gamma$.
THEIA is a new detector concept with water-based liquid scintillator
(10\% liquid scintillator and 90\% water) that will be able to detect and distinguish
between the Cerenkov and the scintillation light \cite{Theia:2019non}.
Here, we project sensitivities for both THEIA-25 and THEIA-100
with fiducial masses 17 and 80 ktons, respectively, that were considered in ref.~\cite{Theia:2019non}.
Due to the ability to detect scintillation signals
from charged particles such as $K^+$ produced below its Cerenkov threshold,
and Cerenkov signals,
the THEIA detector aims to have enhanced sensitivity to the
$p \rightarrow \overline{\nu} K^+$ mode \cite{Theia:2019non}
while also being able to probe the $p \rightarrow e^+ \pi^0$ mode \cite{Dev:2022jbf}.
The numbers of protons per kiloton of detector material are
\beq
N_p =
\begin{cases}
2.71 \times 10^{32} \quad \text{(DUNE)},\\[-7pt]
3.38 \times 10^{32} \quad \text{(JUNO)},\\[-7pt]
3.34 \times 10^{32} \quad \text{(Hyper-Kamiokande)},\\[-7pt]
3.35 \times 10^{32} \quad \text{(THEIA)}.
\end{cases}
\eeq
For the purposes of projecting sensitivities for THEIA and JUNO, we took the liquid scintillator in both detectors to have
$6.75 \times 10^{33}$ protons per 20 kilotons based on ref.~\cite{JUNO:2015zny}.

\begin{figure}
  \includegraphics[width=13.0cm]{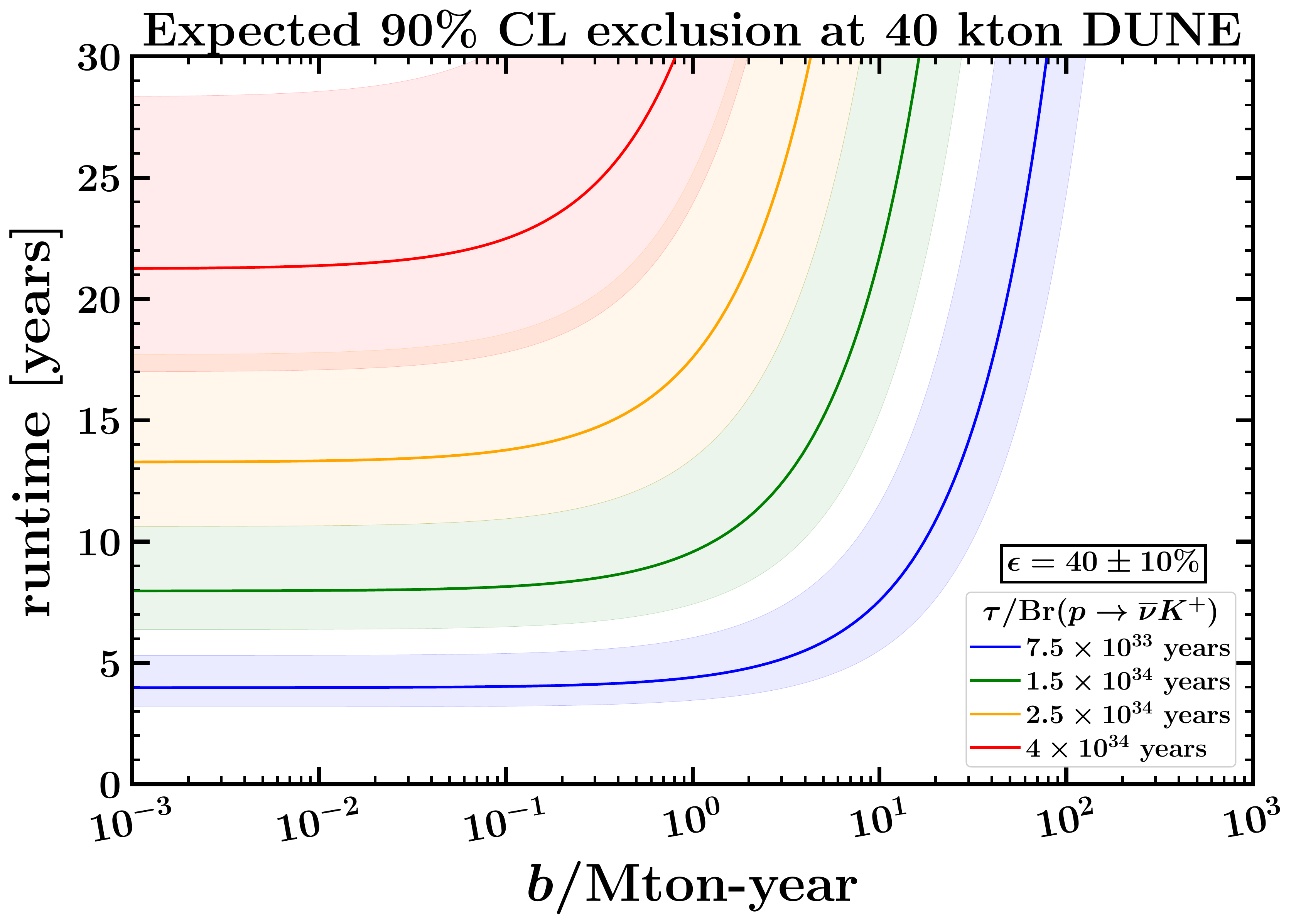}
  \includegraphics[width=13.0cm]{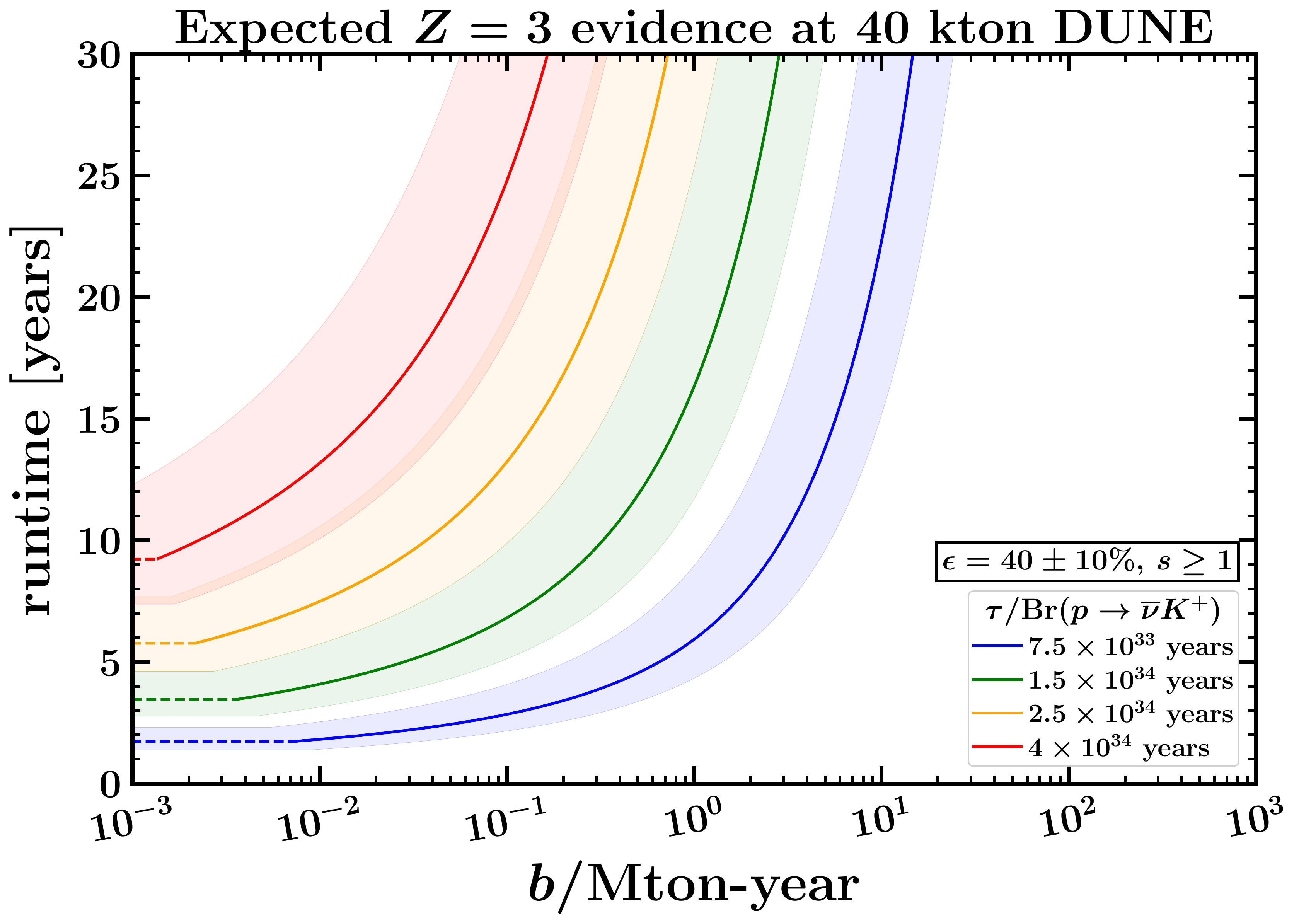}
\caption{The required runtimes at DUNE (with 40 kiloton fiducial mass of liquid argon) for an expected $90\%$ CL exclusion (top panel) and $Z=3$ evidence (bottom panel) as a function of background rate per megaton-year of exposure, for various proton partial lifetimes in the 
$p \rightarrow \overline{\nu} K^+$ channel, as labeled. The runtimes are computed using eq.~(\ref{eq:protondecay_exptruntime}) where the signal needed for 90\% CL exclusion ($Z=3$ evidence) is obtained from setting eq.~(\ref{eq:CLsAsimov}) (eq.~(\ref{eq:CLdiscAsimov})) equal to 0.1 (0.00135). We also require $s \geq 1$ in the bottom panel, which yields the horizontal dashed lines for very small $b$ in the lower left corner. The solid lines (and dashed lines in the bottom panel) assume the signal selection efficiency $\epsilon$ to be 40\%, and the shaded bands encompassing them correspond to varying $\epsilon$ by $\pm 10\%$. 
\label{fig:DUNE_runtime}}
\end{figure}

Figure~\ref{fig:DUNE_runtime} shows the runtimes at DUNE that are required for an
expected 90\% CL exclusion (first panel) and $Z=3$ evidence (second panel),
in the $p \rightarrow \overline{\nu} K^+$ decay mode, as a function of
the background rate per megaton-year of exposure.
The colored lines and bands correspond to various choices of proton partial lifetimes.
For the purposes of illustration, we chose a signal selection efficiency
$\epsilon = 40 \pm 10 \%$
that is plausible,
based on various signal selection efficiencies that are considered in
refs.~\cite{Alt:2020blf,Alt:2020rii,Alt:2020gog,DUNE:2020ypp,DUNE:2020fgq,DUNE:2022aul}.
The solid lines in the figure assume $\epsilon = 40 \%$, and the shaded bands surrounding them
vary $\epsilon$ by $\pm 10 \%$.
The required runtimes $\Delta t$  in the figure are obtained using eq.~(\ref{eq:lifetime reach/limit for one search channel}), which gives
\beq
\Delta t &=& \frac{s \tau_p}{N_p N_{\rm kton} \epsilon},
\label{eq:protondecay_exptruntime}
\eeq
where $N_{\rm kton}$ is the number of kilotons of detector material,
and $s$ is the upper limit on signal for 90\% CL exclusion obtained
from setting $\CLexcl^A$ (as in eq.~(\ref{eq:CLsAsimov})) equal to 0.1,
or the signal needed for $Z=3$ evidence
obtained from setting $\CLdisc^A$ (as in eq.~(\ref{eq:CLdiscAsimov})) equal to 0.00135.
As discussed at the end of Section~\ref{subsec:singlechannel}, the zero background limit for the discovery case
is not well defined, in a sense that at $b=0$, any non-zero signal, albeit arbitrarily small,
would yield an infinite significance.
Therefore, to be conservative, we require that the mean expected number of signal events $s$ is at least 1 in order to have an expected discovery.
The dashed lines for very small $b$/Mton-year in the lower left corner of the bottom panel (for discovery case) of Figure~\ref{fig:DUNE_runtime}
correspond to this additional requirement that $s\geq 1$.
It is clear from the figure that if the estimated background per megaton-year of exposure at DUNE increases,
the required runtime increases more steeply for discovery than for exclusion.

\begin{figure}
  \includegraphics[width=13.0cm]{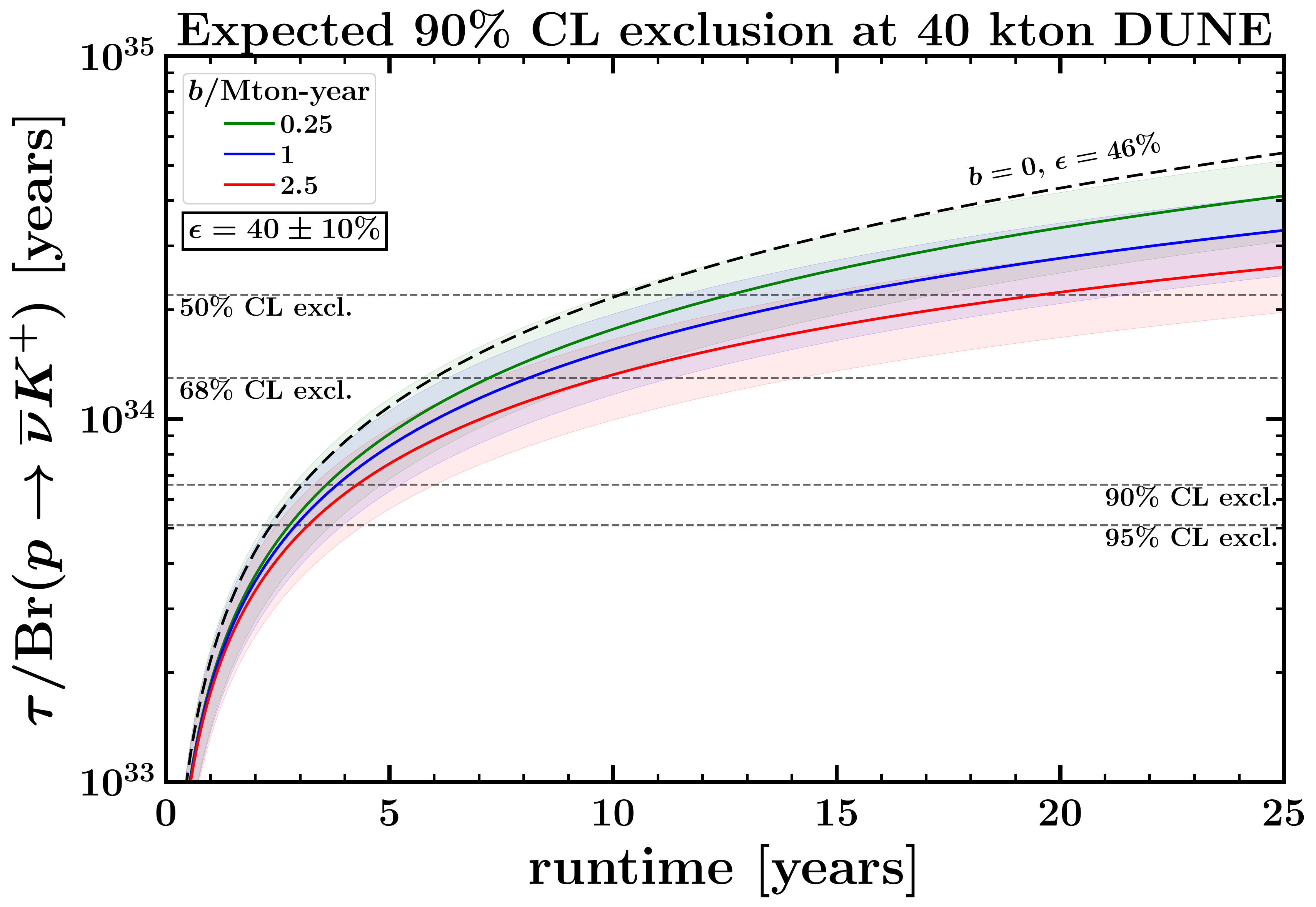}
  \includegraphics[width=13.0cm]{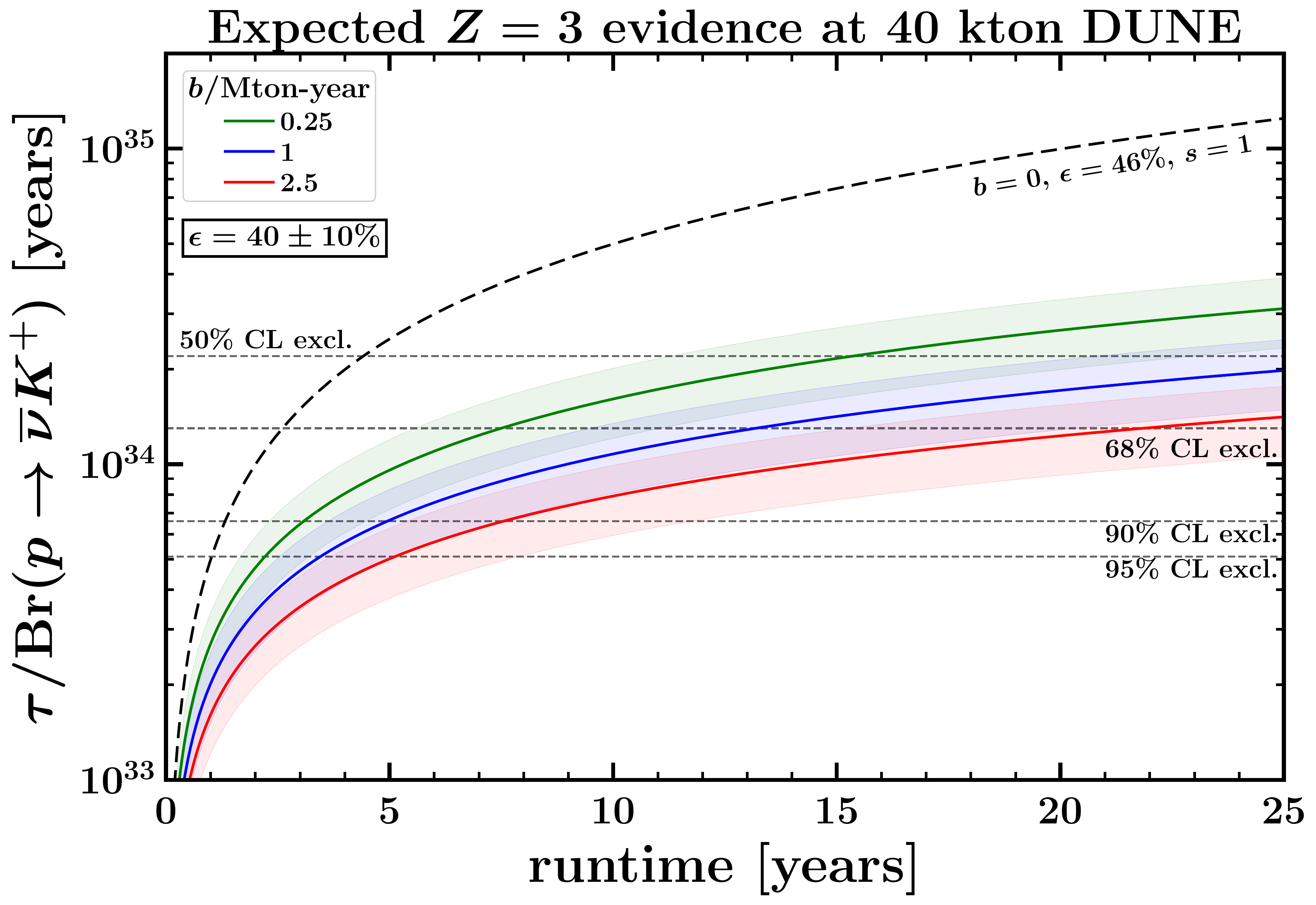}
\caption{Proton partial lifetimes in $p \rightarrow \overline{\nu} K^+$ channel that are expected to be excluded at $90\%$ CL [top panel, from eqs.~(\ref{eq:CLsAsimov}) and (\ref{eq:lifetime reach/limit for one search channel})] or discovered at $Z=3$ significance [bottom panel, from eqs.~(\ref{eq:Zfromp}), (\ref{eq:CLdiscAsimov}) and (\ref{eq:lifetime reach/limit for one search channel})] at 40 kton DUNE, as a function of runtime for various background rates per megaton-year of exposure, as labeled. The signal selection efficiency $\epsilon$ is taken to be 40\% (solid lines) $\pm$ $10\%$ (shaded bands). The long dashed black line in each panel shows the idealized optimistic case of no background and $\epsilon = 46\%$ \cite{Alt:2020blf}, with the expected mean number of events required to be $s=1$
in the second panel. Our estimates of the current 95\%, 90\%, 68\%, and 50\% CL exclusion limit on 
proton partial lifetime, based on Super-Kamiokande's data from 
2014 \cite{Super-Kamiokande:2014otb}, are shown as
horizontal dashed lines.
\label{fig:DUNE_tau}}
\end{figure}

In Figure~\ref{fig:DUNE_tau}, we show the expected 90\% CL exclusion reach (first panel)
and the expected $Z=3$ evidence reach (second panel) for proton partial lifetime
in $p \rightarrow \overline{\nu} K^+$ decay channel at DUNE as a function of
the runtime in years.
The three colored lines/bands correspond to various assumed background rates per
megaton-year of exposure taken
from refs.~\cite{DUNE:2015lol,Alt:2020blf,Alt:2020rii,Alt:2020gog,DUNE:2020ypp,DUNE:2020fgq,DUNE:2022aul}.
The signal selection efficiency is again taken to be 
$\epsilon = 40\%$ (solid colored lines) $\pm$ $10 \%$ (shaded bands). 
The signals computed from setting eq.~(\ref{eq:CLsAsimov}) equal to 0.1, and eq.~(\ref{eq:CLdiscAsimov}) equal to $0.00135$, are plugged into eq.~(\ref{eq:lifetime reach/limit for one search channel}) to obtain the expected 90\% CL exclusion, and $Z=3$ evidence, reaches for proton partial lifetime, respectively. The black dashed curves correspond to a very optimistic scenario with $b=0$ and $\epsilon = 46\%$ \cite{Alt:2020blf}, and using the requirement $s=1$ in the discovery case (bottom panel). Also shown in Figure~\ref{fig:DUNE_tau} and other figures below are horizontal lines at our previously mentioned estimates of the current 95\%, 90\%, 68\%, and 50\% CL
exclusion limit based on Super-Kamiokande's data from 2014 \cite{Super-Kamiokande:2014otb}.

\begin{figure}
  \begin{center}
    \includegraphics[width=12.0cm]{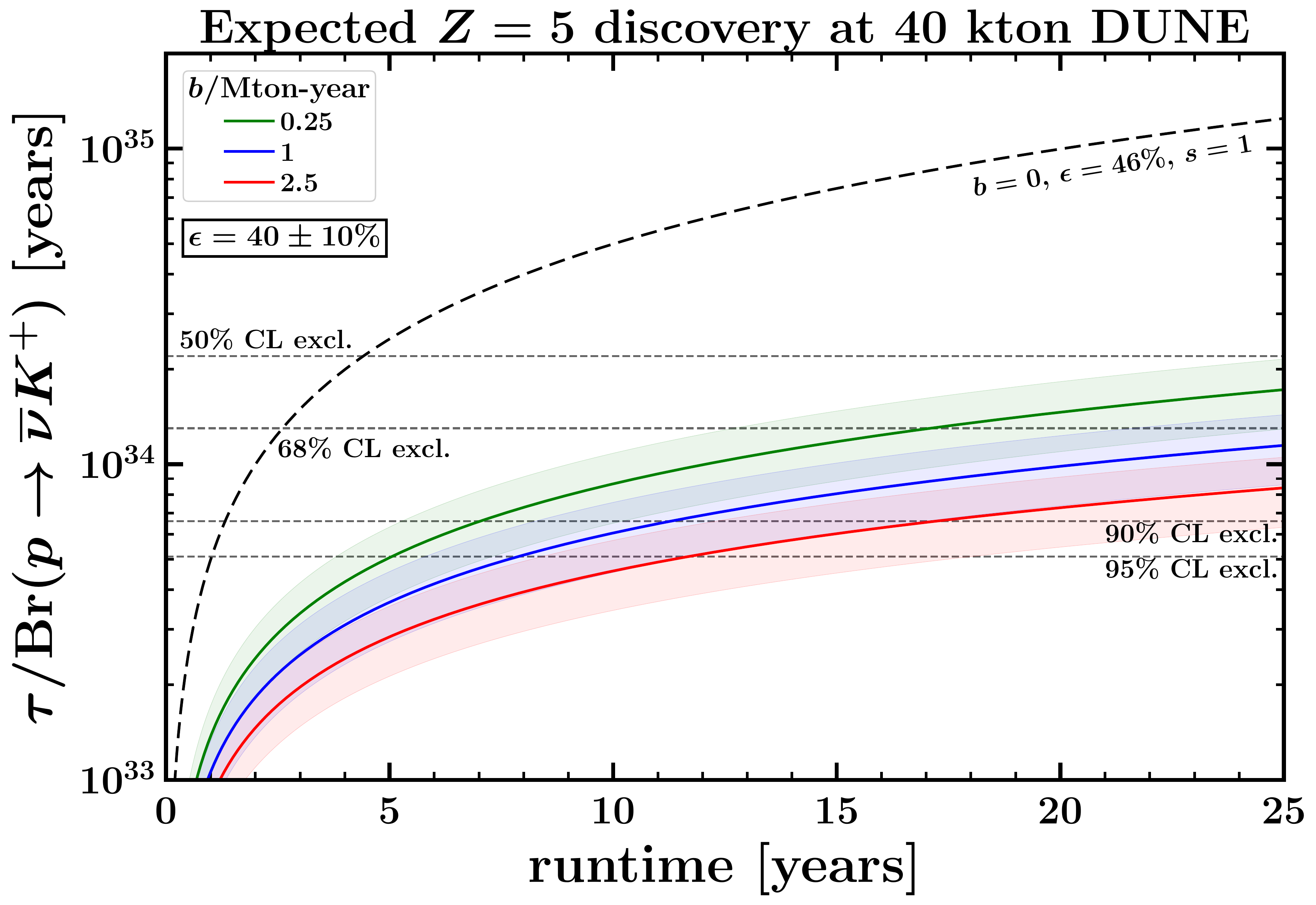}
  \end{center}
\vspace{-0.5cm}
\caption{Proton partial lifetimes in the $p \rightarrow \overline{\nu} K^+$ channel
that are expected to be discovered with a significance $Z=5$ at 40 kton DUNE, as a function
of the runtime, for various background rates per megaton-year of exposure, as labeled. The results are obtained from
eqs.~(\ref{eq:Zfromp}), (\ref{eq:CLdiscAsimov}) and 
(\ref{eq:lifetime reach/limit for one search channel}).
The horizontal dashed lines shown are our estimates of the current 95\%, 90\%, 68\%, and 50\% CL
exclusion limit on proton partial lifetime,
based on Super-Kamiokande's data from 2014 \cite{Super-Kamiokande:2014otb}.
\label{fig:DUNE_disc_Z5}}
\end{figure}

The usual standard for discovery in particle physics is a significance of $Z=5$. Therefore, we show in Figure \ref{fig:DUNE_disc_Z5} the expected reach for $Z=5$ in the $p \rightarrow \overline \nu K^+$ channel at 40 kton DUNE, as a function of the runtime. We note that
even after 25 years, the discovery reach in this channel with nominal background rates remains below the value of $\tau_p(p \rightarrow \overline \nu K^+)$ that we estimate to be excluded at 50\% CL by the Super-Kamiokande data already published in 2014. Of course, a 50\% CL exclusion is far from definitive, but this indicates the challenge being faced. This could change if the background can be reduced to near 0, as indicated by the dashed line, while maintaining a high efficiency for the signal.
\begin{figure}
  \begin{center}
    \includegraphics[width=12.0cm]{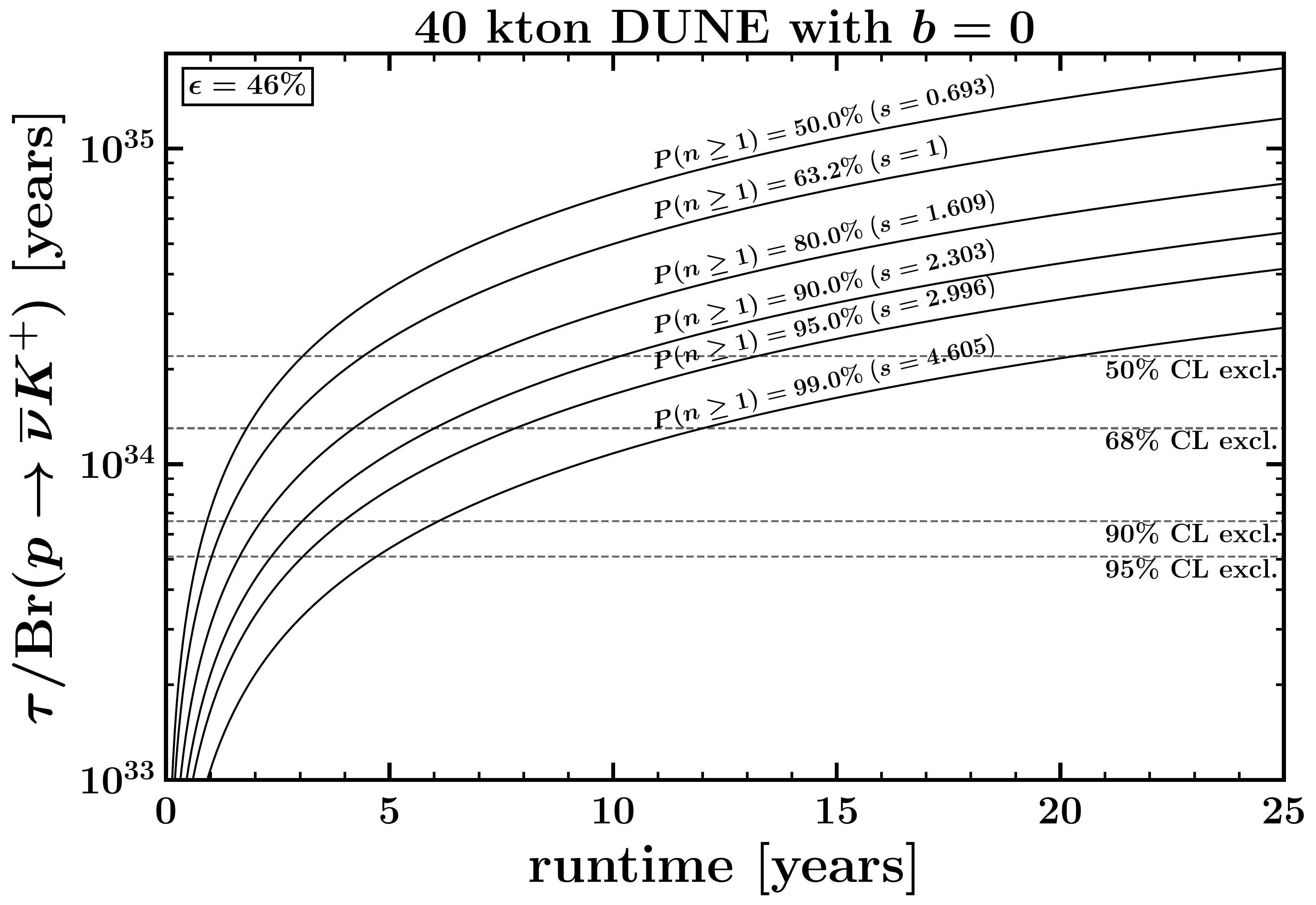}
  \end{center}
\vspace{-0.5cm}  
\caption{Proton partial lifetimes in the $p \rightarrow \overline{\nu} K^+$ channel
that give different probabilities of observing at least one event
from eq.~(\ref{eq:P_n_ge_1_disc}), which in turn correspond to different values of the 
expected signal, as labeled.
The signal selection efficiency $\epsilon$ is taken to be 46\% based on \cite{Alt:2020blf}.
The horizontal dashed lines shown are our estimates of the current 95\%, 90\%, 68\%, and 50\% CL
exclusion limit on proton partial lifetime,
based on Super-Kamiokande's data from 2014 \cite{Super-Kamiokande:2014otb}.
\label{fig:DUNE_disc_b0}}
\end{figure}

As noted above in Section~\ref{subsec:singlechannel},
if the mean expected number of signal
events is $s=1$, and one makes the optimistic assumption that the background is completely negligible ($b=0$), then the probability of obtaining at least one event is about 63.2\%.
Figure~\ref{fig:DUNE_disc_b0} shows the value of $\tau$/Br$(p \rightarrow \overline \nu K^+)$,
as a function of the runtime, that would give various other probabilities of obtaining at least one event, again with the very optimistic assumption of absolutely no background $b=0$ and $\epsilon = 46\%$ \cite{Alt:2020blf}. Each of these choices for $P(n\geq 1)$ is equivalent to a requirement on the signal $s$, as labeled in the figure.

Ref.~\cite{DUNE:2020ypp} also provided a preliminary estimate for the background and signal efficiency for proton decay search in $p \rightarrow e^+ \pi^0$ mode at DUNE. Although DUNE is most sensitive to $p \rightarrow \overline{\nu} K^+$ mode, for completeness, we will also show our expected reach estimates for proton partial lifetime in $p \rightarrow e^+ \pi^0$ mode at DUNE after 10 years and 20 years of runtime in our summary plots in Figures~\ref{fig:Reach_excl} and \ref{fig:Reach_disc} in the Outlook section below.

\begin{figure}
  \includegraphics[width=13.0cm]{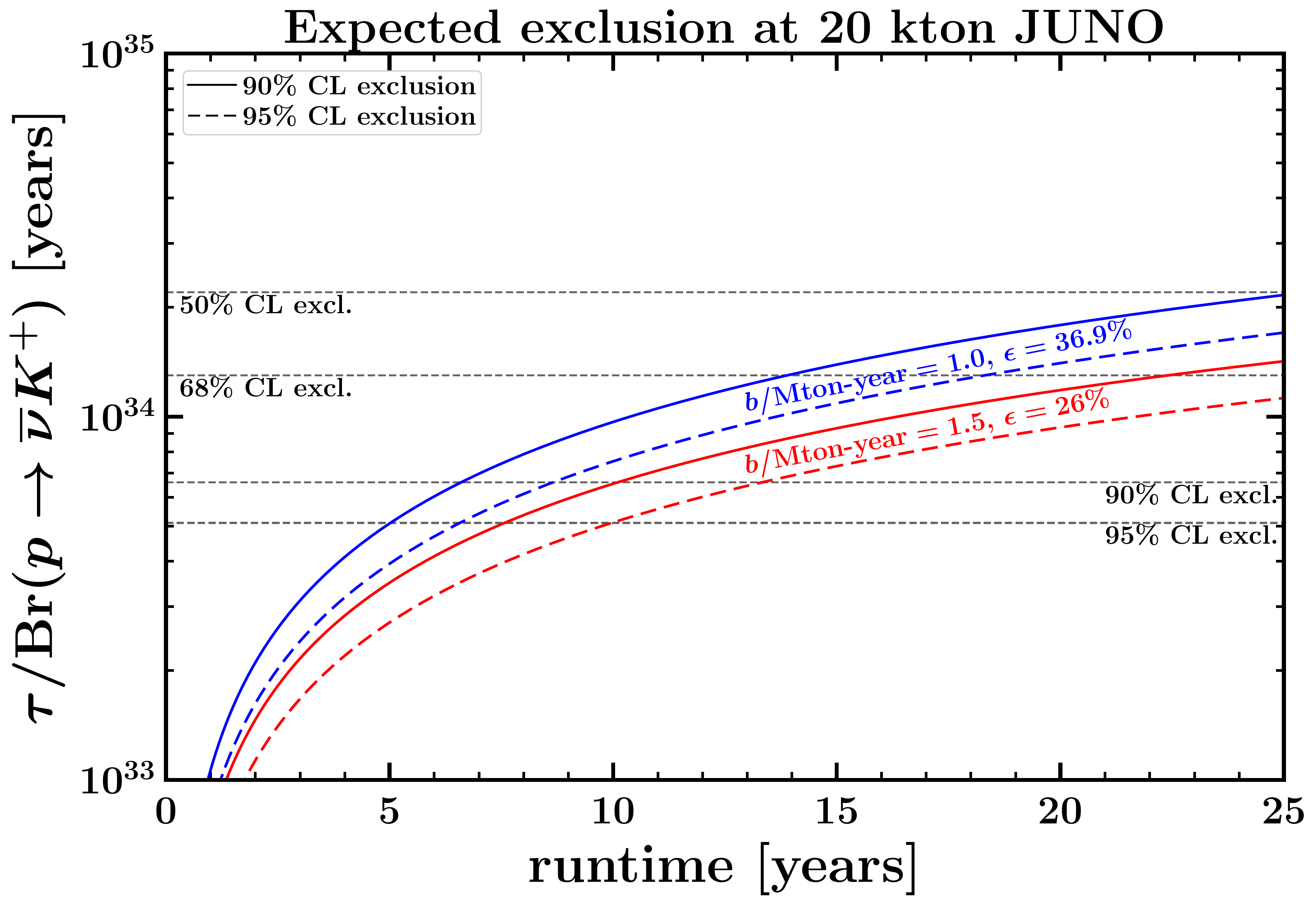}
  \includegraphics[width=13.0cm]{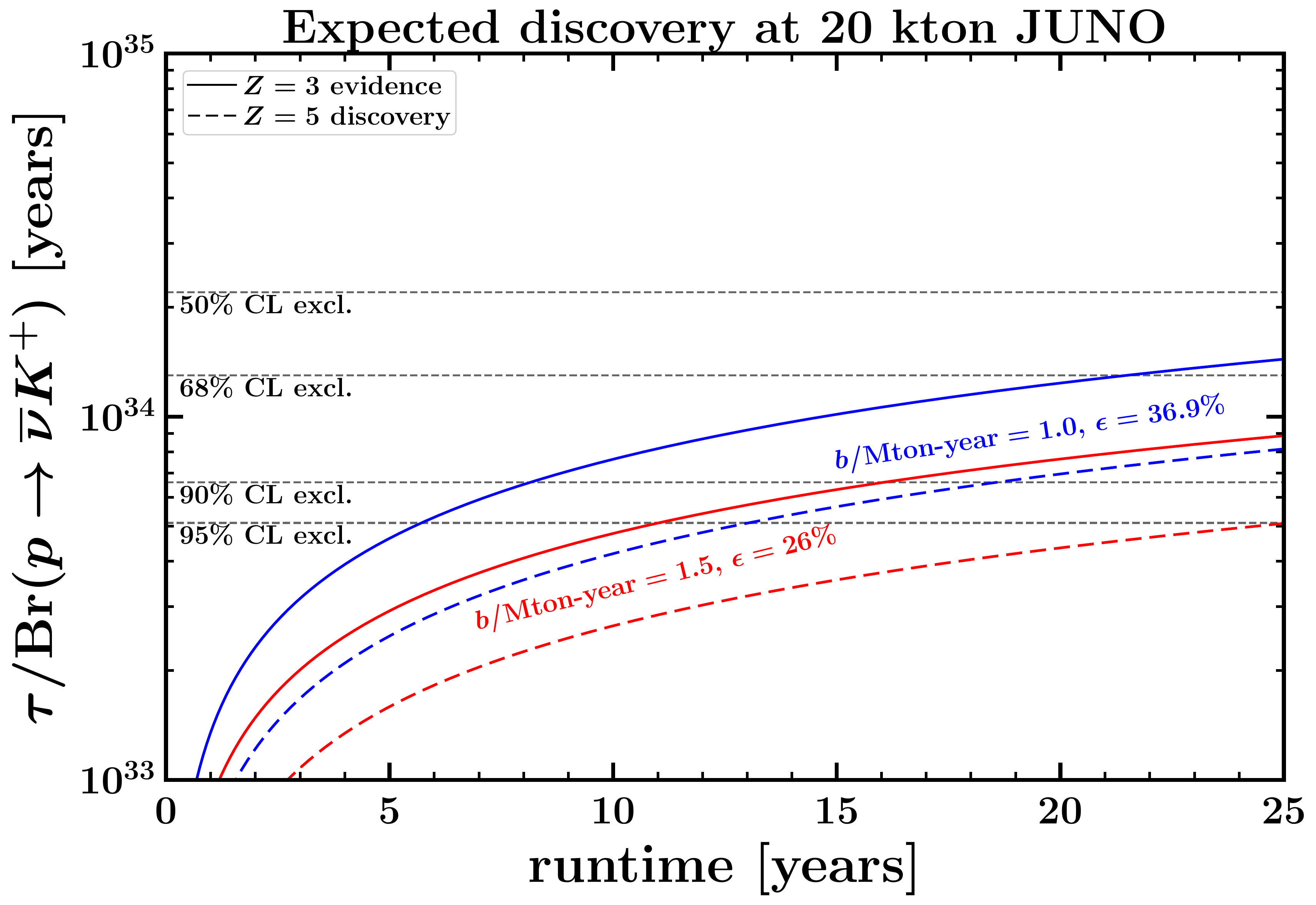}
\caption{Proton partial lifetime in the $p \rightarrow \overline{\nu} K^+$ channel that is expected to be excluded at $90\%$ or 95\% CL [top panel, from eqs.~(\ref{eq:CLsAsimov}) and (\ref{eq:lifetime reach/limit for one search channel})] or discovered at $Z=3$ or $Z=5$ significance [bottom panel, from eqs.~(\ref{eq:Zfromp}), (\ref{eq:CLdiscAsimov}) and (\ref{eq:lifetime reach/limit for one search channel})] at 20 kton JUNO, as a function of runtime, for two different estimated 
\cite{JUNO:2022qgr,JUNO:2021vlw}
combinations of background rates per year and signal selection efficiencies $(b/{\rm year}, \epsilon)$, as labeled.
Our estimates of the current 95\%, 90\%, 68\%, and 50\% CL exclusion limit on proton partial lifetime, based on Super-Kamiokande's data from 2014 \cite{Super-Kamiokande:2014otb}, are shown as horizontal dashed lines.\label{fig:JUNO}}
\end{figure}

We now turn to projections for JUNO with 20 ktons of a liquid scintillator.
We again obtain the upper limit
on the signal using eq.~(\ref{eq:CLsAsimov}) for exclusion reach, and the signal needed
for discovery using eq.~(\ref{eq:CLdiscAsimov}) for discovery reach,
then applying eq.~(\ref{eq:lifetime reach/limit for one search channel}).
Figure~\ref{fig:JUNO} shows the proton lifetime in $p \rightarrow \overline{\nu} K^+$ decay channel
that is expected to be excluded at 90\% or 95\% CL (top panel) or discovered at $Z=3$ or $Z=5$ significance (bottom panel)
at JUNO, as a function of the runtime.
The two curves correspond to two different estimates of the background accumulated per megaton-year of exposure
and the signal selection efficiency as labeled,
taken from
ref.~\cite{JUNO:2022qgr} ($b$/Mton-year = 1.0, $\epsilon = 36.9\%$; upper curve) and
ref.~\cite{JUNO:2021vlw}\footnote{In order to obtain the expected reaches for proton partial lifetime at JUNO using eq.~(\ref{eq:lifetime reach/limit for one search channel}), we redefined the signal efficiencies by multiplying the signal
efficiencies given in ref.~\cite{JUNO:2021vlw}
with the branching ratio of about 84.5\% of the $K^+$ decays that is included in JUNO's analysis.} ($b$/Mton-year = 1.5, $\epsilon = 26\%$; lower curve).
For comparison, our estimates of the current 95\%, 90\%, 68\%, and 50\% CL exclusion limit on proton partial lifetime, based on Super-Kamiokande's data from 2014 \cite{Super-Kamiokande:2014otb}, are shown as horizontal dashed lines.

For projected exclusion sensitivities,
both DUNE \cite{Alt:2020blf} and JUNO \cite{JUNO:2015zny} experiments made use of the Feldman-Cousins (FC)
method \cite{Feldman:1997qc} to obtain the upper limit on the signal assuming a fixed number of observed events,
e.g., $n=0$.
This approach can be problematic for projections because the FC upper limits at a fixed $n$ decrease with $b$
(as can be seen from Figure~\ref{fig:sULexcl_vs_b_fix_n}),
and for projections
it can imply that the expected sensitivity of the experiment gets better
if the background increases.
Also considered in ref.~\cite{Alt:2020blf} is the usage of the FC method with $n=b$.
For integer values of $b$, the FC upper limit with $n=b$ sensibly increases as the background increases.
But for non-integer $b$, $n$ is still an integer,
and the FC upper limit with $n = {\rm round}(b)$
does not always increase with $b$, as shown above in Figure~\ref{fig:sULexcl_vs_b}.
As a result, the projected sensitivity does not always decrease with $b$.
This is why we chose to use the $\CLexcl$ (= ${\rm CL}_s$ for single-channel counting experiments)
upper limit with the exact Asimov approximation
given by eq.~(\ref{eq:CLsAsimov}) for DUNE and JUNO.
While the FC sensitivity of ref.~\cite{Feldman:1997qc} from eq.~(\ref{eq:FCsensitivity}) also gives sensible projections for exclusion,
we note that it is computationally more intense to evaluate (and gives only
slightly more conservative results)
than the exact Asimov expected $\CLexcl$ upper limits.

\begin{figure}
  \includegraphics[width=13.0cm]{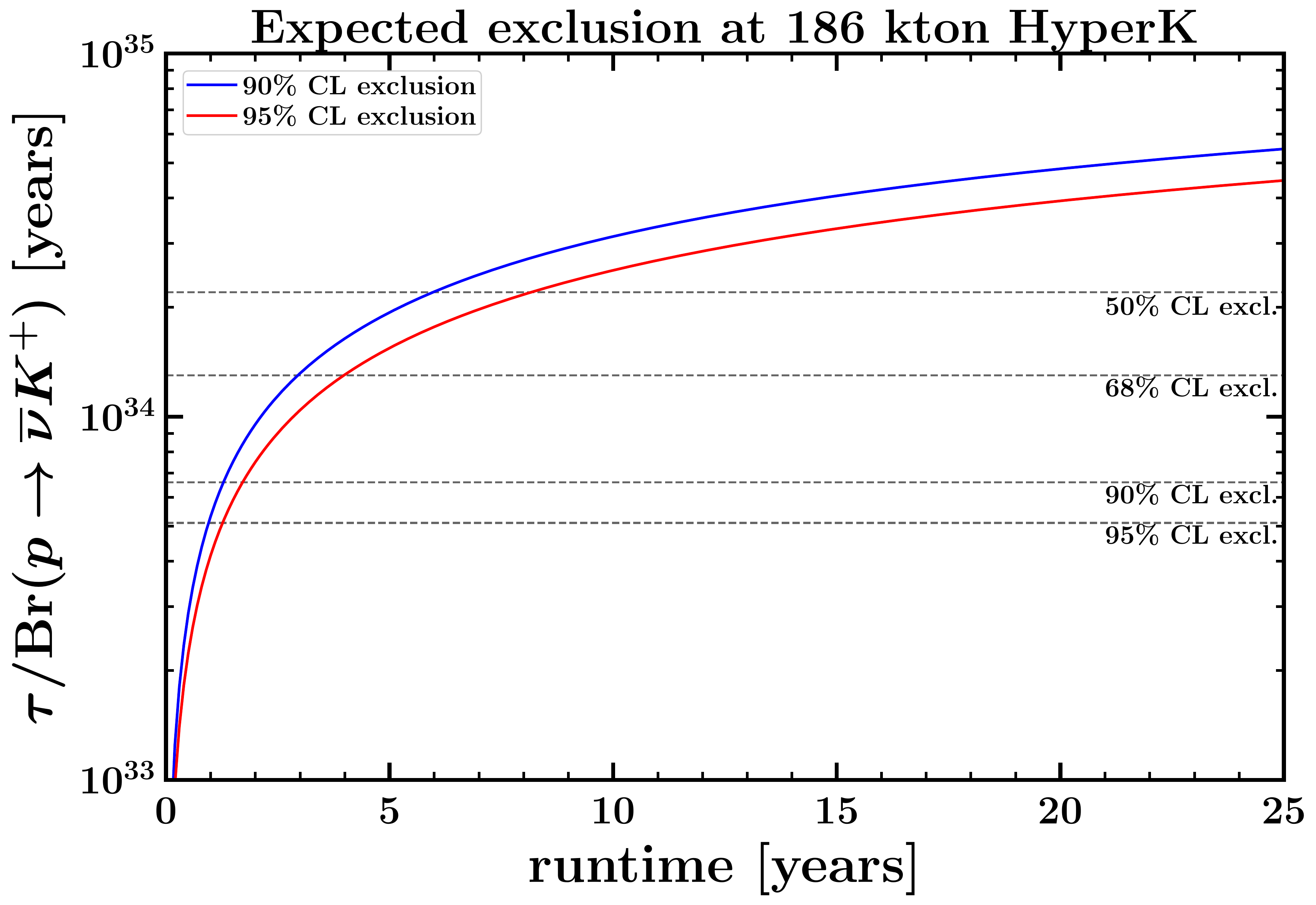}
  \includegraphics[width=13.0cm]{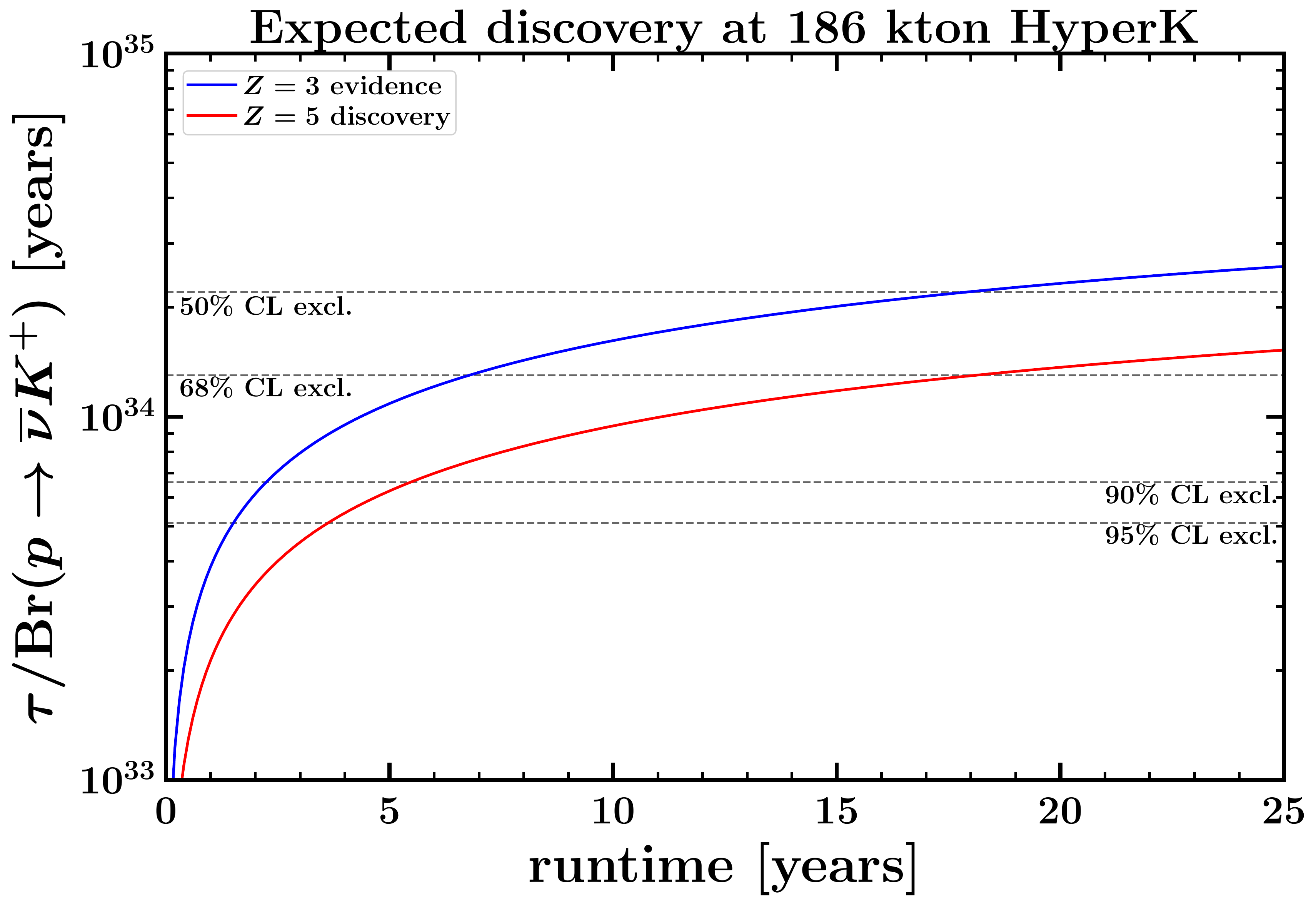}
 \caption{Proton partial lifetime in the $p \rightarrow \overline{\nu} K^+$ channel that
 is expected to be excluded at $90\%$ or 95\% CL [top panel; from eq.~(\ref{eq:Bayesian projections for exclusion})] or discovered at $Z=3$ or $Z=5$ significance [bottom panel; from eq.~(\ref{eq:Bayesian projections for discovery})] at Hyper-Kamiokande with 186 kilotons of water, as a function of runtime, with the uncertainties in background and signal selection efficiency listed in Table \ref{tab:HyperK_data}, taken from ref.~\cite{Hyper-Kamiokande:2018ofw}. Our estimates of the current 95\%, 90\%, 68\%, and 50\% CL exclusion limit on proton partial lifetime, based on Super-Kamiokande's data from 2014 \cite{Super-Kamiokande:2014otb}, are shown as horizontal dashed lines.
\label{fig:HyperK_nuK}}
\end{figure}

\begin{figure}
  \includegraphics[width=13.0cm]{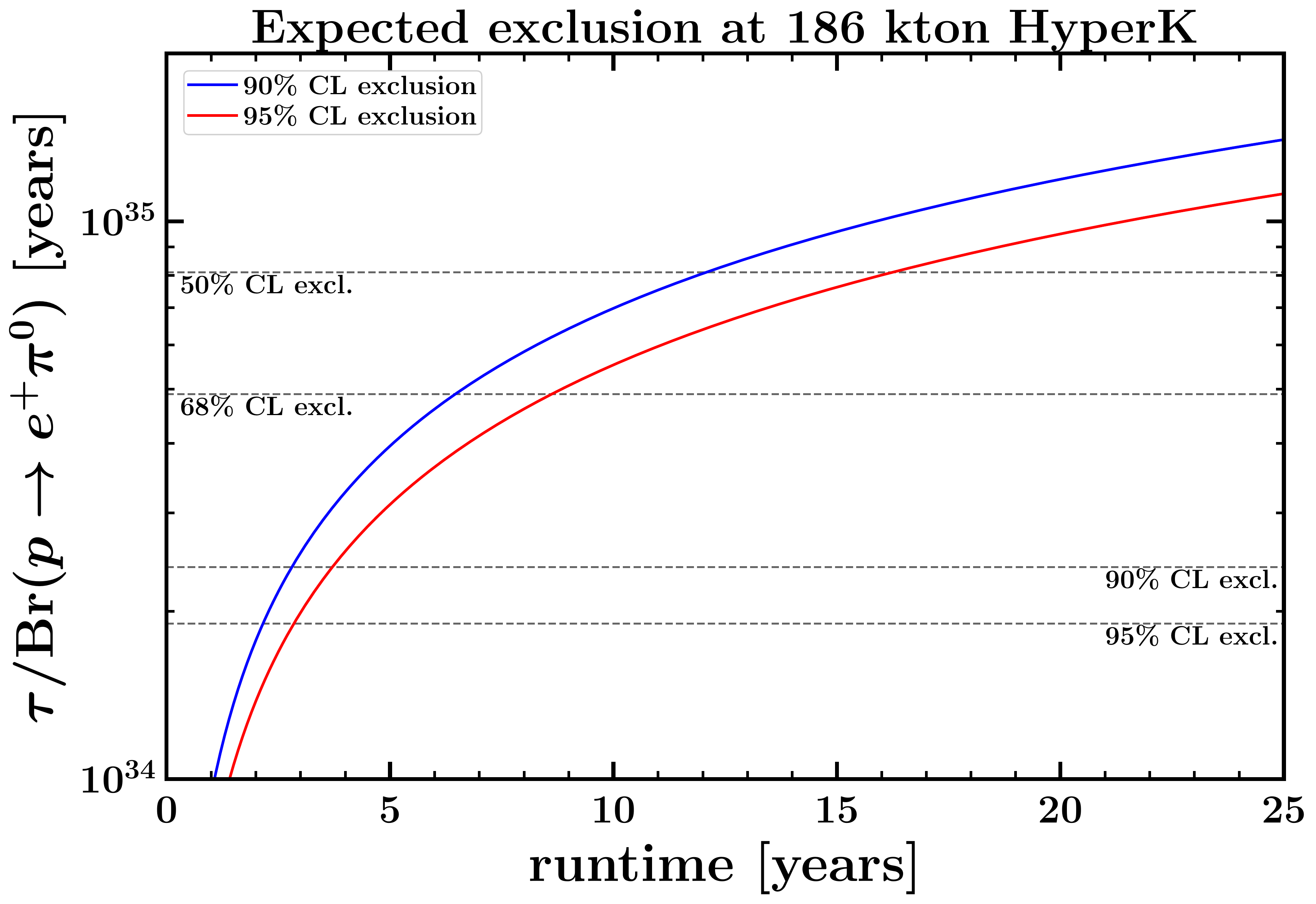}
  \includegraphics[width=13.0cm]{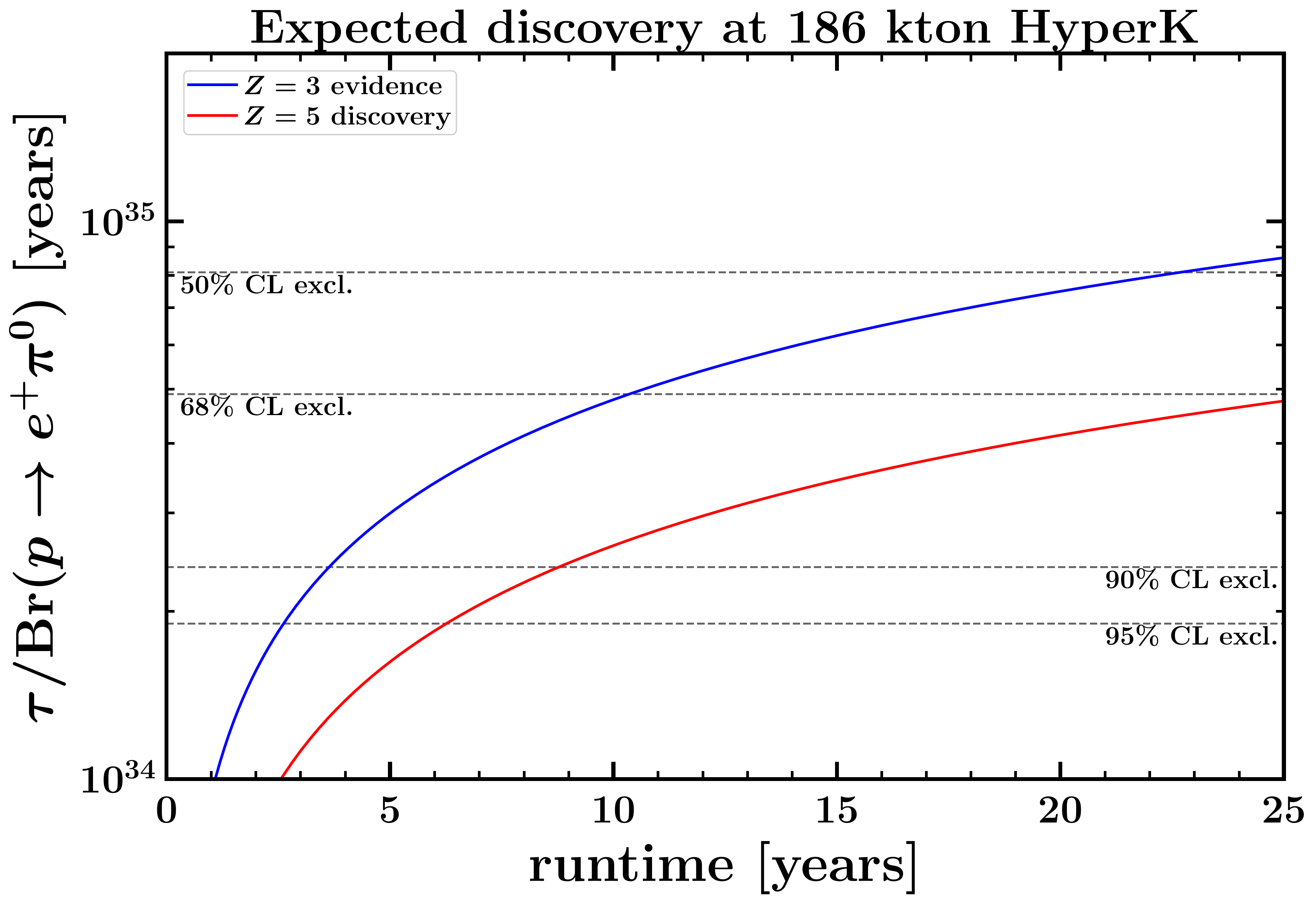}
 \caption{Proton partial lifetime in the $p \rightarrow e^+ \pi^0$ channel that
is expected to be excluded at $90\%$ or 95\% CL [top panel; from eq.~(\ref{eq:Bayesian projections for exclusion})] or discovered at $Z=3$ or $Z=5$ significance [bottom panel; from eq.~(\ref{eq:Bayesian projections for discovery})] at Hyper-Kamiokande with 186 kilotons of water, as a function of runtime, with the uncertainties in background and signal selection efficiency listed in Table \ref{tab:HyperK_data}, taken from ref.~\cite{Hyper-Kamiokande:2018ofw}.
Our estimates of the current 95\%, 90\%, 68\%, and 50\% CL exclusion limit on proton partial lifetime, based on Super-Kamiokande's data from 2020 \cite{Super-Kamiokande:2020wjk}, are shown as
horizontal dashed lines.
\label{fig:HyperK_epi}}
\end{figure}

We next turn to projections for Hyper-Kamiokande.
Figures~\ref{fig:HyperK_nuK} and \ref{fig:HyperK_epi} show our estimates for the
proton partial lifetimes in $p \rightarrow \overline{\nu} K^+$ and $p \rightarrow e^+ \pi^0$
decay channels, respectively, that are
expected to be excluded at 90\% or 95\% CL (top panels) or
discovered at $Z=3$ or $Z=5$ significance (bottom panels), as a function of runtime at Hyper-Kamiokande.%
In order to obtain the exclusion and discovery reaches for $\tau_p$, the upper limit on partial width and the partial width needed for discovery are solved from eqs.~(\ref{eq:Bayesian projections for exclusion}) and (\ref{eq:Bayesian projections for discovery}), respectively. These equations are used to combine the independent search channels in each decay mode,
based on the background means and the signal selection efficiencies, along with their
uncertainties, given in ref.~\cite{Hyper-Kamiokande:2018ofw} and summarized in our Table~\ref{tab:HyperK_data}.
\begin{table}
\begin{minipage}[]{0.95\linewidth}
\caption{Estimated backgrounds $\hat b_i \pm \Delta_{b_i}$ per megaton-year of exposure
and signal efficiencies $\hat \epsilon_i \pm \Delta_{\epsilon_i}$ at Hyper-Kamiokande,
taken from ref.~\cite{Hyper-Kamiokande:2018ofw},
for $p \rightarrow \overline{\nu} K^+$ and
$p \rightarrow e^+ \pi^0$ decay modes.
The last column gives a brief description of each of the channels referring
to the name of the search method
used in ref.~\cite{Hyper-Kamiokande:2018ofw}.
Exposure in each channel for a 186 kton Hyper-Kamiokande is given by $\lambda_i = 0.186 \text{ Mton} \times \text{runtime in years}$.
\label{tab:HyperK_data}}
\end{minipage}
\begin{center}
\begin{tabular}{|c || c | c || c |}
\hline
~Decay mode~ & ~$\hat b_i \pm \Delta_{b_i}$ [/Mton-year]~  &  ~$\hat \epsilon_i \pm \Delta_{\epsilon_i}$ [\%]~  & ~Comment~\\
\hline
\hline
    ~$p \rightarrow \overline{\nu} K^+$~ & ~$0.9 \pm 0.2$~ & ~$12.7 \pm 2.4$~ & ~prompt $\gamma$~
    \\[1pt]
    \cline{2-4}
    ~~ & ~$0.7 \pm 0.2$~ & ~$10.8 \pm 1.1$~ & ~$\pi^+ \pi^0$~
    \\[1pt]
    \cline{2-4}
    ~~ & ~$1916$~ & ~$31$~ & ~$p_\mu$ spectrum~
    \\[1pt]
    \hline
\hline
    ~$p \rightarrow e^+ \pi^0$~ & ~$0.06 \pm 0.02$~ & ~$18.7 \pm 1.2$~ & ~$0 < p_{\rm tot} < 100$ MeV/c~
    \\[1pt]
    \cline{2-4}
    ~~ & ~$0.62 \pm 0.20$~ & ~$19.4 \pm 2.9$~ & ~$100 < p_{\rm tot} < 250$ MeV/c~
    \\[1pt]
\hline
\end{tabular}
\end{center}
\end{table}
Figures~\ref{fig:HyperK_nuK} and \ref{fig:HyperK_epi} also show our previously discussed estimates of the current exclusion limits at 95\%, 90\%, 68\%, 50\% CL in $p \rightarrow \overline{\nu} K^+$ and $p \rightarrow e^+ \pi^0$ decay modes based on the data from refs.~\cite{Super-Kamiokande:2014otb} and \cite{Super-Kamiokande:2020wjk}, respectively.

\begin{figure}
  \includegraphics[width=13.0cm]{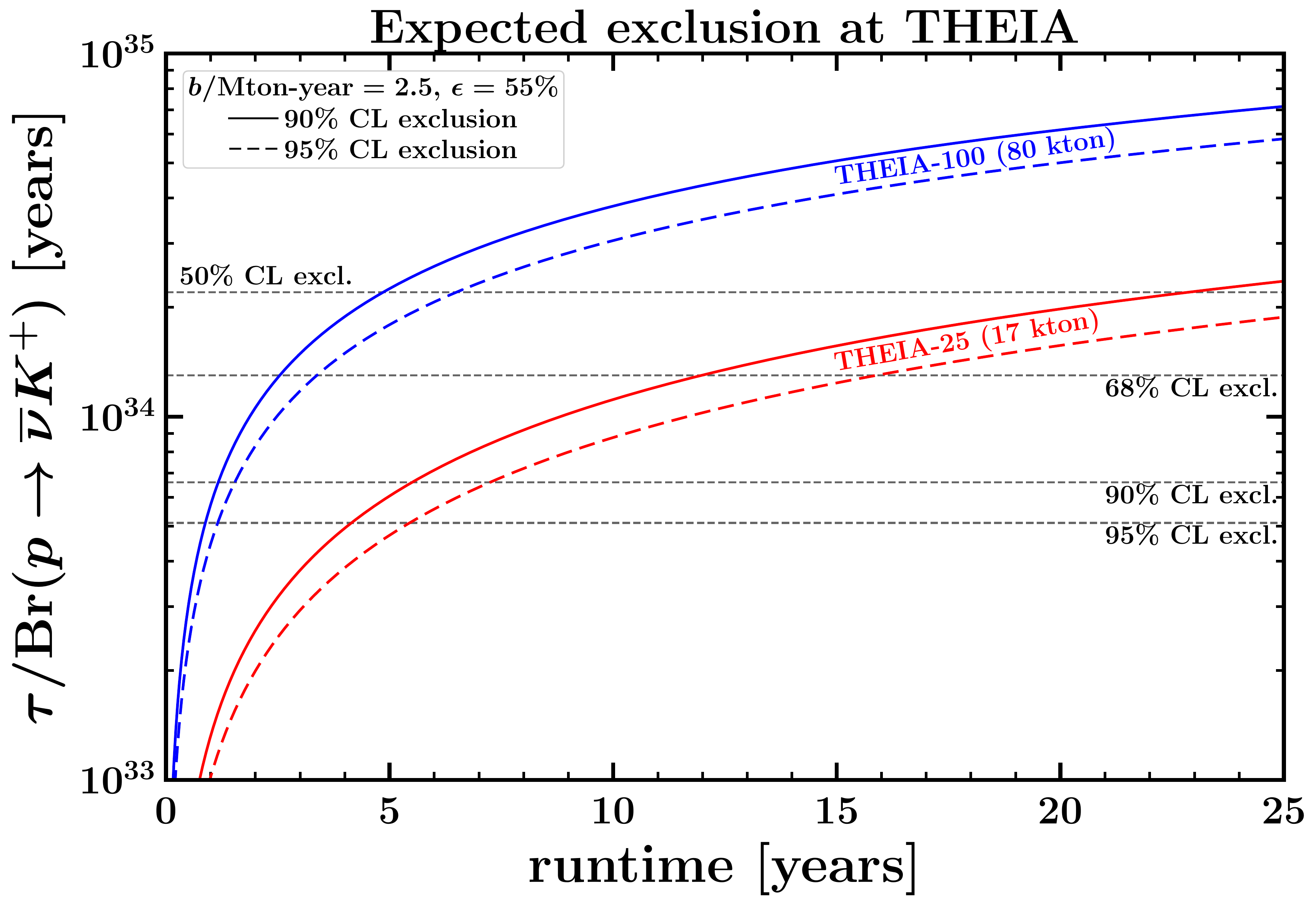}
  \includegraphics[width=13.0cm]{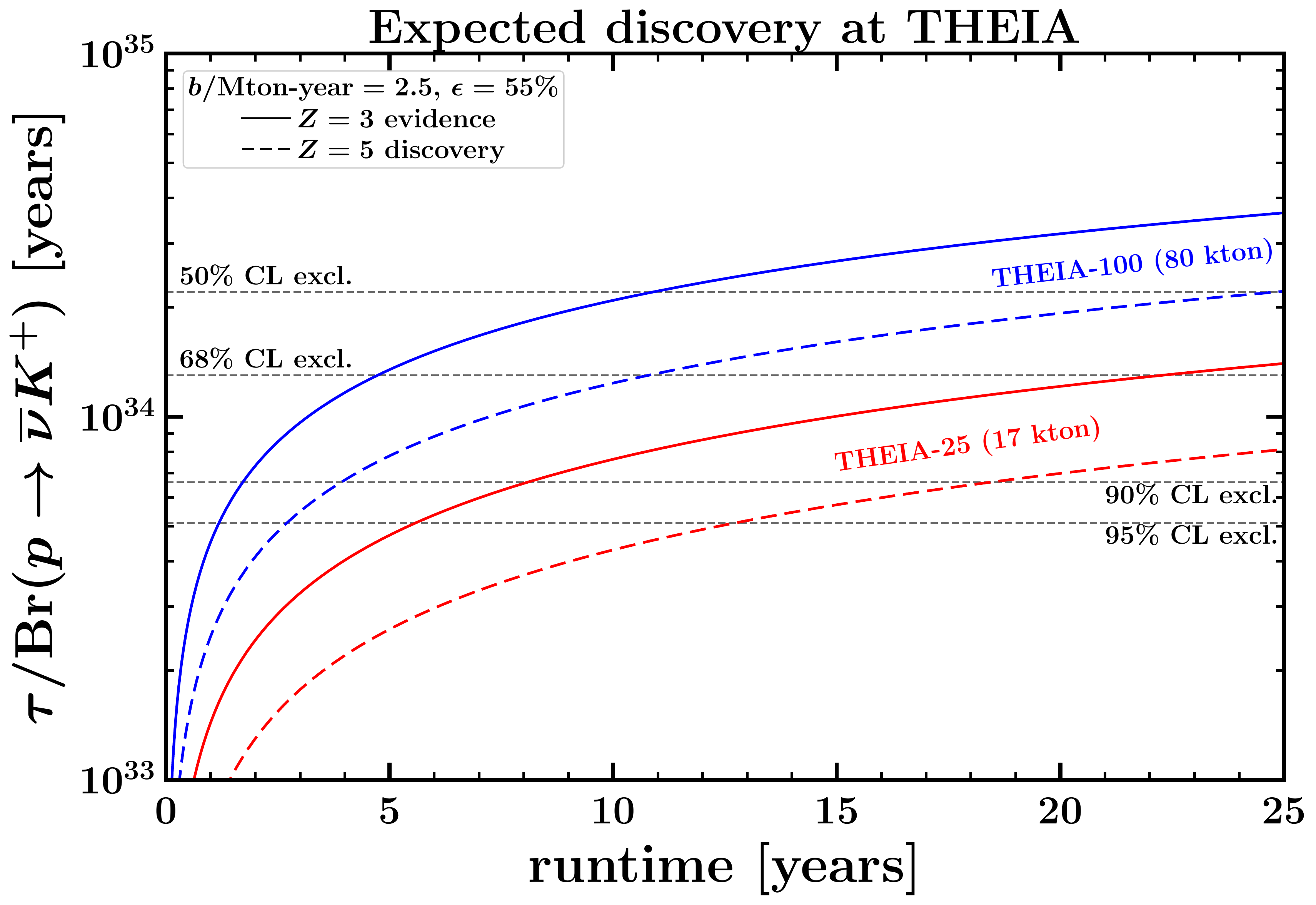}
\caption{Expected 90\% or 95\% CL exclusion reaches [top panel; from eqs.~(\ref{eq:CLsAsimov}) and (\ref{eq:lifetime reach/limit for one search channel})] and $Z=3$ or $Z=5$ discovery reaches [bottom panel; from eqs.~(\ref{eq:Zfromp}), (\ref{eq:CLdiscAsimov}) and (\ref{eq:lifetime reach/limit for one search channel})] for proton partial lifetime in $p \rightarrow \overline{\nu} K^+$ with THEIA-25 (red lines) and THEIA-100 (blue lines) with 17 and 80 ktons of water based liquid scintillator, respectively, as a function of runtime. The estimates for the background (per megaton-year of exposure) and the signal efficiencies are taken from ref.~\cite{Theia:2019non}. Our estimates of the current 95\%, 90\%, 68\%, and 50\% CL exclusion limit on proton partial lifetime, based on Super-Kamiokande's data from 2014 \cite{Super-Kamiokande:2014otb}, are shown as horizontal dashed lines.
\label{fig:THEIA_nuK}}
\end{figure}

\begin{figure}
  \includegraphics[width=13.0cm]{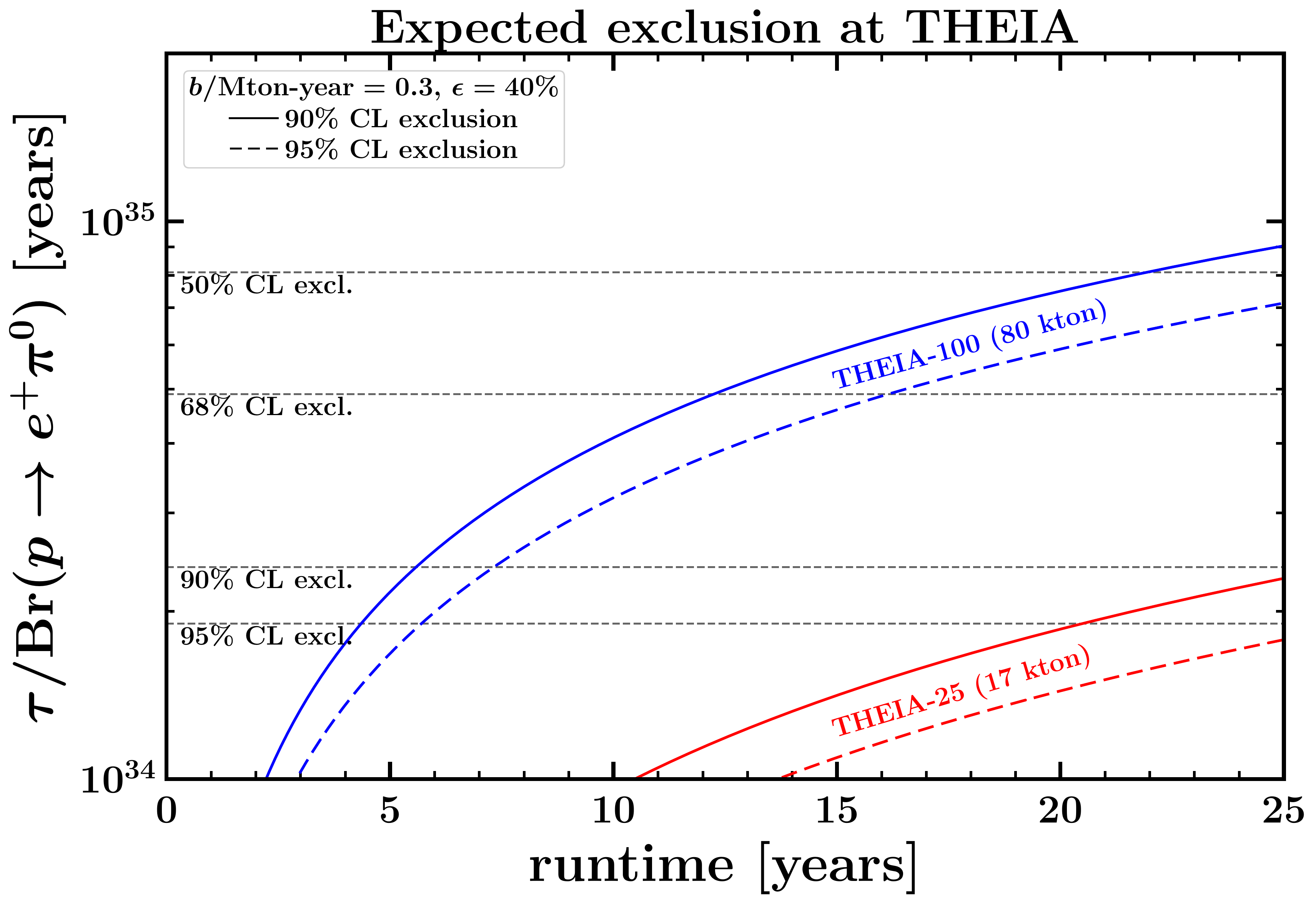}
  \includegraphics[width=13.0cm]{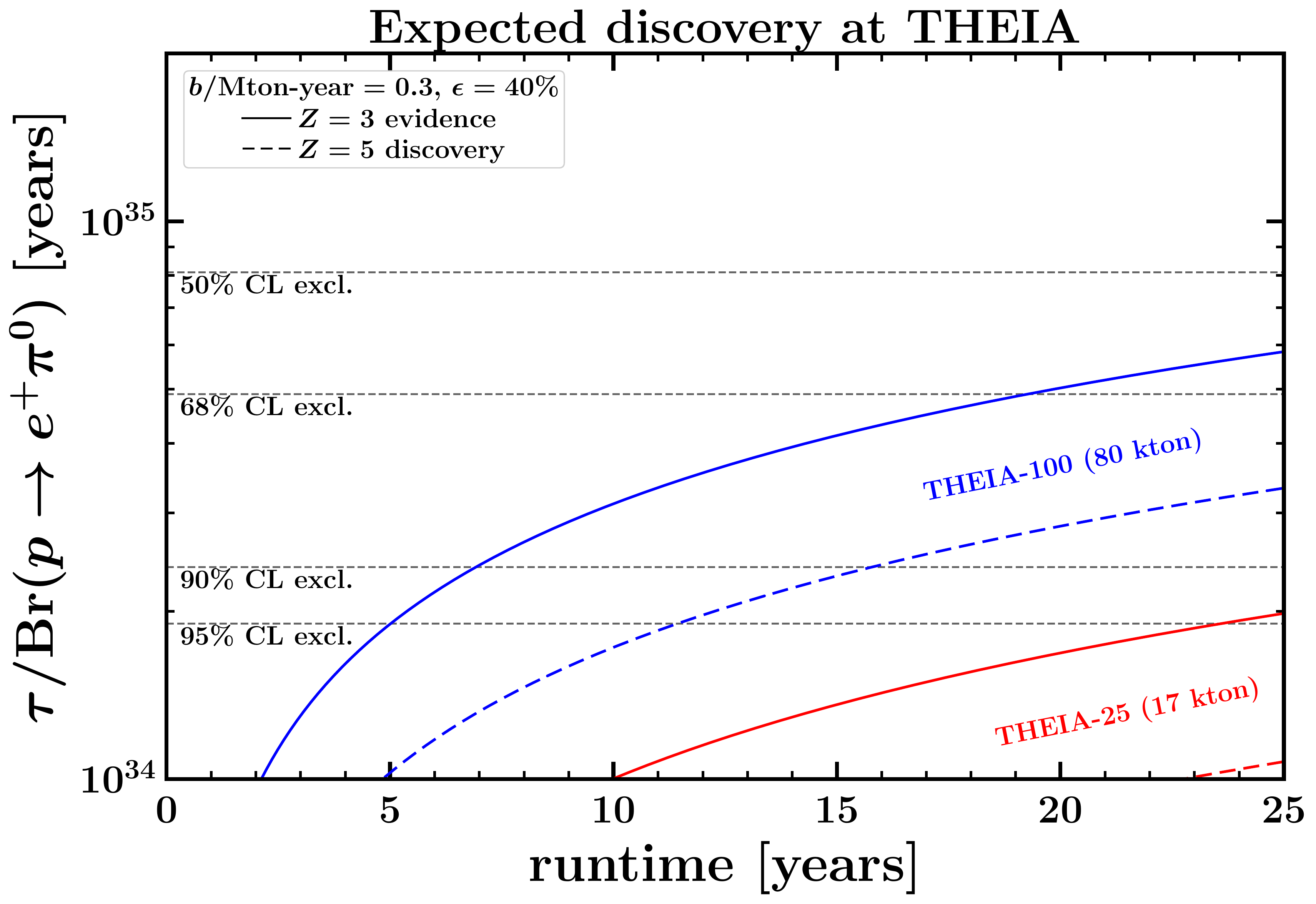}
\caption{Expected 90\% or 95\% CL exclusion reaches [top panel; from eqs.~(\ref{eq:CLsAsimov}) and (\ref{eq:lifetime reach/limit for one search channel})] and $Z=3$ or $Z=5$ discovery reaches [bottom panel; from eqs.~(\ref{eq:Zfromp}), (\ref{eq:CLdiscAsimov}) and (\ref{eq:lifetime reach/limit for one search channel})] for proton partial lifetime in $p \rightarrow e^+ \pi^0$ with THEIA-25 (red lines) and THEIA-100 (blue lines) with 17 and 80 ktons of water based liquid scintillator, respectively, as a function of runtime. The estimates for the background (per megaton-year of exposure) and the signal efficiencies are taken from ref.~\cite{Dev:2022jbf}. Our estimates of the current 95\%, 90\%, 68\%, and 50\% CL exclusion limit on proton partial lifetime, based on Super-Kamiokande's data from 2020 \cite{Super-Kamiokande:2020wjk}, are shown as
horizontal dashed lines.
\label{fig:THEIA_epi}}
\end{figure}

Finally, we turn to projections for THEIA.
In Figures~\ref{fig:THEIA_nuK}~and~\ref{fig:THEIA_epi}, we show the expected reaches,
as a function of runtime, for proton partial lifetime in $p \rightarrow \overline{\nu} K^+$
and $p \rightarrow e^+ \pi^0$ decay modes, respectively, for 90\% or 95\% CL exclusion (top panels)
and discovery at $Z=3$ or $Z=5$ significance (bottom panels).
The lower (red) lines show the results for THEIA-25 with 17 ktons of fiducial mass
of water based liquid scintillator, while the upper (blue) lines are for THEIA-100 with 80 ktons fiducial mass. The expected reach for proton partial lifetime is computed using eq.~(\ref{eq:lifetime reach/limit for one search channel}), where the expected signal for 90\% CL exclusion ($Z=3$ evidence) is obtained from setting eq.~(\ref{eq:CLsAsimov}) (eq.~(\ref{eq:CLdiscAsimov})) to 0.1 (0.00135). The estimates for the background rate per megaton-year of exposure and the signal selection efficiency for the decays modes $p \rightarrow \overline{\nu} K^+$ and $p \rightarrow e^+ \pi^0$ are taken from refs.~\cite{Theia:2019non} and \cite{Dev:2022jbf}, respectively. As before, we also show our estimates for the current lower limits at various confidence levels based on the data from Super-Kamiokande \cite{Super-Kamiokande:2014otb,Super-Kamiokande:2020wjk}.


\section{Outlook\label{sec:outlook}}
\setcounter{equation}{0}
\setcounter{figure}{0}
\setcounter{table}{0}
\setcounter{footnote}{1}

\begin{figure}
  \includegraphics[width=16.0cm]{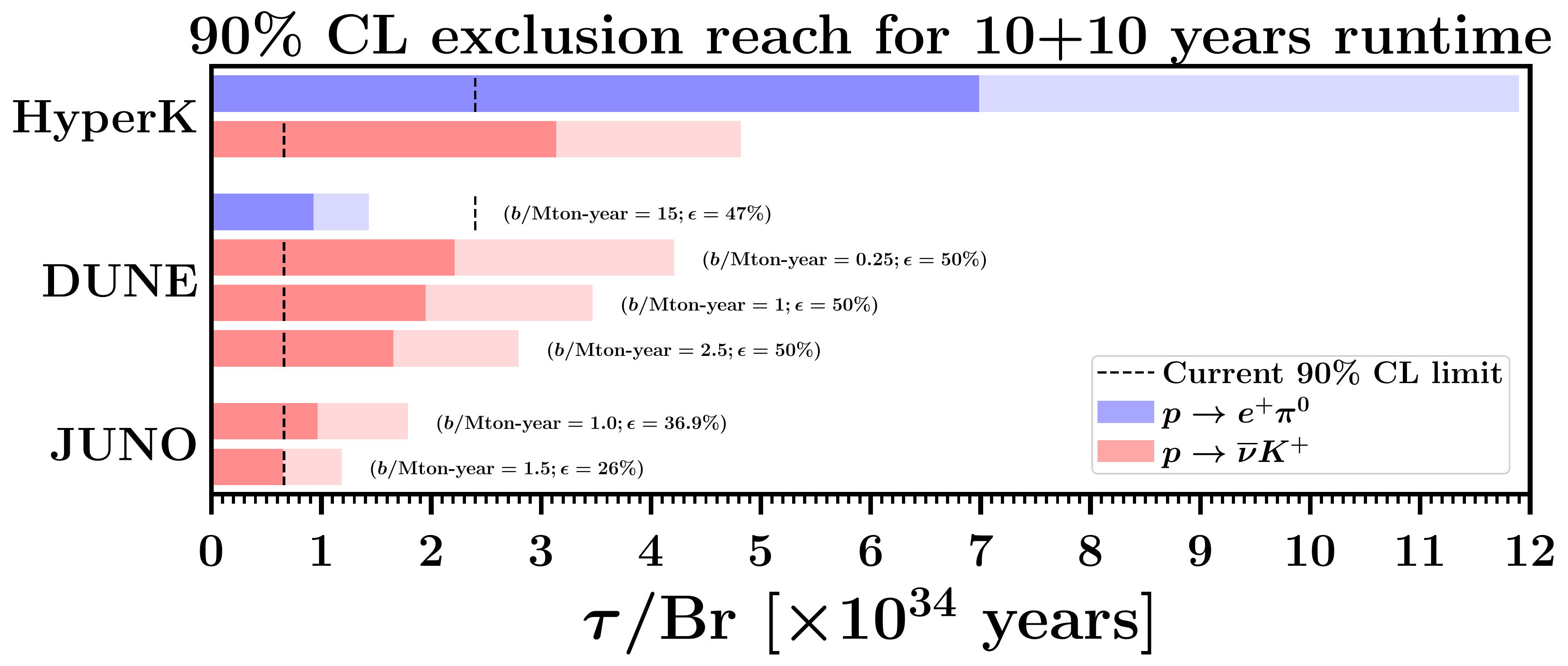}
  \includegraphics[width=16.0cm]{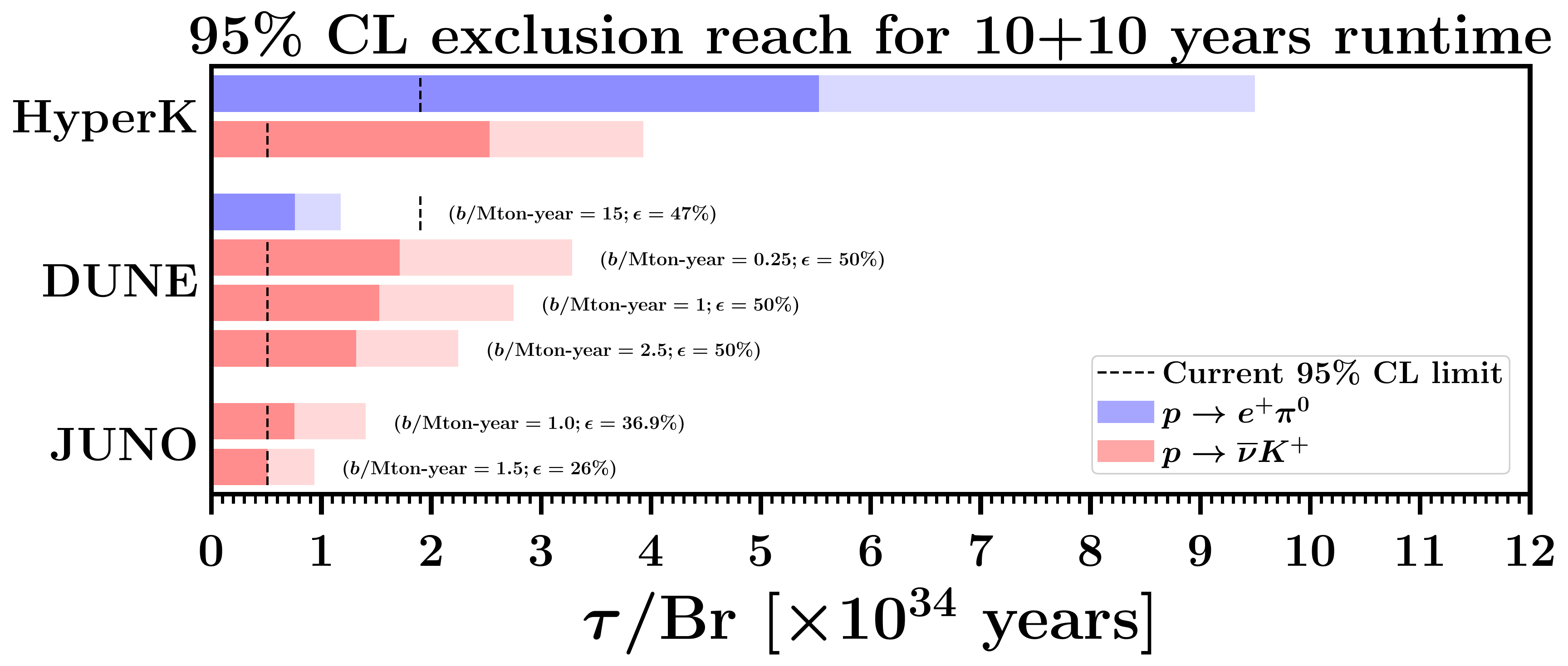}
\caption{Expected exclusion reaches at 90\% CL (top panel) and 95\% CL (bottom panel) for proton partial lifetime in 
$p \rightarrow e^+ \pi^0$ (blue bars) 
and 
$p \rightarrow \overline{\nu} K^+$ (red bars)
decay channels at JUNO, DUNE, and Hyper-Kamiokande after 10 years (darker shading) and 20 years (lighter shading) of runtime. The assumed backgrounds and signal efficiencies for JUNO and DUNE are labeled in the plots, and for Hyper-Kamiokande, the corresponding information is given in Table~\ref{tab:HyperK_data}, quoted from ref.~\cite{Hyper-Kamiokande:2018ofw}. These results are based on preliminary estimates of the backgrounds and signal efficiencies, which are likely to change as the experiments progress, and therefore should be viewed with some caution as comparisons. 
The vertical dashed lines are our estimates of the current 90\% CL (top panel) and 95\% CL (bottom panel)
lower limits based on 
Super-Kamiokande's data from 2014 \cite{Super-Kamiokande:2014otb} and 2020 
\cite{Super-Kamiokande:2020wjk}.
\label{fig:Reach_excl}}
\end{figure}

\begin{figure}
  \includegraphics[width=16.0cm]{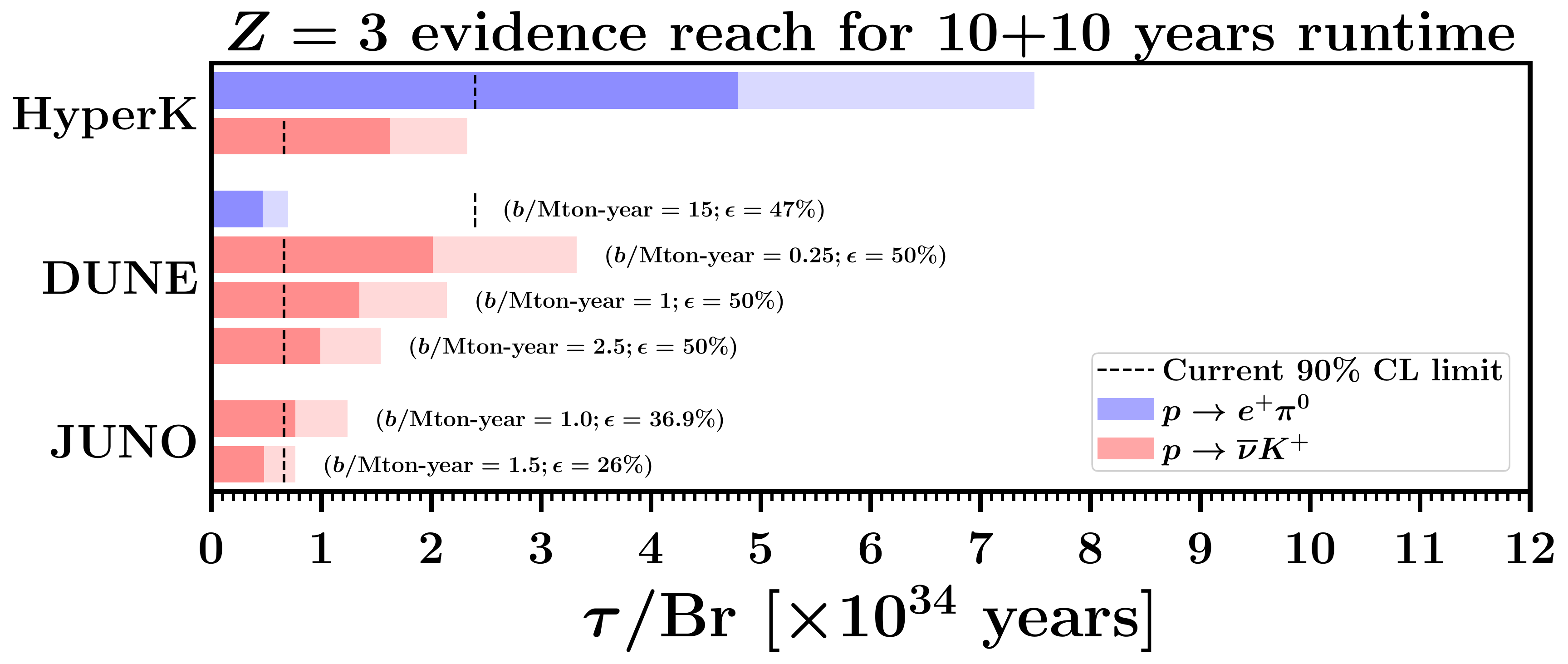}
  \includegraphics[width=16.0cm]{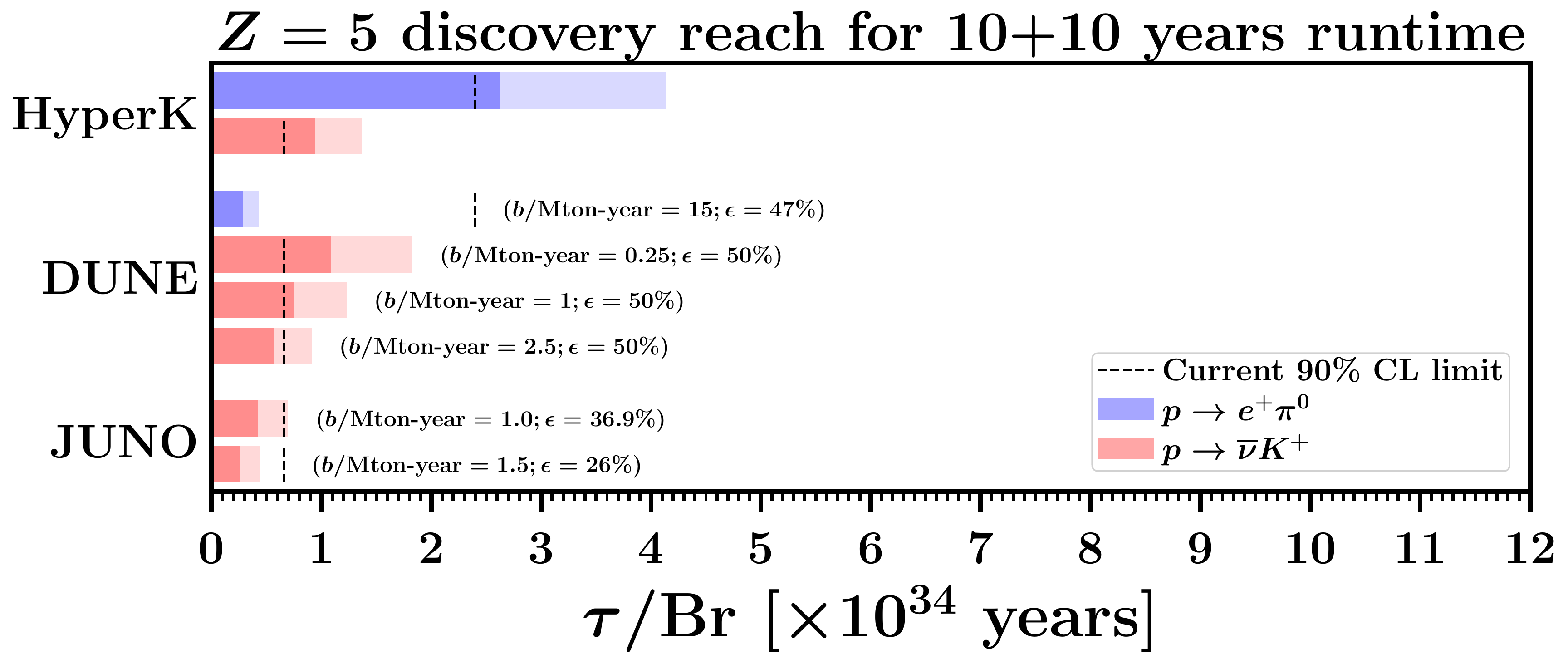}
\caption{Expected reaches for $Z=3$ evidence (top panel) and $Z=5$ discovery (bottom panel) for proton partial lifetime in 
$p \rightarrow e^+ \pi^0$ (blue bars) 
and 
$p \rightarrow \overline{\nu} K^+$ (red bars) 
decay channels, at JUNO, DUNE, and Hyper-Kamiokande after 10 years (darker shading) and 20 years (lighter shading) of runtime. The assumed backgrounds and signal efficiencies for JUNO and DUNE are labeled in the plots, and for Hyper-Kamiokande, the corresponding information is given in Table~\ref{tab:HyperK_data}, quoted from ref.~\cite{Hyper-Kamiokande:2018ofw}. These results are based on preliminary estimates of the backgrounds and signal efficiencies, which are likely to change as the experiments progress, and therefore should be viewed with some caution as comparisons. The vertical dashed lines are our estimates of the current 90\% CL lower limits based on 
Super-Kamiokande's data from 2014 \cite{Super-Kamiokande:2014otb} and 2020 
\cite{Super-Kamiokande:2020wjk}.
\label{fig:Reach_disc}}
\end{figure}

\begin{figure}
  \includegraphics[width=16.0cm]{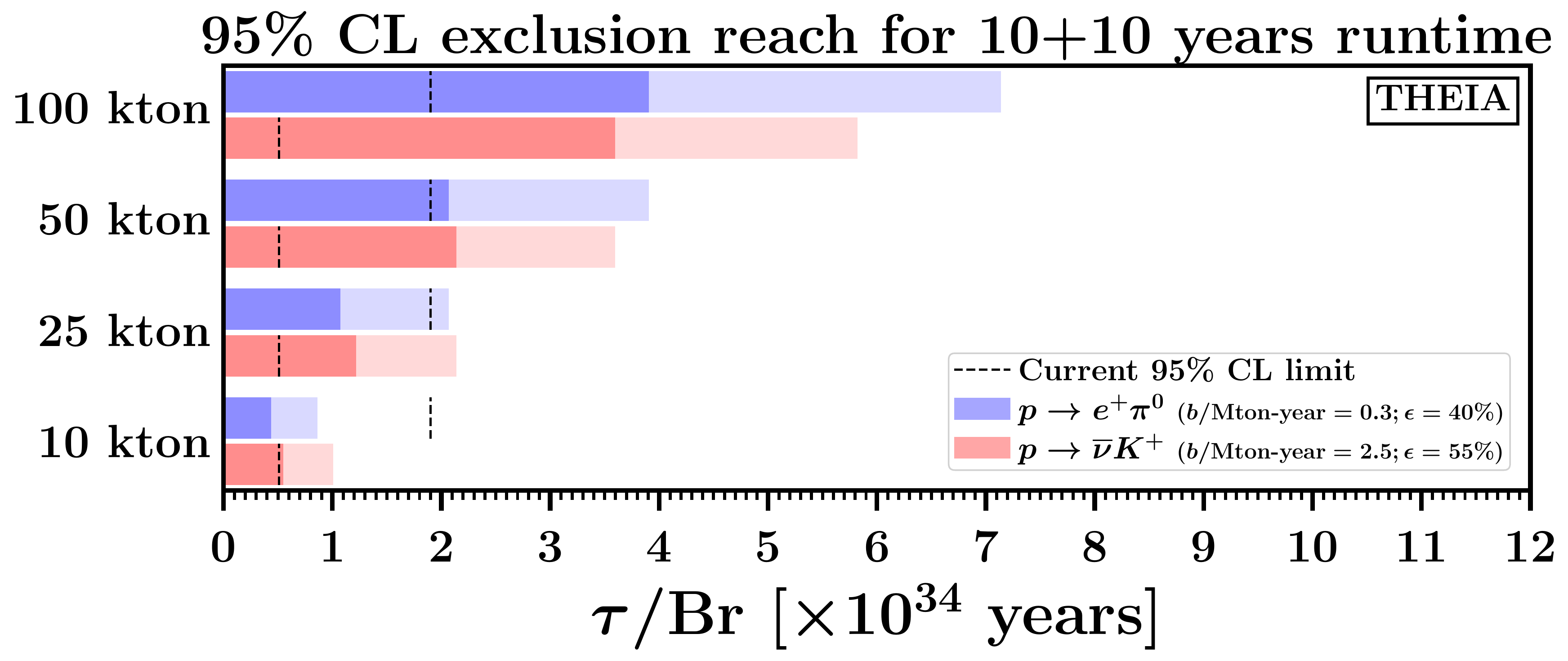}
  \includegraphics[width=16.0cm]{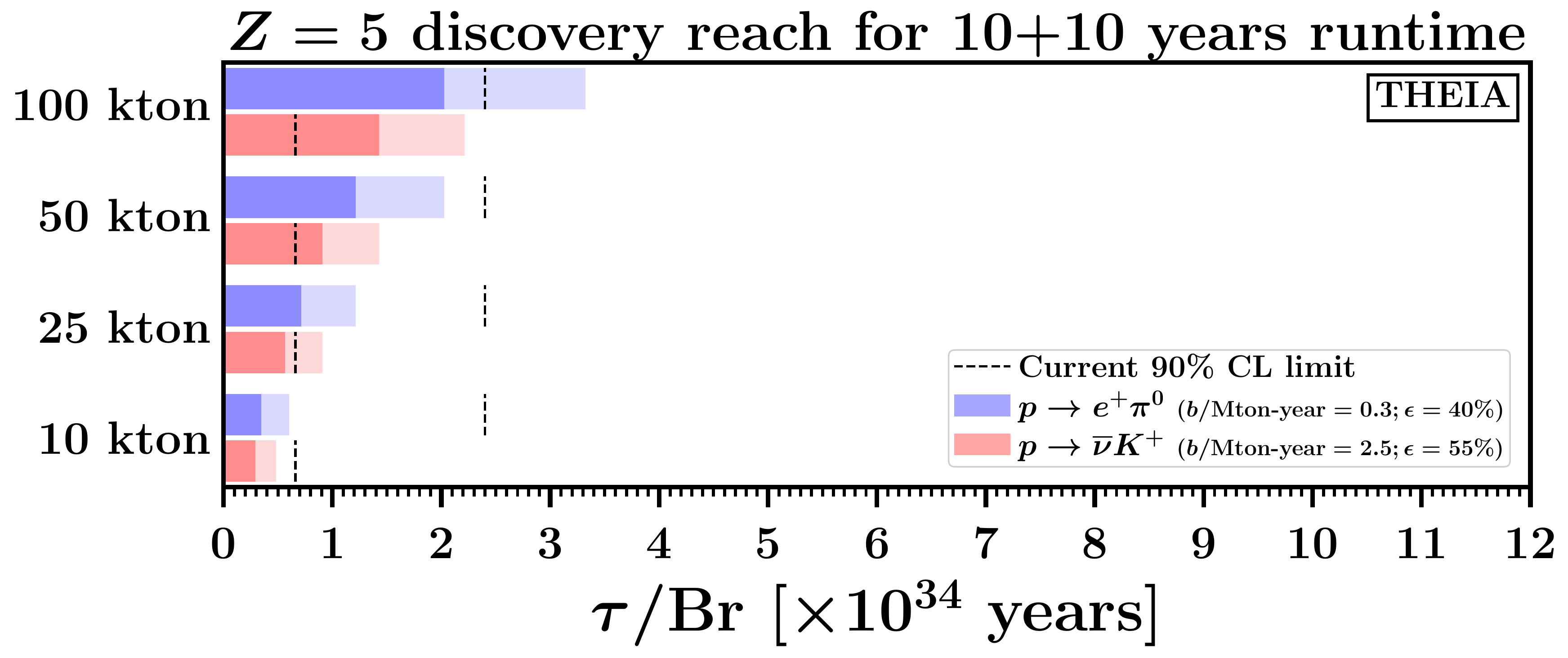}
\caption{Expected reaches at THEIA for 95\% CL exclusion (top panel) and
$Z=5$ discovery (bottom panel) for proton partial lifetime in
$p \rightarrow e^+ \pi^0$ (blue bars) 
and 
$p \rightarrow \overline{\nu} K^+$ (red bars)
decay channels with various fiducial masses, as labeled, after 10 years (darker shading) and 20 years (lighter shading) of runtime. The assumed background rates and signal efficiencies for THEIA are labeled in the plots. These results are based on preliminary estimates of the backgrounds and signal efficiencies, which are likely to change as the experiment progresses.
The vertical dashed lines are our estimates of the current 95\% CL (top panel) and 90\% CL (bottom panel)
lower limits based on 
Super-Kamiokande's data from 2014 \cite{Super-Kamiokande:2014otb} and 2020 
\cite{Super-Kamiokande:2020wjk}.
\label{fig:THEIA_Reach}}
\end{figure}

We summarize our projections for future proton decay searches in the final states
$e^+ \pi^0$ and $ \overline \nu K^+$ at
DUNE, JUNO, and Hyper-Kamiokande
in Figure \ref{fig:Reach_excl} for exclusion (assuming the signal is indeed absent), 
and in Figure \ref{fig:Reach_disc} for discovery (assuming the signal is actually present).
And in Figure \ref{fig:THEIA_Reach} we summarize our projections at THEIA for 95\% CL exclusion and $Z=5$ discovery for various fiducial masses $N_{\rm kton} = (10, 25, 50, 100)$ kton.
In each case, we show results for 10 years and 20 years of runtime.
The assumed backgrounds and signal efficiencies for DUNE\footnote{For projections at DUNE in $p \rightarrow \overline{\nu} K^+$ channel, we are using the optimistic choices based on ref.~\cite{Alt:2020gog}. More pessimistic choices from refs.~\cite{Alt:2020rii,DUNE:2020ypp,DUNE:2020fgq,DUNE:2022aul} will of course lead to lower reach estimates.}, JUNO, and THEIA in each proton decay mode are labeled in the plots, while the corresponding information for the multi-channel Hyper-Kamiokande searches was given in Table~\ref{tab:HyperK_data} above, quoted from ref.~\cite{Hyper-Kamiokande:2018ofw}.
The vertical dashed lines correspond to our estimate of the current 90\% CL
(Fig.~\ref{fig:Reach_disc}, top panel of Fig.~\ref{fig:Reach_excl}, and bottom panel of Fig.~\ref{fig:THEIA_Reach})
or 95\% CL
(bottom panel of Fig.~\ref{fig:Reach_excl} and top panel of Fig.~\ref{fig:THEIA_Reach})
lower limit on proton partial lifetime in the respective decay channels, based on the published Super-Kamiokande data \cite{Super-Kamiokande:2014otb,Super-Kamiokande:2020wjk}.%

As noted above, our projections here are based on the exact Asimov evaluation of the Bayesian statistics $\CLexcl$ and $\CLdisc$. Our results are somewhat more conservative than previous projections appearing in refs.~\cite{Hyper-Kamiokande:2018ofw} and the Snowmass report \cite{Dev:2022jbf}, which we have generalized to include 90\% CL exclusion and $Z=3$ evidence reach estimates as a function of runtime
(for various estimates of backgrounds and signal efficiencies, notably for DUNE and JUNO)
as well as estimates for 95\% CL exclusion and $Z=5$ discovery.
In the cases of single-channel searches for DUNE, JUNO, and THEIA, we have also investigated the use of the exact Asimov frequentist $p$-value measures $p_{\rm excl}$ and $p_{\rm disc}$. These results are not shown in the figures; we find that they are only slightly less conservative than the estimates shown.

The two panels of Figure \ref{fig:Reach_excl} show the projected exclusion reaches at 
90\% and 95\% confidence level, while the two panels of Figure \ref{fig:Reach_disc} give the projected reaches for $Z=3$ evidence and $Z=5$ discovery at DUNE, JUNO, and Hyper-Kamiokande.
And the top (bottom) panel of Figure \ref{fig:THEIA_Reach} shows the projected 95\% CL exclusion ($Z=5$ discovery) reaches at THEIA with various fiducial masses of the detector material.
As expected, for each planned experiment the reaches for exclusion are substantially higher than the corresponding reaches for a possible discovery.
We note that the prospects for a definitive $Z=5$ discovery are particularly modest after one takes into account the limits already obtained by Super-Kamiokande. 

The results shown in Figures~\ref{fig:Reach_excl}, \ref{fig:Reach_disc}, and \ref{fig:THEIA_Reach}
are preliminary estimates, as the presently available  background and signal efficiency estimates vary significantly in their reliability, and
more robust estimates will become available only when the experiments are closer to collecting data. For the same reason, the results should be viewed with some caution as a direct comparison of the different experiments, which are at very different stages of planning and development.

Proton decay experiments prior to Super-Kamiokande have ruled out
the the simplest variations of minimal $SU(5)$ GUT
\cite{Georgi:1974sy}, and 
Super-Kamiokande has seemingly ruled out the minimal supersymmetric $SU(5)$ GUT
\cite{Dimopoulos:1981zb,Dimopoulos:1981dw,Sakai:1981pk,Hisano:1992jj}
with sfermion masses less than around the TeV scale.
However there are many other well-motivated GUT models
that predict proton partial lifetimes well beyond the current lower limits
(see summary tables in refs.~\cite{Dev:2022jbf,Bueno:2007um} and references therein).

For example,
non-supersymmetric GUTs such as
some minimally extended $SU(5)$ models \cite{Dorsner:2019vgf,FileviezPerez:2016sal} and
minimal $SO(10)$ model \cite{Babu:2015bna}
predict $p \rightarrow e^+ \pi^0$ to be the dominant decay mode
with partial lifetimes of order $10^{32} - 10^{36}$ years and $\lesssim 5 \times 10^{35}$ years, respectively.
Supersymmetric $SU(5)$ GUTs predict
the proton partial lifetime for the leading
mode $p \rightarrow \overline{\nu} K^+$ to be
$3 \times 10^{34} - 2 \times 10^{35}$ years in
minimal supergravity framework (MSUGRA) and $3 \times 10^{34} - 10^{36}$ years
in supergravity models with non-universal gaugino masses (NUSUGRA),
as discussed in ref.~\cite{Liu:2013ula} in the light of the observed Higgs
mass.
Ref~\cite{Ellis:2019fwf} revisited the minimal supersymmetric $SU(5)$ GUT
and obtained
$\tau_p/{\rm Br}(p \rightarrow \overline{\nu} K^+)
\lesssim
(2 - 6) \times 10^{34}$
years
assuming universality of the soft supersymmetry breaking parameters at the GUT scale with sfermion masses less than around $\mathcal{O}(10)$ TeV.
There are also supersymmetric GUTs such as
the split $SU(5)$ supersymmetry \cite{Arkani-Hamed:2004zhs}
and flipped $SU(5)$ supersymmetric GUTs \cite{Ellis:2002vk,Ellis:2021vpp,Ellis:2020qad},
where the dominant decay mode can be $p \rightarrow e^+ \pi^0$
with lifetimes of order $10^{35} - 10^{37}$ years.

From our estimates of the reaches summarized in Figures~\ref{fig:Reach_excl}, \ref{fig:Reach_disc}, and
\ref{fig:THEIA_Reach}, we can see that DUNE, JUNO, Hyper-Kamiokande,
and THEIA can probe a significant fraction of the parameter space of various presently viable supersymmetric and non-supersymmetric GUTs and could eventually
lead the way to a more complete theory.

The existing code repository {\sc Zstats} \cite{Zstats:v2}
is updated with
various statistical measures of significance for counting experiments
with multiple independent search channels
as investigated in this paper.
The updates include the significances based on our proposed Bayesian-motivated
measures $\CLdisc$ and $\CLexcl$,
and their application to study the statistical significances for proton decay at
current and future neutrino detectors.
To demonstrate the usage of the code, the repository also contains some code snippets
in a Python notebook that generate the data in each of the figures in this paper. 

{\it Acknowledgments:}
We thank Jennifer Raaf and Edward Kearns for helpful discussions.
P.N.B. thanks Tracy Slatyer for helpful discussions on Bayes factors and Zirui Wang for useful discussions on statistical combinations using RooStats package \cite{Moneta:2010pm}.
P.N.B. and J.D.W. thank Jianming Qian for useful discussions.
P.N.B. and J.D.W. also thank John Thiels for his help and support in using
the Great Lakes cluster at University of Michigan.
This research was supported in part through computational resources and services provided
by Advanced Research Computing (ARC), a division of Information and Technology Services
(ITS) at the University of Michigan, Ann Arbor.
This research was also supported in part through computational resources and services provided
by NICADD compute cluster at Northern Illinois University.
This work is supported in part by the National Science Foundation under grant number 2013340,
and by the Department of Energy under grant number DE-SC0007859.


\end{document}